\documentclass[acmsmall, screen]{acmart}
\usepackage{booktabs} % For formal tables
\usepackage{makecell}
\usepackage{lipsum}
\usepackage[utf8]{inputenc}
\usepackage{array}
\usepackage{wrapfig}
\usepackage{multirow}
\usepackage{tabularx}
\usepackage{pifont}
\usepackage{lineno}
\usepackage[table]{xcolor}
\usepackage{subfig}
\usepackage{enumerate}
\usepackage{longtable}
\usepackage{balance}
\usepackage{xcolor,pifont}

\usepackage{soul}

\usepackage[linesnumbered, ruled,vlined]{algorithm2e}
\RequirePackage{pdflscape}
\usepackage{flushend}
\usepackage{multicol}
\usepackage{placeins}
\usepackage{booktabs}
\usepackage{indentfirst}
\usepackage{fancybox}
\usepackage[most]{tcolorbox}
\usepackage{fontawesome}
\usepackage{mathtools}
\usepackage{MnSymbol,wasysym}
\usepackage{graphicx}

\usepackage[linesnumbered, ruled, vlined]{algorithm2e} 
\usepackage{algorithmic}

\usepackage{rotating}
\usepackage{adjustbox}
\usepackage{colortbl}
\usepackage{color}
\usepackage{tcolorbox}
\newtcolorbox{mybox}[2][]
{colback = white, colframe = black, fonttitle = \bfseries,
    colbacktitle = gray, enhanced,
    attach boxed title to top left={yshift=-3mm, xshift=3mm},
    title=#2, #1}

\usepackage{framed}
\definecolor{Gray}{gray}{0.9}
\definecolor{shadecolor}{gray}{0.95}
\usepackage{tikz}
\usetikzlibrary{trees,positioning,shapes,shadows,arrows}
\tikzset{
  basic/.style  = {draw, text width=2cm, drop shadow, font=\sffamily, rectangle},
  root/.style   = {basic, rounded corners=2pt, thin, align=center, fill=white},
  level-2/.style = {basic, rounded corners=6pt, thin,align=center, fill=white, text width=3cm},
  level-3/.style = {basic, thin, align=center, fill=white, text width=1.8cm}
}
\newcommand{\todo}[1]{}
\renewcommand{\todo}[1]{{\color{red} TODO: {#1}}}
\newcommand{\fakesection}[2][1em]{\vspace{#1}\noindent\textit{\textbf{#2}}}

\AtBeginDocument{%
  }

\setcopyright{acmlicensed}
\copyrightyear{2025}
\acmYear{2025}
\acmDOI{XXXXXXX.XXXXXXX}

\acmJournal{TOSEM}
\acmVolume{0}
\acmNumber{0}
\acmArticle{0}
\acmMonth{0}

\begin{document}

%\title{ArchISMiner: Automatic Mining of Architectural Issue-Solution Pairs from Online Developer Communities}
\title{ArchISMiner: A Framework for Automatic Mining of Architectural Issue-Solution Pairs from Online Developer Communities}
\author{Musengamana Jean de Dieu}
\email{mjados@outlook.com}
\affiliation{
  \institution{School of Computer Science, Wuhan University}
  \city{Wuhan}
  \country{China}
}

\author{Ruiyin Li}
\email{ryli_cs@whu.edu.cn}
%\authornote{Corresponding author}
\affiliation{
% \institution{School of Computer Science, Wuhan University}
\institution{School of Computer Science, Wuhan University}
  \city{Wuhan}
  \country{China}
}

\author{Peng Liang}
\email{liangp@whu.edu.cn}
%\authornote{Corresponding author}
\affiliation{
% \institution{School of Computer Science, Wuhan University}
\institution{School of Computer Science, Wuhan University}
  \city{Wuhan}
  \country{China}
}

\author{Mojtaba Shahin}
\email{mojtaba.shahin@rmit.edu.au}
\affiliation{
% \institution{School of Computing Technologies, RMIT University}
\institution{ School of Computing Technologies, RMIT University}
  \city{Melbourne}
  \country{Australia}
}

\author{Muhammad Waseem}
\email{muhammad.waseem@tuni.fi}
\affiliation{
%  \institution{Faculty of Information Technology and Communication Sciences, Tampere University}
 \institution{Faculty of Information Technology and Communication Sciences, Tampere University}
  \city{Tampere}
  \country{Finland}
}

\author{Zengyang Li}
\email{zengyangli@ccnu.edu.cn}
\affiliation{
% \institution{School of Computer Science, Central China Normal University}
\institution{School of Computer Science, Central China Normal University}
  \city{Wuhan}
  \country{China}
}

\author{Bangchao Wang}
\email{wangbc@whu.edu.cn}
\affiliation{
% \institution{School of Computer Science and Artificial Intelligence, Wuhan Textile University}
\institution{School of Computer Science and Artificial Intelligence, Wuhan Textile University}
  \city{Wuhan}
  \country{China}
}

\author{Arif Ali Khan}
\email{arif.khan@oulu.fi}
\affiliation{
% \institution{M3S Empirical Software Engineering Research Unit, University of Oulu}
\institution{M3S Empirical Software Engineering Research Unit, University of Oulu}
  \city{Oulu}
  \country{Finland}
}

\author{Mst Shamima Aktar}
\email{shamima@whu.edu.cn}
\affiliation{
% \institution{School of Computer Science, Wuhan University}
\institution{School of Computer Science, Wuhan University}
  \city{Wuhan}
  \country{China}
}

%% headers. Often, this list is too long and will overlap
\renewcommand{\shortauthors}{Jean de Dieu et al.}

\begin{abstract}
Stack Overflow (SO), a leading online community forum, is a rich source of software development knowledge. However, locating architectural knowledge, such as architectural solutions remains challenging due to the overwhelming volume of unstructured content and fragmented discussions. Developers must manually sift through posts to find relevant architectural insights, which is time-consuming and error-prone. This study introduces \textbf{ArchISMiner}, a framework for mining architectural knowledge from SO. The framework comprises two complementary components: \textit{ArchPI} and \textit{ArchISPE}. ArchPI trains and evaluates multiple models, including conventional ML/DL models, Pre-trained Language Models (PLMs), and Large Language Models (LLMs), and selects the best-performing model to automatically identify Architecture-Related Posts (ARPs) among programming-related discussions. ArchISPE employs an indirect supervised approach that leverages diverse features, including BERT embeddings and local TextCNN features, to extract architectural issue-solution pairs. Our evaluation shows that the best model in ArchPI achieves an F1-score of 0.960 in ARP detection, and ArchISPE outperforms baselines in both SE and NLP fields, achieving F1-scores of 0.883 for architectural issues and 0.894 for solutions. A user study further validated the quality (e.g., relevance and usefulness) of the identified ARPs and the extracted issue-solution pairs. Moreover, we applied ArchISMiner to three additional forums, releasing a dataset of over 18K architectural issue-solution pairs. Overall, ArchISMiner can help architects and developers identify ARPs and extract succinct, relevant, and useful architectural knowledge from developer communities more accurately and efficiently. The replication package of this study has been provided at \url{https://github.com/JeanMusenga/ArchISPE}.
\end{abstract}




\maketitle

\section{Introduction}\label{introduction}
Online developer communities provide a platform for developers to exchange ideas, discuss challenges, foster community engagement, and seek technical support for development-related issues \cite{naghshzan2021leveraging}. These communities have become a vital element of contemporary software development, serving both open-source communities with globally distributed contributors \cite{chen2024empirical, huang2022towards, alnusair2018utilizing} and private companies aiming to enhance in-house team productivity \cite{soliman2021exploring}. As the most prominent development community, Stack Overflow (SO) has attracted the attention of software engineers to learn, practice, and utilize development knowledge. As of June 2025, this platform owns over 29 million registered users\footnote{\url{https://stackexchange.com/sites?view=list\#users}}, and has received over 24 million questions and 36 million answers, solidifying its role as a valuable knowledge repository.

\begin{figure}[h]
\centering
\subfloat[An architecture-related post on SO]{%
    \includegraphics[width=0.53\linewidth, height=0.34\linewidth]{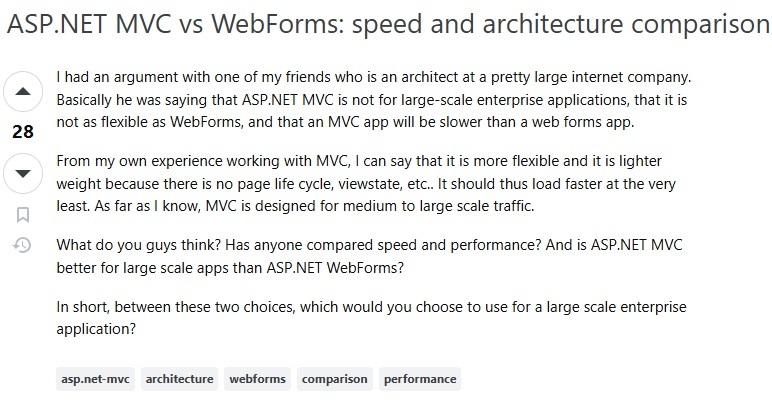}
}\hfill
\subfloat[A programming-related post on SO]{%
    \includegraphics[width=0.42\linewidth, height=0.34\linewidth]{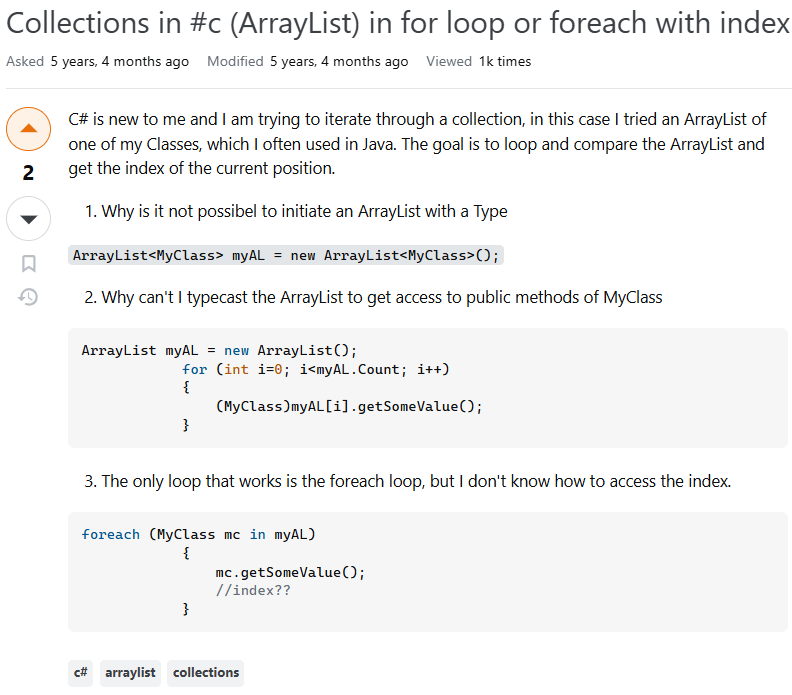}
}
\caption{Examples of an ARP and a programming-related post on SO.}
\label{ExamplesOfARPs_and_PRP}
\end{figure}

Previous studies have shown that developers often rely on SO to acquire development knowledge by searching through its vast repository of existing answers/solutions, such as learning API usage~\cite{uddin2020mining, silva2019recommending}, fixing bugs~\cite{mahajan2020recommending}, or discovering technology trends~\cite{barua2014developers}. SO is primarily used by developers for resolving coding-related problems~\cite{treude2011programmers}. For instance, the post \href{https://stackoverflow.com/q/58671354/12381813}{SO\#58671354} titled “\textit{Collections in \#c (ArrayList) in for loop or foreach with index}” (see Figure~\ref{ExamplesOfARPs_and_PRP}(b)) addresses a programming-level concern that is typically less relevant to software architects, as it focuses on low-level implementation details. However, architects can also benefit from online developer communities like SO when addressing higher-level architectural issues/problems. Recent studies~\cite{de2023characterizing, soliman2017developing, de2024users, bi2021mining, jean2024mining} confirm that SO hosts a wide range of architectural problems and corresponding solutions, which have proven useful in practice~\cite{de2023characterizing}. An example of such a post is \href{https://stackoverflow.com/q/1035642/12381813}{SO\#1035642} “\textit{ASP.NET MVC vs WebForms: speed and architecture comparison}” (see Figure~\ref{ExamplesOfARPs_and_PRP}(a)), which compares two application frameworks from the architecture concerns perspective.

In this study, we specifically focus on Architecture-Related Posts (ARPs) shared on SO. These posts contain high-level knowledge essential to software development. Examples include architectural issues (e.g., cyclic dependencies \cite{feng2024empirical}, modularity violations \cite{baldwin2000design}, and architectural configuration issues \cite{de2023characterizing}) and solutions (e.g., architectural patterns and tactics \cite{de2023characterizing, soliman2017developing, de2024oss, jean2024mining, de2024users}) that represent critical development-related knowledge \cite{SA2012}. However, the variety of SO posts, ranging from general programming queries to in-depth architectural discussions, makes it challenging to navigate and compare different answers and solutions. Developers often invest considerable time and effort in manually extracting architectural knowledge from unstructured and noisy data. 

To address the above challenges, we proposed \textbf{ArchISMiner} (\textbf{Arch}itectural \textbf{I}ssue \textbf{S}olution \textbf{Miner}), an automatic framework designed to mine architectural knowledge from Q\&A platforms. \textit{ArchISMiner} comprises two main components: (1) \textbf{ArchPI} (\textbf{Arch}itectural Post \textbf{I}dentifier), leverages a suite of models, including traditional Machine Learning (ML), Deep Learning (DL), Pre-trained Language Models (PLMs), and Large Language Models (LLMs), to automatically identify ARPs and distinguish them from programming-related posts; (2) \textbf{ArchISPE} (\textbf{Arch}itectural \textbf{I}ssue \textbf{S}olution \textbf{P}air \textbf{E}xtractor), is an indirectly supervised method that extracts architectural issue-solution pairs from ARPs. It combines BERT embeddings, local features derived from TextCNN, linguistic patterns, and heuristic features to address the diverse characteristics of Q\&A content and enrich data representation. We evaluated \textit{ArchISMiner} (including both ArchPI and ArchISPE) through a combination of automated and user-based evaluations. For the automated evaluation, a benchmark dataset of architectural issue-solution pairs from developer communities is essential. However, to the best of our knowledge, no such dataset currently exists. To address this gap, we constructed \textbf{ArchISPBench}, a benchmark dataset comprising sentences extracted from ARPs on SO. The evaluation results show that the best-performing model within \textit{ArchPI} achieves a Precision of 0.963, a Recall of 0.956, an F1-score of 0.960, and an Accuracy of 0.960 for ARP identification. Similarly, \textit{ArchISPE} achieves a Precision of 0.884, a Recall of 0.885, and an F1-score of 0.883 for architectural issue extraction, surpassing baseline methods from both the Software Engineering (SE) and Natural Language Processing (NLP) fields. For architectural solution extraction, \textit{ArchISPE} obtains a Precision of 0.898, a Recall of 0.892, and an F1-score of 0.894, again outperforming the baselines in both SE and NLP fields. Moreover, findings from the user study demonstrate that the ARPs identified by \textit{ArchPI} and the architectural issue-solution pairs extracted by \textit{ArchISPE} are perceived as both relevant and useful in supporting software development tasks. ArchISMiner aims to help architects capture concise, relevant, and actionable architectural knowledge shared in online developer communities more accurately and efficiently. It also supports architects' decision-making by utilizing architectural knowledge shared by the community. 
The main \textbf{contributions} of our study are:

\begin{enumerate}
\item We defined the problem of mining architectural knowledge from online Q\&A community data and introduced \textbf{ArchISMiner}, a framework that incorporates two specialized learning methods to effectively address this problem.

\item We developed \textbf{ArchISPBench}, a high-quality benchmark dataset curated to enable the automatic evaluation of methods for extracting architectural issue-solution pairs from Q\&A communities. It contains 367 architectural issues and 1,967 solutions extracted from SO, with 5,234 sentence candidates (1,970 from questions and 3,264 from answers). This benchmark provides a reusable resource for future studies, enabling consistent evaluation of new approaches or tools. 

\item We conducted a comprehensive evaluation of ArchISMiner using both automated and user-based assessments. The results confirm the effectiveness and practical utility of ArchISMiner. Furthermore, we release an open-source replication package that includes a benchmark dataset of architectural issue–solution sentences and a large dataset comprising over 18K architectural issue–solution pairs extracted from Q\&A sites \cite{datasetTOSEM}. 
\end{enumerate}

The remainder of this paper is organized as follows. Section~\ref{Research_Gap_Challenges} presents the research gaps and challenges. Section~\ref{relatedwork} reviews the related work, while Section~\ref{Methodology} outlines the research methodology. Section~\ref{Our_Proposed_ArchISMiner_Framework} introduces the proposed framework. Section~\ref{Experiment_Results} reports the study results, followed by further discussion and implications in Section~\ref{Implications}. Section~\ref{ThreatsValidity} discusses potential threats to validity, and Section~\ref{Conclusion} concludes the paper with future research directions. %Finally, Section~\ref{Data_Availability} describes the data availability.

\section{Research Gaps and Challenges}\label{Research_Gap_Challenges}
In this section, we discuss the current research gaps and challenges in extracting architectural issue–solution pairs from SO posts. Developers often turn to SO to seek architectural knowledge, as it hosts a wealth of architecture-related discussions. However, the unstructured nature of SO posts and the overwhelming volume of questions and answers make it challenging to locate relevant insights. Many posts are lengthy, requiring developers to sift through extensive content to identify useful information. In a survey of 72 developers from two IT companies, Xu \textit{et al.}~\cite{xu2017answerbot} found that ``\textit{answers in long posts are hard to find}''. They also reported that nearly 6.5 million questions (37\% of all questions) have multiple answers, with an average answer length of 789 characters. Additionally, our previous work~\cite{de2023characterizing, de2024users} revealed that not all sentences within a post contribute meaningfully to the discussion. For example, the sentence \textit{``This is an interesting question, (...)''} in \href{https://stackoverflow.com/a/58509043/12381813}{SO \#58509043} adds little value to the answer, while \textit{``Firstly, I apologize for the rather basic question (...)''} in \href{https://stackoverflow.com/q/73646438/12381813}{SO \#73646438} is similarly unhelpful in clarifying the issue. Existing studies (e.g., \cite{xu2017answerbot, nadi2020essential, yang2023techsumbot}) have primarily focused on the summarization of development knowledge at a low level of abstraction (e.g., code-level solutions), rather than targeting at higher levels, such as architectural problems and solutions~\cite{de2023characterizing, de2024oss, de2024users}. Architectural knowledge shared on Q\&A sites like SO is important, and developers often reuse it to reason about architectural design decisions, select suitable architectural patterns or tactics, and understand trade-offs among architectural alternatives. However, little attention has been paid to systematically identifying and extracting architectural knowledge, leaving a gap in supporting developers' access to high-level design insights from Q\&A discussions.

\textbf{Scenario: Architectural Decision-Making for Authentication in Microservice Systems}. Consider Augustin, a software architect designing a microservices-based system, who must choose a secure and scalable authentication mechanism that fits microservices principles. To explore viable architectural tactics, he searches for \textit{``Microservice Authentication Strategy''} on SO and discovers the ARP \href{https://stackoverflow.com/q/29644916/12381813}{SO \#29644916}, which discusses various authentication strategies. The ARP contains five answers, each proposing different authentication solutions. To optimize his decision-making process, Augustin begins by inspecting the accepted answer \href{https://stackoverflow.com/a/29675246/12381813}{SO \#29675246}, which recommends adopting \textit{OAuth 2} protocol to structure authentication within a microservices architecture. While this solution is widely recognized, the post lacks architectural depth, as highlighted by a commenter who asks: \textit{``Your answer is great, but how does the token generated from the API Gateway (from inside of it, or in an AuthMicroService) can be handled by a random microservice? (...)?}'', indicating missing details on the authentication flow and token lifecycle management. Seeking a more comprehensive architectural perspective, Augustin examines other highly voted responses. However, he quickly encounters challenges: some answers are lengthy, introduce tangential discussions, or lack a structured explanation of architectural trade-offs. For instance, the third most-voted answer presents an authentication model based on \textit{JWT tokens}, {OAuth 2.0}, and {OpenID Connect}, outlining key architectural components such as authentication services, token generation, and client authentication flows. However, the architectural rationale is embedded within multiple paragraphs and an accompanying diagram, making it difficult to extract core insights efficiently. Due to time constraints, like many architects, Augustin might only skim a few answers, potentially missing valuable architectural alternatives in lower-ranked responses.

\begin{figure}[!t]
 \centering
 \includegraphics[width=0.85\textwidth]{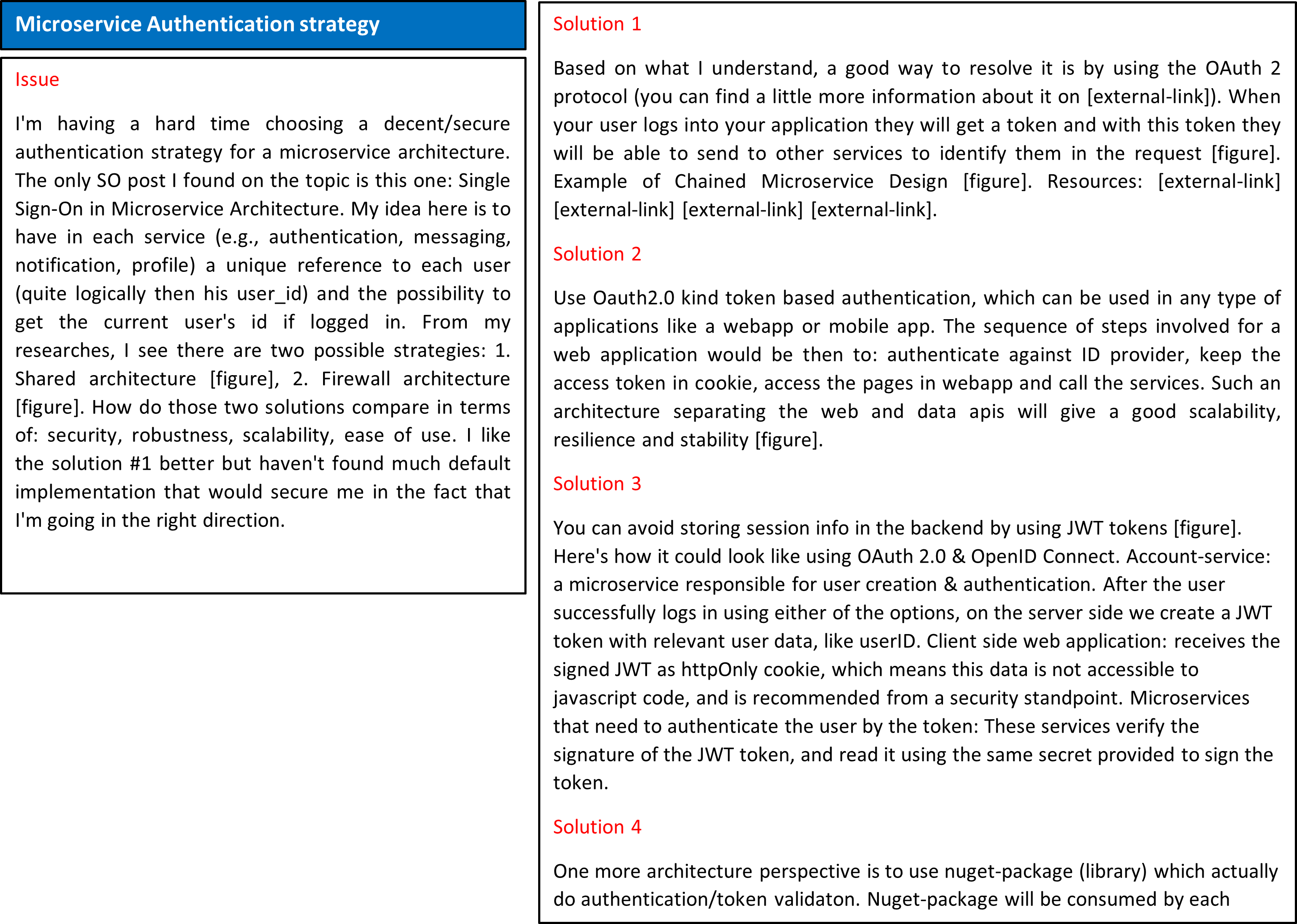}
  \caption{An example of architectural issues and solutions extracted by our proposed approach from the \href{https://stackoverflow.com/q/29644916/12381813}{SO post \#29644916}}
 \label{Issue_Solution_Extraction_By_ArchISPE}
\end{figure}

\textbf{Benefits of the Proposed Framework in Addressing Challenges Faced by Augustin}. To address the aforementioned research gap and challenges, our framework assists Augustin by automatically extracting key sentences that describe architectural solutions from each answer post and linking them to relevant sentences that articulate the corresponding architectural issue. For instance, Figure \ref{Issue_Solution_Extraction_By_ArchISPE} illustrates an example of the architectural issue–solution pair extracted from the same SO post that Augustin discovered. The extracted pair is concise, self-contained, and structures critical information from both questions and answers, enabling Augustin to quickly review multiple posts and prioritize those worth deeper investigation. Consequently, he can efficiently analyze design options, assess trade-offs, and make well-informed architectural decisions. Ultimately, the framework supports Augustin in selecting the most appropriate architectural tactics (e.g., authentication strategies) and supporting technologies (e.g., frameworks, protocols, and tools) that align with his system’s quality requirements, such as security and scalability.

\section{Related Work}\label{relatedwork}
This section reviews the related work on mining architectural knowledge (Section \ref{Mining_ArchitecturalInformationOnlineCommunities}) and issue-solution extraction from online community forums (Section \ref{IssueSolution_Extraction_from_Online_Communities}).

\subsection{Mining Architectural knowledge from Online Community Forums}\label{Mining_ArchitecturalInformationOnlineCommunities}
Several studies have explored architectural knowledge provided in ARPs on SO from various perspectives. Bi \textit{et al.} \cite{bi2021mining} employed a semi-automatic method to extract discussions on Quality Attributes (QAs) and Architecture Tactics (ATs) in SO posts. They utilized a dictionary-based classifier, to automatically identify QA-AT-related discussions and subsequently structured the design relationships between ATs and QAs used in practice. Their results assist architects in making informed AT design decisions. Chinnappan \textit{et al.} \cite{chinnappan2021architectural} mined architectural tactics for energy-efficient robotics software by analyzing data from five open-source software repositories (including SO). To promote the broad applicability of the identified tactics, they described the tactics in a generic, implementation-independent manner using diagrams inspired by UML component and sequence notations. These energy-aware tactics provide valuable guidance for roboticists and other developers designing energy-efficient software. Wijerathna \textit{et al}. \cite{wijerathna2022mining} examined the relationship between design contexts and patterns on SO, introducing a mapping framework to match patterns with design contexts based on developer discussions. Soliman~\textit{et al}.~\cite{soliman2017developing} developed an ontology that covers architecture knowledge concepts in SO. The ontology provides a description of architecture-related information to represent and structure architectural knowledge in SO. Tian \textit{et al.} \cite{tian2019developers} analyzed SO discussions to investigate users' perceptions of architectural smells. Their study revealed that users often describe architectural smells with broad terms such as “bad”, “wrong”, and “brittle”, or as violations of architectural patterns. Li \textit{et al.} \cite{li2021understanding} studied architecture erosion by analyzing data from eight popular online developer communities (including SO). Their findings indicate that developers perceive architecture erosion either through its structural manifestations or its impact on runtime qualities, maintenance, and evolution. Musengamana \textit{et al}. \cite{de2023characterizing} analyzed ARPs, including questions and solutions, on SO. They categorized these posts into various types, such as architectural configuration, architecture implementation, architecture evolution. Additionally, they assessed the usefulness of these posts from the perspective of SO users. Aktar \textit{et al}.~\cite{aktar2025architecture} extracted SO posts and GitHub issues to investigate architectural decisions as well as the challenges developers face when making such decisions in quantum software development.

While these studies have provided valuable insights into architectural discussions on SO, none explicitly address two fundamental challenges. First, no prior work automatically distinguishes architectural-related posts (ARPs) from general programming content - a prerequisite for large-scale architectural knowledge mining. Second, most prior approaches focus solely on extracting architectural solutions while neglecting the underlying problems they address. Without this context, the extracted knowledge remains incomplete and less actionable. Our work bridges this gap by automatically identifying ARPs and mining architectural issue-solution pairs, thereby linking architectural problems with their corresponding solutions. This dual perspective provides a more holistic and context-aware understanding of architectural knowledge in developer communities, complementing and extending previous efforts.
%While these studies provide valuable insights into specific architectural discussions on SO, our work diverges in two key ways. First, to the best of our knowledge, no prior work has explicitly addressed the challenge of automatically distinguishing ARPs from other posts, such as general programming-related content, in online developer communities, an essential precursor for accurate mining of architectural knowledge at scale. Second, although existing studies emphasize extracting one asset of architectural knowledge (e.g., architectural solutions), they often overlook the problem space, that is, the identification and extraction of architectural issues that solutions aim to address. Without this context, the practical relevance and applicability of solutions may be diminished. Our study fills this critical gap by automatically identifying ARPs and further mining issue-solution pairs from developer discussions. This dual focus enhances the utility of mined knowledge by pairing architectural solutions with the problems they address, thereby offering more actionable insights for software architects and developers. In doing so, our approach complements and extends prior work by advancing toward a more complete and context-aware understanding of architectural knowledge shared in online developer communities. 

\subsection{Issue-Solution Extraction from Online Community Forums}\label{IssueSolution_Extraction_from_Online_Communities}
\textbf{Emerging issue detection from short-text platforms}. Several studies have focused on identifying emerging issues from app stores (e.g., Google Play Store) and short-text social media platforms (e.g., Twitter) using traditional anomaly detection methods. Guo \textit{et al}. \cite{guo2020caspar} proposed a method for extracting and synthesizing user-reported mini-stories regarding app problems from user reviews. Leveraging NLP techniques, their approach classifies ordered events extracted from app reviews as either user actions or app problems, synthesizing them into action-problem pairs. Their evaluation demonstrated the method's effectiveness in identifying pairs from app reviews. Similarly, Gao \textit{et al}.~\cite{gao2018online} introduced a topic-labeling approach, \textbf{IDEA}, to automatically detect emerging issues in app versions based on statistical patterns from previous releases. They evaluated IDEA on six popular apps from Google Play and Apple’s App Store, using official app changelogs as ground truth. Results showed that IDEA effectively identified emerging app issues, with 88.9\% of the detected issues deemed valuable for app development according to feedback from engineers and product managers. To overcome the inherent limitations of topic modeling (e.g., the need to predefine the number of topics), their subsequent work~\cite{gao2019emerging} proposed \textbf{DIVER}, which integrates a depth-first pattern mining strategy to identify emerging issues by comparing changes across app versions and time periods. Shrestha \textit{et al}. \cite{shrestha2004detection} proposed a rule-based method for identifying issues/questions in email threads. By training a set of if-then rules to predict questions and corresponding answers, they demonstrated that structural features of email threads could enhance the lexical similarity of discourse segments, improving issue-solution pairing accuracy. Henß \textit{et al}. \cite{henss2012semi} introduced an approach for semi-automatically extracting frequently asked programming questions from development mailing lists. They conducted both qualitative and quantitative evaluations of the proposed approach, and the evaluation results confirm the promising quality of the generated frequently asked programming questions.

Previous work has focused on (1) low-level, code-related issues and (2) short-textual datasets (e.g., social media, app stores, emails), using methods tailored to those domains. While effective for concise and domain-specific texts, such approaches are ill-suited for extracting high-level architectural discussions that are longer, context-rich, and semantically nuanced. In contrast, our study focuses on extracting architectural issue-solution pairs from SO posts, which capture real-world design concerns and rationale. To facilitate reproducible evaluation, we introduce \textit{ArchISPBench}, a curated benchmark dataset of architecturally significant issue-solution pairs. This benchmark bridges a critical gap in the literature by enabling systematic comparison of automated extraction techniques and advancing the broader goal of mining actionable architectural knowledge from large-scale developer discussions. Ultimately, our work broadens the scope of software engineering research by surfacing emerging architectural challenges and their remedies, thereby accelerating issue resolution, promoting knowledge reuse, and empowering developers to make well-informed architecture decisions.
%Previous work has focused on (1) low-level, code-related issues and (2) short-textual datasets (e.g., social media, app stores, emails). In contrast, our research targets high-level architectural artifacts by extracting architectural issue–solution pairs from a novel source: SO posts on architectural design. Prior methods were tailored to specific datasets and thus ill-suited for this task. These approaches utilize the characteristics of their corpora and are best fit for their specific tasks, so they cannot directly transform their methods to the task of extracting architectural issue-solution pairs from SO. Our approach generalizes across multiple online Q\&A forums (e.g., Stack Overflow, ServerFault, and Software Engineering Stack Exchange) to automatically extract architectural problems and their solutions. To support this, we introduce ArchISPBench, a curated benchmark dataset of architecturally significant issue–solution pairs derived from real-world developer discussions. ArchISPBench enables the evaluation of automated techniques in extracting architectural issue-solution pairs, bridging a critical gap in the literature. This broadens the scope of software-engineering research and surfaces emerging architectural challenges alongside effective remedies. By highlighting resolved architectural issues, we accelerate issue resolution, promote reuse of architectural knowledge, and empower developers to make well-informed design decisions.

\textbf{Community-based programming post summarization}. Several approaches have been proposed to summarize programming questions and answers on SO into concise texts to facilitate post navigation~\cite{xu2017answerbot, nadi2020essential, wang2021automatic, yang2023techsumbot}. AnswerBot extracts summative paragraphs from Java-related posts based on features such as information entropy and paragraph position~\cite{xu2017answerbot}. Similarly, CraSolver employs a multi-factor ranking mechanism to summarize bug solutions provided in Java and Android-related questions on SO~\cite{wang2021automatic}. Both AnswerBot and CraSolver generate summaries at the paragraph level. Nadi and Treude~\cite{nadi2020essential} experimented with four different IR-based techniques to select essential sentences from JSON-tagged posts on SO. Through a survey with 43 developers, they found that although participants desired navigation support for browsing SO, the IR-based approaches failed to provide such support. 

Prior summarization approaches primarily target low-level, code-oriented content and rely on IR-based or heuristic techniques~\cite{nadi2020essential}. Such methods struggle to capture the semantic richness and ambiguity of architectural discussions. Unlike these efforts, our work moves beyond summarization toward structured knowledge extraction, leveraging a multi-feature, deep learning-based framework to accurately identify and pair architectural issues and solutions within complex, natural language discussions.
%Unlike prior work, which primarily focuses on summarizing low-level code-related content, our research targets higher-level artifacts by emphasizing the extraction of architectural knowledge. These existing approaches cannot be directly applied to the extraction of architectural issue-solution pairs, which presents a fundamentally different problem setting. Moreover, the prior studies relied on techniques based solely on IR methods or heuristics (e.g.,~\cite{nadi2020essential}) for text summarization, which are insufficient to capture the ambiguity and sophistication of natural language text shared on SO. Motivated by these limitations, we propose a novel multi-features and DL-based framework that can more accurately extract important sentences from SO ARPs. 

\section{Methodology}\label{Methodology}
\begin{figure}[!t]
 \centering
\includegraphics[width=0.95\textwidth]{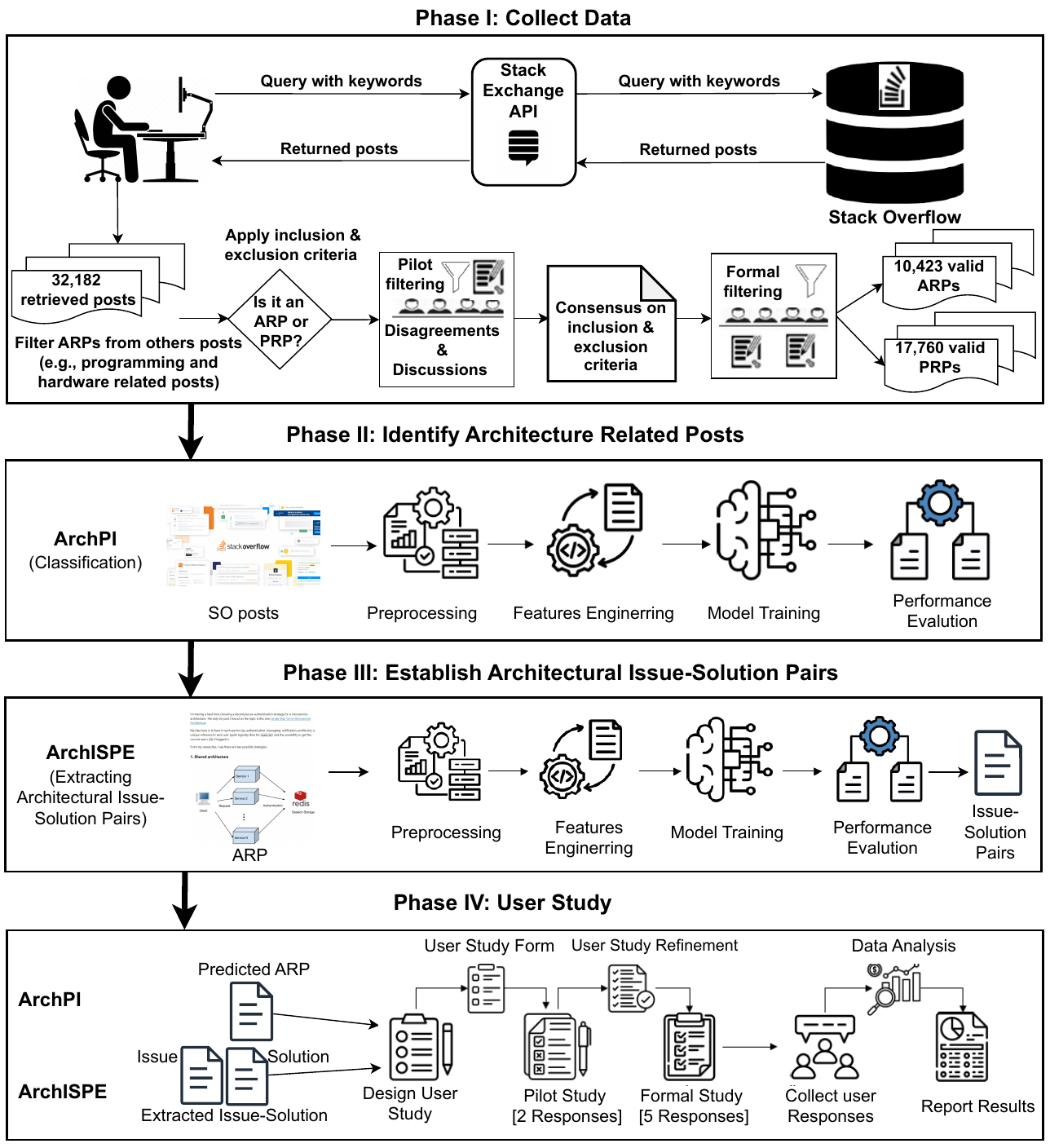}
\caption{Overview of the research methodology}
\label{fig_OverviewOftheReseachMethod}
\end{figure}

This section presents the study goals, Research Questions (RQs), data collection, and an overview of the research methodology (Figure~\ref{fig_OverviewOftheReseachMethod}).

\subsection{Goal and Research Questions}\label{Goal_ResearchQuestions}
The \textbf{goal} of this study is to propose an automatic framework to identify architecture-related posts and establish architectural issue-solution pairs from the identified posts for task-specific architecture problems (e.g., performance bottleneck in real-time systems) from the online developer community, i.e., SO. This approach aims to assist engineers, especially architects and developers, in capturing relevant architectural knowledge accurately and efficiently. Additionally, it supports engineers in architecture decision-making by leveraging architectural knowledge shared by their peers. To this end, we formulate the following three RQs:

\begin{tcolorbox} \textbf{RQ1.} How can we automatically identify architecture-related posts in online developer communities and distinguish them from programming-related posts?
\end{tcolorbox}

\textbf{Motivation}: Online developer communities contain rich development-related information, but their unstructured nature makes it difficult to distinguish between content types, such as architectural and programming-related discussions. The lack of explicit and automated methods for identifying ARPs and differentiating them from programming-related posts remains an unresolved challenge~\cite{de2023characterizing}. Although the manual classification of ARPs has provided valuable insights into architectural discussions in these communities~\cite{tian2019developers, de2023characterizing}, such methods are not scalable for large datasets. Recent studies~\cite{bi2021mining, wijerathna2022mining} have explored automated machine learning and deep learning techniques to mine architectural knowledge from developer communities, but they do not specifically address the distinction between ARPs and programming posts. Building on insights from our recent exploratory study~\cite{de2023characterizing}, we propose \textit{ArchPI}, an automated approach for identifying ARPs and distinguishing them from programming-related content. Automating this process lays the foundation for more advanced mining techniques, such as extracting architectural issue-solution pairs, ultimately improving developers' access to and use of architectural knowledge.

\begin{tcolorbox} \textbf{RQ2.} How can we automatically establish architectural issue-solution pairs in online developer communities?
 \end{tcolorbox}

\textbf{Motivation}: Developers frequently share architectural knowledge in online community forums by posting architectural issues. However, these posts can be lengthy, requiring developers to review multiple discussions to find relevant information. Given the vast amount of development-related content on SO, quickly identifying key architectural insights is challenging. A survey by Xu \textit{et al}. \cite{xu2017answerbot} underscores this challenge, as participants reported that sifting through numerous posts is cognitively demanding and time-consuming. %A recent survey of 72 professional developers \cite{xu2017answerbot} highlights this challenge, with participants reporting that sifting through numerous posts is cognitively demanding and time-consuming. 
Many expressed a need for tools to streamline this process. To address these challenges, we propose an automated approach, \textit{ArchISPE}, to extract architectural issue-solution pairs. This approach aims to enhance navigation and improve developer productivity by efficiently surfacing relevant architectural insights.

\begin{tcolorbox} 
\textbf{RQ3.} How do practitioners perceive automatically identified ARPs and extracted issue-solution pairs?
\end{tcolorbox}

\textbf{Motivation}: Automated methods for identifying ARPs and extracting architectural issue-solution pairs from SO have the potential to assist practitioners by surfacing actionable architectural knowledge from vast amounts of discussions. However, the practical value of such methods hinges on how practitioners perceive their quality and effectiveness in real-world contexts. This RQ seeks to assess our proposed framework through a user study, focusing on practitioners' perceptions of the relevance, comprehensiveness, and usefulness of the identified ARPs and extracted issue-solution pairs in supporting development tasks.

\subsection{Data Collection}\label{Data_Collection}
A dataset was required to develop and evaluate our proposed framework. To construct the dataset, we utilized posts curated in our recent study \cite{de2023characterizing}, as summarized in \textbf{Phase I} of the research method and illustrated in Figure~\ref{fig_OverviewOftheReseachMethod}. Specifically, we followed a two-step process to build a dataset of ARPs and programming-related posts from SO. First, we retrieved data using SQL queries via the StackExchange Data Explorer\footnote{\url{https://data.stackexchange.com/stackoverflow/query/new}}, a Web API that allows the execution of SQL queries across various Stack Exchange Q\&A sites (including SO). More specifically, we extracted 32,182 posts from SO. Subsequently, we manually identified 10,423 candidate ARPs through a combination of thorough pilot and formal post-filtering processes. Details of the retrieval and filtering procedures (e.g., pilot and formal keyword searches) are described in our previous study \cite{de2023characterizing}. 

The remaining 21,759 posts contained considerable noise and many invalid or non-informative posts. In this study, we revisited these posts to identify potential programming-related posts for our experimental setup (see Section~\ref{ARPs_Indetification_Component}). Following our previous work \cite{de2023characterizing}, we defined and applied specific inclusion and exclusion criteria to further filter the collected posts and isolate programming-related posts:

\begin{enumerate}[(i)]
    \item Posts discussing low-level design or code implementation (e.g., “sorted array vs unsorted array” (\href{https://stackoverflow.com/q/11227809/12381813}{SO \#11227809})) were included.
    \item Posts with a score lower than 1 (i.e., insufficient attention from the community \cite{UnderstQuestQuali2014}) were excluded.
\end{enumerate}

Two annotators (the first and ninth authors) conducted the data filtering process. They manually reviewed 21,759 posts according to the above criteria. Initially, the two authors performed a pilot ARP filtering. They independently and manually examined a random sample of 1,000 posts from the 21,759 posts. To measure the inter-rater agreement, we calculated Cohen's Kappa coefficient \cite{cohen1960}, yielding a substantial agreement of 0.887. Any disagreements during the filtering process were resolved through a negotiated agreement approach \cite{campbell2013coding} to ensure the reliability of the data filtering. Following the pilot, the two authors sequentially reviewed the remaining posts in batches of 1,000 until all were examined. Ultimately, we collected 17,760 programming-related posts (see \textbf{Phase I} in Figure \ref{fig_OverviewOftheReseachMethod}). The third author verified the final results, and the data filtering process took approximately 264 hours. 

Building on a previous study \cite{uddin2020mining}, we selected subsets of ARPs and programming-related posts via random sampling to ensure a diverse representation of development scenarios in online forums, while avoiding bias toward any single architectural or programming context. Specifically, we randomly selected 7,466 ARPs from the original set of 10,423 ARPs identified in our previous study \cite{de2023characterizing}. This sample size is statistically representative, exceeding the requirement for a 99\% confidence level with a 3\% margin of error \cite{israel1992determining}. To construct a balanced dataset, essential for achieving robust prediction performance with standard machine learning algorithms \cite{he2009imbalance}, we also randomly selected 7,466 programming-related posts from the pool of 17,760. This balance mitigates generalization issues that arise from class imbalance \cite{he2009imbalance}. In total, our final dataset includes 14,932 posts (7,466 ARPs and 7,466 programming-related posts), which we used to train and evaluate the techniques supporting our proposed framework (see Section~\ref{ARPs_Indetification_Component} and Section~\ref{AutomaticExtractionArchitecturalIssue_solution}).

\section{ArchISMiner Framework}\label{Our_Proposed_ArchISMiner_Framework}
We developed a framework to mine architectural knowledge by analyzing ARPs from SO. Our framework comprises two complementary components, \textit{ArchPI} and \textit{ArchISPE}, which operate in two main phases: 1) \textit{Automatic Identification of Architecture-Related Posts}, and 2) \textit{Automatic Extraction of Architectural Issue-Solution Pairs}. The following sections describe the two key components of the framework and the experimental details.

\subsection{Automatic Identification of Architecture-Related Posts}\label{ARPs_Indetification_Component}
In this section, we present \textit{ArchPI}, the first component of our framework, and describe the experimental setup used to train and evaluate models for identifying ARPs from SO (Figure~\ref{fig_ARPs_Indetification_Component}). This step corresponds to \textbf{Phase II} of our research method (Figure~\ref{fig_OverviewOftheReseachMethod}).

\subsubsection{Potential Models for Identifying ARPs from Stack Overflow}\label{Potential_Techniques_for_Identifying_ARPs}
We employed a diverse set of models, including traditional ML, DL, state-of-the-art PLMs, and LLMs. We experimented with multiple models for several reasons: (1) the optimal model for our task was initially unknown, and prior studies indicate that feature-based ML techniques can sometimes outperform state-of-the-art PLMs or LLMs ~\cite{ezzini2022automated}; (2) while ML and DL models have demonstrated promising results in classification tasks, their effectiveness varies across different datasets~\cite{abdalkareem2020machine, aggarwal2012survey, caruana2006empirical}; and (3) models differ in their handling of overfitting, explainability, and run time~\cite{abdalkareem2020machine, kotsiantis2006machine}.  

\begin{figure}[!t]
 \centering
 \includegraphics[width=0.99\textwidth]{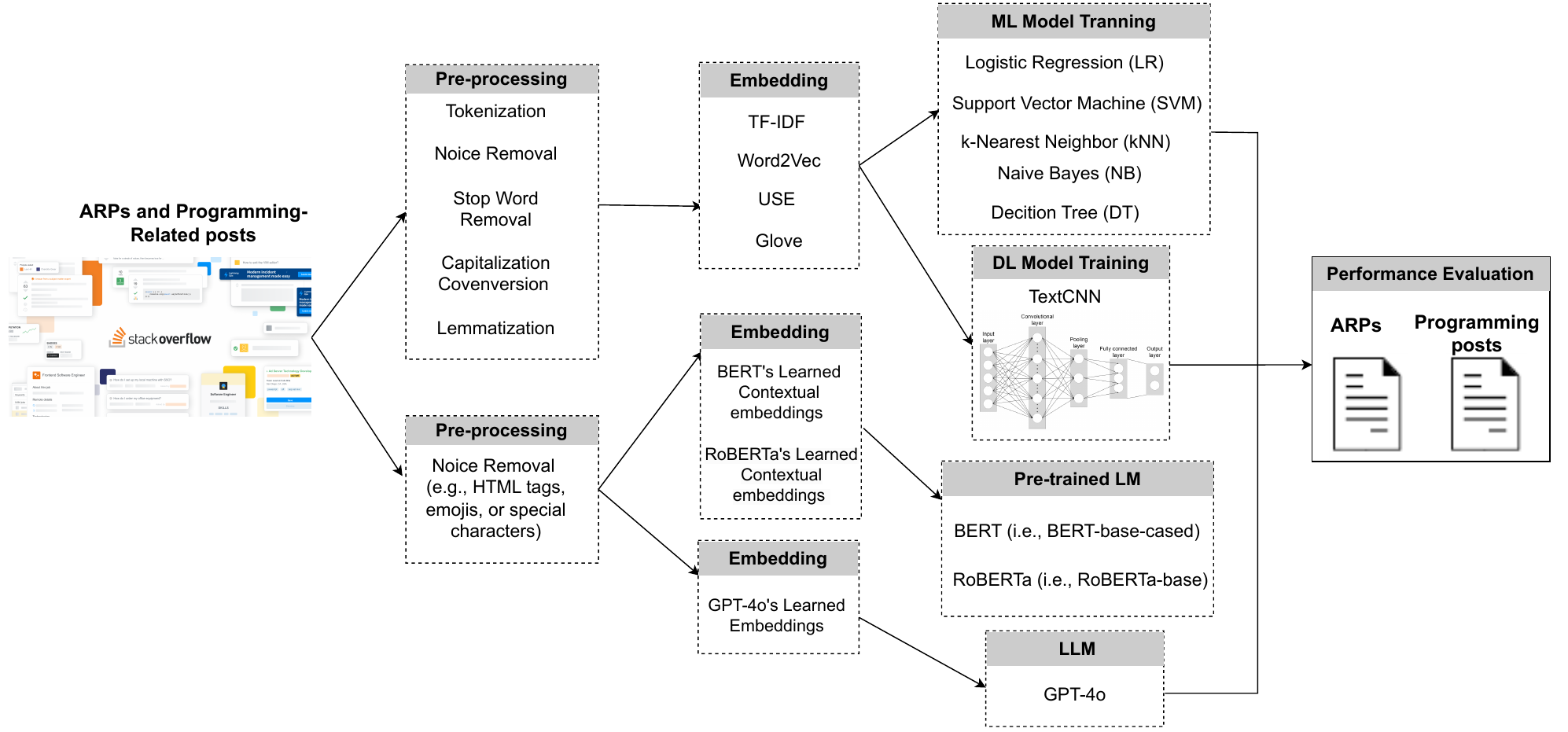}
 \caption{The structure of the experimental setup for identifying ARPs from SO}
 \label{fig_ARPs_Indetification_Component}
\end{figure}

Figure \ref{fig_ARPs_Indetification_Component} shows the experimental setup of model evaluations. First, we preprocess the collected posts into word- and sentence-level representations. Second, we generate dense vector embeddings from these representations. Third, we train and evaluate four classes of models, ML, DL, PLMs, and LLMs, to classify each post as ARP-related or programming-related. %The following subsections describe each step in detail for the different model types.

\textbf{A. Data Preprocessing}:\label{Data_Preprocessing} Before training traditional ML and DL models, we transformed the original text into a structured format suitable for model training through a preprocessing pipeline, including five key steps:

\textit{Step 1. Tokenization}: Tokenization is the process of splitting a stream of text into individual components, such as words, punctuation, and other meaningful elements known as tokens. According to each model specifications, we utilized libraries such as NLTK \cite{bird2009natural} and Hugging Face Transformers \cite{wolf2020transformers} to tokenize the posts, breaking the text into words, sub-words, or sentences as required by the models.

\textit{Step 2. Noise Removal}: Posts on SO often include extraneous elements such as whitespace, typos, punctuation, and special characters (e.g., ``/'', ``*'') that do not contribute valuable semantic information. To remove the noise, we applied heuristic techniques and retained only meaningful text.

\textit{Step 3. Stop Word Removal}: Stop words, such as ``the'', ``and'', ``or'', and ``if'', are common in text but contribute little semantic value. As they do not meaningfully differentiate content, they are typically excluded from analysis. We removed stop words using NLTK~\cite{bird2009natural} as well, following the requirements of each model.

\textit{Step 4. Capitalization Conversion}: To ensure consistency in word forms and avoid redundant counting of identical words with different cases, all textual data were converted to lowercase, standardizing word representation across the dataset.

\textit{Step 5. Lemmatization}: Lemmatization and stemming are both techniques used to reduce words to their root form. However, lemmatization is preferred over stemming because it considers the context and part of speech to ensure that the base form (the lemma) is a valid word \cite{balakrishnan2014stemming}. Stemming, in contrast, simply removes prefixes and suffixes to form a “stem”. Since lemmatization offers higher accuracy and more meaningful results, we used this technique to map words to their correct base form. For example, the word “better” would be lemmatized to “good” based on the context.

\textbf{B. Feature Engineering}: Feature engineering and embedding methods are crucial for converting unstructured text into structured representations, such as matrices, vectors, or encodings, as the inputs for models (e.g., ML models). We employed four widely adopted embedding methods to extract textual features and produce interpretable inputs for both traditional ML and deep models: TF-IDF, Word2Vec, Universal Sentence Encoder (USE), and Global Vectors for Word Representation (GloVe). These methods are commonly applied in NLP tasks due to their ability to capture both semantic and syntactic relationships among words. Below, we briefly describe each method and outline the configurations used to generate feature vectors from our dataset.

\begin{enumerate}
\item\textit{TF-IDF}: It calculates the importance of words within a text corpus \cite{ramos2003using}. Following the approach of Uddi \textit{et al.} \cite{uddin2017mining}, we configured TF-IDF to capture n-grams from SO posts, with \( n = 1, 2, 3 \) for unigrams (single words), bigrams (two consecutive words), and trigrams (three consecutive words). We applied TF-IDF to normalize the importance of both frequent and specific n-grams across the dataset.

\item \textit{Word2Vec}: Word2Vec is a word embedding technique to capture semantic and syntactic relationships between words \cite{mikolov2013distributed}. It takes words as input and outputs a word embedding matrix, where each word is represented by a high-dimensional vector.

\item \textit{GloVe}: The GloVe method \cite{pennington2014glove} generates word embeddings based on aggregated word co-occurrence statistics, capturing global relationships among words. Unlike Word2Vec, which focuses on local word co-occurrence, GloVe benefits from leveraging global co-occurrence data. We used the pre-trained GloVe word vectors derived from Twitter data (two billion tweets), with 200-dimensional embeddings.

\item \textit{USE}: The USE method operates at the sentence level, producing vector representations of sentences by utilizing pre-trained sentence embedding models \cite{cer2018universal}. We employed the Deep Averaging Network (DAN) encoder, which encodes posts into 512-dimensional vectors. The DAN encoder is particularly suited for large-scale tasks and is efficient in computational resources, making it an ideal choice for analyzing SO posts containing substantial amounts of text.
\end{enumerate}

\textbf{C. Models}. As mentioned above, we experimented with four types of models: ML, DL, state-of-the-art PLMs and LLMs. Below, we briefly describe each model type and the configurations applied.

\textit{i. ML Models}. We selected five widely used traditional ML models for the binary classification task in this study: Logistic Regression (LR), Decision Tree (DT), Support Vector Machine (SVM), Bernoulli Naive Bayes (NB), and k-Nearest Neighbor (kNN). These models have been commonly applied in classification tasks across both software engineering \cite{yang2022predictive} and other domains \cite{han2022data}. Among other libraries, we implemented these models using the scikit-learn library in Python 3.11.7. Following prior work \cite{bi2021mining, wijerathna2022mining, treude2016augmenting}, we used the default settings (including hyperparameters) for each ML model. Due to the large number of hyperparameter settings, we omit the specific values in the paper; all details have been provided in our replication package \cite{datasetTOSEM}.

\textit{ii. DL Models}. Deep learning models, particularly Convolutional Neural Networks (CNNs), are widely used in text classification tasks \cite{minaee2021deep, guo2019improving}. We adopted TextCNN \cite{kim2014cnn} due to its effectiveness and frequent use as a baseline model \cite{guo2019improving, liu2018recurrent}. Implemented in TensorFlow \cite{abadi2016tensorflow}, our model classifies ARPs by distinguishing them from programming-related discussions. The architecture comprises several layers, including an input layer for tokenized word sequences, an embedding layer to learn dense word representations, and three convolutional blocks (each consisting of a \texttt{Conv1D} and \texttt{MaxPooling1D} layer) with kernel sizes of 3, 4, and 5 for multi-scale n-gram feature extraction. The outputs are concatenated and passed through a flattening layer, followed by a dense layer (128 units, ReLU activation), dropout (rate = 0.5), and a sigmoid output layer for binary classification. We used the \texttt{Adam} optimizer and trained the model for 10 epochs with a batch size of 32, applying early stopping with a patience of 8 epochs. This model architecture enables effective extraction of contextual features and supports robust identification of ARPs from Q\&A content.

\textit{iii. Pre-trained Language Models (PLMs)}. In this study, we utilized two widely recognized PLMs to automatically identify ARPs from SO posts and differentiate them from programming-related posts.

\begin{itemize}
    \item \textit{BERT (Bidirectional Encoder Representations from Transformers)} is a neural network pre-trained on a large English corpus in a self-supervised manner \cite{devlin2018bert}. It learns rich contextual representations through two objectives: (1) Masked Language Modeling (MLM), where 15\% of words are masked and predicted, enabling bidirectional learning; and (2) Next Sentence Prediction (NSP), where the model determines whether two sentences are sequential. In our study, we employed the \texttt{BertForSequenceClassification} model based on the ``\texttt{bert-base-uncased}'' configuration\footnote{\url{https://huggingface.co/transformers/v3.0.2/model_doc/bert.html}} to distinguish ARPs. Posts were tokenized using \textit{BertTokenizer} (WordPiece tokenization)\footnote{\url{https://huggingface.co/docs/transformers/tokenizer_summary}} and encoded into contextual embeddings. Following BERT's guidelines, we experimented with a batch size of 32, a learning rate of \texttt{3e-5 (Adam optimizer)}, and \texttt{3 training epochs}.

    \item \textit{RoBERTa (Robustly Optimized BERT Pretraining Approach)} is an enhanced variant of BERT that improves performance on various NLP tasks by eliminating the next-sentence prediction objective, using larger mini-batches and learning rates, and applying dynamic masking during pretraining \cite{liu2019roberta}. It also adopts sentence packing to optimize token utilization and leverages byte-level Byte Pair Encoding (BPE) for full Unicode compatibility. In our study, we utilized the \texttt{RobertaForSequenceClassification} model based on the ``\texttt{roberta-base}'' configuration\footnote{\url{https://huggingface.co/transformers/v3.0.2/model_doc/roberta.html}} to distinguish ARPs from programming posts. textual data was tokenized using the \texttt{RobertaTokenizer}\footnote{\url{https://huggingface.co/docs/transformers/en/model_doc/roberta}}, and converted into contextualized embeddings. Following RoBERTa's standard training practices, we used a learning rate of \texttt{3e-5} with the \texttt{Adam optimizer}, trained for 3 epochs with a batch size of 32. Both RoBERTa and BERT models were implemented using TensorFlow \cite{abadi2016tensorflow}. %Complete details of the PLM configurations can be found in our replication package \cite{datasetTOSEM}.
\end{itemize}

\textit{iv. Large Language Models (LLMs)}. We utilized GPT-4o, a state-of-the-art multimodal LLM from OpenAI \cite{hurst2024gpt}, as its demonstrated prevalence in various SE tasks \cite{tanzil2024chatgpt, ding2023parameter, Li2025urc}. GPT-4o processes text, images, and audio in one model, delivering faster inference, lower latency, and richer contextual representation compared to GPT-3 and GPT-4 \cite{hurst2024gpt}. These performance gains make it especially effective for complex tasks such as ARP classification, where nuanced architectural semantics must be captured. We invoked the GPT-4o API with \texttt{temperature=0} and \texttt{max\_tokens=5} (other settings at default). Our prompt is as follows:

\begin{tcolorbox}[colback=white,colframe=black!75,boxrule=0.5pt, left=2pt,right=2pt,top=2pt,bottom=2pt,]
\textit{Tell me if the following text is related to a software architecture discussion or a software programming discussion. Just say 1 for the software architecture discussion or 0 for the software programming discussion.}
\end{tcolorbox}

In total, we developed 27 models to automatically identify and distinguish ARPs from programming-related posts on SO, including 20 ML models, 4 DL models, 2 PLMs, and 1 LLM. Detailed implementation and configuration are available in our replication package \cite{datasetTOSEM}.

\subsubsection{Evaluation Metrics} Following the practices of prior studies \cite{uddin2020mining, uddin2019automatic, bi2021mining}, we utilized four widely used metrics to assess model performance in identifying ARPs from development forums: Precision, Recall, F1-score, and Accuracy. 
\begin{comment}
These metrics are defined as follows:
\textit{Precision} is the proportion of correctly classified ARPs to all posts classified as ARPs (see Equation \ref{equ_Precision}).
\textit{Recall} is the proportion of actual ARPs that were correctly identified (see Equation \ref{equ_Recall}).
\textit{F1-score} represents the harmonic mean of Precision and Recall, balancing the trade-off between the two (see Equation \ref{equ_F1}).
Accuracy measures the overall percentage of correct classifications made by the model (see Equation \ref{equ_Accuracy}).

\begin{multicols}{2}
\begin{equation}\label{equ_Precision}
\text{Precision} = \frac{TP}{TP + FP}
\end{equation}

\begin{equation}\label{equ_Recall}
\text{Recall} = \frac{TP}{TP + FN}
\end{equation}
\columnbreak

\begin{equation}\label{equ_F1}
\text{F}_1\text{-score} = 2 \times \frac{\text{Precision} \times \text{Recall}}{\text{Precision} + \text{Recall}}
\end{equation}

\begin{equation}\label{equ_Accuracy}
\text{Accuracy} = \frac{TP + TN}{TP + TN + FP + FN}
\end{equation}
\end{multicols}

To compute these metrics, four statistics were used:  
\begin{itemize}
    \item \textit{True Positives (TP)}: Number of ARPs correctly classified as ARPs.
    \item \textit{False Positives (FP)}: Number of non-ARPs incorrectly classified as ARPs.
    \item \textit{False Negatives (FN)}: Number of ARPs incorrectly classified as non-ARPs.
    \item \textit{True Negatives (TN)}: Number of non-ARPs correctly classified as non-ARPs.
\end{itemize}
\end{comment}

\subsection{Automatic Extraction of Architectural Issue-Solution Pairs}\label{AutomaticExtractionArchitecturalIssue_solution}

\begin{figure}[ht]
 \centering
\includegraphics[width=\textwidth]{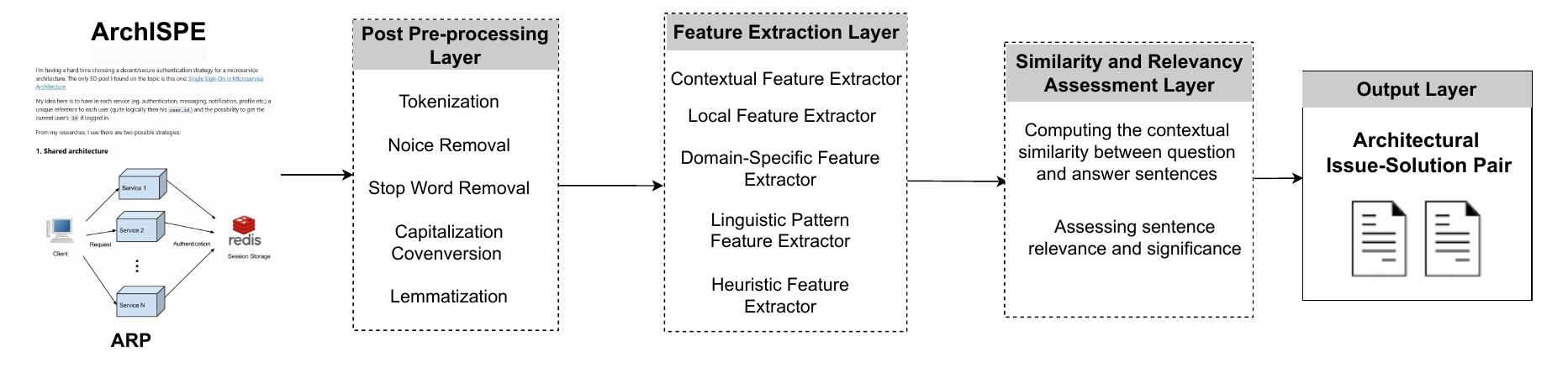}
\caption{ArchISPE architecture for extracting architectural issue-solution pairs from ARPs}
\label{fig_ArchictIssue_SolutioComponent}
\end{figure}

In this section, we present \textit{ArchISPE}, the second component of our framework, and detail its architecture. This step corresponds to \textbf{Phase III} in Figure~\ref{fig_OverviewOftheReseachMethod}. As illustrated in Figure~\ref{fig_ArchictIssue_SolutioComponent}, ArchISPE takes an ARP (i.e., a question and its corresponding answer(s)) as input, automatically extracts key sentences that convey architectural issues and solutions, and generates a concise, self-contained issue-solution pair for each post. The remainder of this section defines the extraction task and explains the structure of the ArchISPE component.

\subsubsection{Task Definition} 
To clearly define the task of issue-solution pair extraction and establish a consistent framework, we first introduce key concepts and notations used throughout this section and subsequent discussions.

\begin{figure}[h]
\centering
\subfloat[A set of sentences describing an architecture-related question]{\includegraphics[width=.5\linewidth]{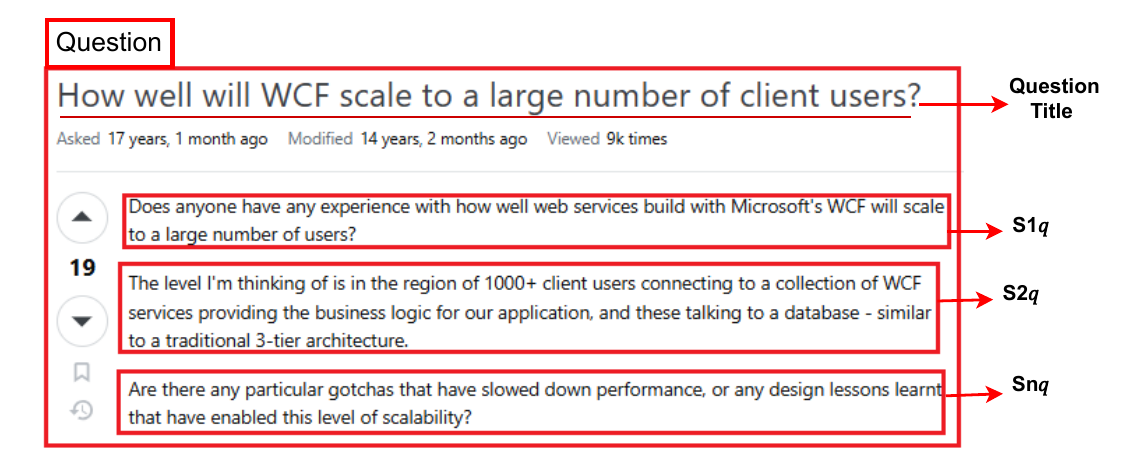}}\hfill 
    \subfloat[A set of sentences describing a potential answer/solution]{\includegraphics[width=.46\linewidth]{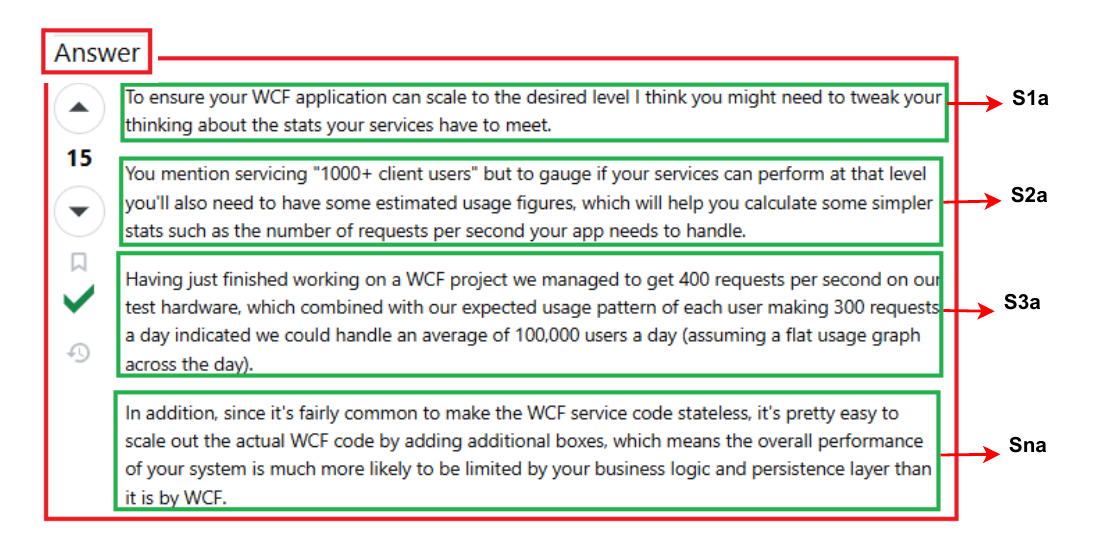}}\hfill
    \caption{The example of question and answer bodies on SO}
    \label{Sentences_In_ARPs}
\end{figure}

Each SO post is structured into threads. Specifically, an SO post includes a \textit{question thread}, comprising the title, body, tag(s), and comment(s); an \textit{answer thread} consists of one or more answers, containing corresponding answer comments. Accordingly, we decompose each ARP in our dataset \( D \) into two main components: \textit{Question\_body} and \textit{Answer\_body} (see Figure \ref{Sentences_In_ARPs}). Each ARP serves as a distinct data point and is represented as follows: %Each ARP serves as a distinct data point \( D \), represented as follows:

\begin{align}
        \text{\textit{Question\_body}} &= \{ \text{s}_{1q}, \text{s}_{2q}, \dots, \text{s}_{nq} \} \notag\\
        \text{\textit{Answer\_body}}  &= \{ \text{s}_{1a}, \text{s}_{2a}, \dots, \text{s}_{na} \} \notag\\
        \text{ARP} &= \{\text{\textit{Question\_body}}, \text{\textit{Answer\_body}}\} \notag\\
        \text{D} &= \{\text{ARP}_1, \text{ARP}_2, \dots, \text{ARP}_m\}
        \label{eq:arp_definition}
\end{align}

\noindent where \textit{Question\_body} represents the set of sentences describing an architecture-related question, \textit{Answer\_body} represents the set of sentences describing a potential answer/solution, and \textit{n} denotes the number of sentences in either question or answer. ARP is the tuple of these two sets, and \( D \) is the collection of all ARPs in the dataset. 

Following the concepts, we define the task of automatically extracting architectural issue-solution pairs as follows: Given an ARP (i.e., a \textit{Question\_body} and its corresponding \textit{Answer\_body}) consisting of \( n \) sentences, the goal is to extract a set of key sentences that explicitly express architectural issues and solutions, and to generate a concise, self-contained issue-solution pair for the post. This task is framed as an indirect supervised, contextualized sentence classification problem, where each sentence is classified as either part of the issue-solution pair or not. We then describe how this task is executed in \textit{ArchISPE}. Building on the above definition, \textbf{ArchISPE} further divides the task into three sub-tasks:

\begin{itemize}
    \item \textbf{Dataset splitting}: Given a dataset \( D \) consisting of \( m \) ARPs, \textit{ArchISPE} splits D into separate ARPs, \( \text{D} = \{ \text{ARP}_1, \text{ARP}_2, \dots, \text{ARP}_m \} \).
    \item \textbf{Issue Extraction}: Given a separate \( ARP_i \), \textit{ArchISPE} defines a function \( t \) that determines whether a sentence in \(\text{\textit{Question\_body}}_i \) is relevant and important for describing an issue. The function \( t \) identifies and extracts such sentences so that: \( t(\text{\textit{Question\_body}}_i) = \{ s_{1q}, s_{2q}, \dots, s_{nq} \} \), where each \( s_{iq} \) represents a sentence within \( \text{\textit{Question\_body}}_i \) that contributes to the issue description.   
    \item \textbf{Solution Extraction}: Given an \( ARP_i \), \textit{ArchISPE} defines a function \( g \) that identifies relevant and important sentences within \( \text{\textit{Answer\_body}}_i \) that suggest a solution. The function \( g \) identifies and extracts such sentences so that: \( g(\text{\textit{Answer\_body}}_i) = \{ s_{1a}, s_{2a}, \dots, s_{na} \} \), where each \( s_{ia} \) represents an sentence within \( \text{\textit{Answer\_body}}_i \) that contributes to the solution description.
\end{itemize}
%Thereafter, the output of ArchISPE is a set of architectural issue-solution pairs. Ideally, end users do not need other information (e.g., the tag(s) and comments on posts) to understand these pairs.

\subsubsection{ArchISPE Architecture}
\textit{ArchISPE} architecture comprises four primary layers (see Figure~\ref{fig_ArchictIssue_SolutioComponent}):

\paragraph{I. Post-Preprocessing Layer}\label{Post_Preprocessing_Layer}

Each ARP is processed by the following steps. Given the unique characteristics of SO data, we applied tailored preprocessing techniques: (i) sentences containing both hyperlinks and text, where we retained the textual content and replaced hyperlinks with the placeholder (\textit{[external-link]}); and (ii) embedded architectural diagrams and tables, which were substituted with special tokens (\textit{[figure]}) and (\textit{[table]}), respectively. Finally, we processed the textual content using NLTK\footnote{\url{https://www.nltk.org/}}, including stop word removal, tokenization, lemmatization, and lowercase conversion.

\paragraph{II. Feature Extraction Layer}\label{Feature_ExtractionLayer} 
This layer includes four feature extractors as follows:

\textbf{1. Contextual Feature Extractor}. We leveraged contextual representations from a BERT model to encode the textual information in ARPs, capturing both their semantics and relational information. These features enable the identification of complex relationships and meanings within the content, facilitating a deeper understanding of the architectural issue-solution context. In this work, we utilized \textit{BERTOverflow}, a BERT-based model fine-tuned on 152 million sentences from SO \cite{tabassum2020code}. This model is specifically trained on SE-related discussions, making it well-suited for capturing domain-specific terms and textual patterns \cite{tabassum2020code}. It has demonstrated good performance across various NLP tasks \cite{zhou2024large, he2022ptm4tag, von2022validity}. The BERTOverflow embedding layer produces a vector of 768 dimensions per token. 

We fine-tuned BERTOverflow on our dataset to generate embedding vectors for each sentence in an  ARP. Specifically, we computed sentence embeddings by averaging the embeddings of all tokens within a sentence, resulting in a 768-dimensional vector per sentence. Given an architectural post \( P \) containing \( n \) sentences, represented as \( P = \left[s_1, s_2, \dots, s_n\right] \), each sentence \( s_i \) is mapped to a 768-dimensional vector using BERTOverflow. The final sentence embedding matrix \( E(P) \) is:

\[
E(P) =
\begin{bmatrix}
e_{1,1} & e_{1,2} & \dots & e_{1,768} \\
e_{2,1} & e_{2,2} & \dots & e_{2,768} \\
\vdots & \vdots & \ddots & \vdots \\
e_{n,1} & e_{n,2} & \dots & e_{n,768}
\end{bmatrix}
\in \mathbb{R}^{n \times 768}
\]

where:
\begin{itemize}
    \item Each row represents the embedding of one sentence.
    \item \( e_{i,j} \) is the embedding value of the \( i \)-th sentence in the \( j \)-th dimension (out of 768 dimensions).
\end{itemize}   

Thus, for an ARP containing \( n \) sentences, the final representation is a 2D matrix with dimensions \( n \times 768 \). Integrating deep contextualized sentence embeddings enables the ArchISPE approach to effectively capture the semantics of a given sentence. This is further validated by the ablation study in Section \ref{AblationStudy}, which shows that excluding BERTOverflow decreases the performance of our model. 

\textbf{2. Local Feature Extractor}. While BERT (BERTOverflow) captures rich contextual dependencies, it may overlook local lexical patterns (e.g., specific word combinations or short phrases) that convolutional models can effectively capture~\cite{dos2014deep, young2018recent}. To address this, we employed a customized TextCNN~\cite{kim2014cnn} to enhance BERT embeddings by modeling local patterns and spatial hierarchies within sentences of the ARPs from our dataset. %The model consists of eight layers: one embedding layer, three convolutional layers, three max-pooling layers, and one fully connected layer.To capture fine-grained local dependencies between adjacent tokens, we employed a customized \textit{TextCNN}. 
Unlike the standard multi-kernel TextCNN~\cite{kim2014cnn}, our configuration employs three parallel convolutional filters with the same kernel size (\(h = 2\)), designed to emphasize bi-gram feature learning from the embedding space. Formally, given a sentence of length \(T\), the embedding layer maps each token to a \(d\)-dimensional vector, forming
\begin{equation}
X = [x_1, x_2, \dots, x_T], \quad X \in \mathbb{R}^{T \times d}.
\end{equation}
The input is reshaped to \(X' \in \mathbb{R}^{1 \times 1 \times T \times d}\) to match the convolutional layer format. 
Each 2D convolutional kernel \(W_i \in \mathbb{R}^{1 \times h \times d}\) slides over the temporal dimension, computing
\begin{equation}
C_i = \mathrm{ReLU}(W_i * X' + b_i),
\end{equation}
where ReLU is the activation function and \(b_i\) is the bias term.
The resulting feature maps \(C_i \in \mathbb{R}^{B \times F \times (T-h+1)}\) (with \(B\) as the batch size and \(F\) as the number of filters) are then processed via one-dimensional max pooling:
\begin{equation}
\hat{c}_i = \mathrm{maxpool}(C_i),
\end{equation}
which selects the strongest activation along the sequence length.
The pooled feature vectors from the three convolutional branches are concatenated and projected through a fully connected layer:
\begin{equation}
z = W_f [\hat{c}_1; \hat{c}_2; \hat{c}_3] + b_f,
\end{equation}
yielding a fixed 256-dimensional sentence representation \(z \in \mathbb{R}^{256}\).

\textbf{3. Linguistic Pattern Extractor}. While labeling ARPs from SO posts for training \textit{ArchISPE}, we identified and summarized phrases and keywords that may indicate important information in a question or an answer. The first twenty-three phrases and keywords are listed in Table \ref{Linguistic_Patterns_for_Sentences}, and the remaining ones are provided in the replication package. By following previous work \cite{silva2019recommending, robillard2015recommending}, we used linguistic pattern features for identifying key sentences from SO posts. Specifically, the first and ninth authors randomly sampled 50 ARPs from the labeled dataset of 367 ARPs, manually analyzing the key sentences of these posts. They identified an initial set of common phrases shared across multiple sentences in questions and answer bodies and then applied these patterns back to the sentences in the remaining posts to assess coverage, refining the patterns iteratively. This procedure is similar to approaches used in prior work for identifying linguistic patterns \cite{silva2019recommending, robillard2015recommending, li2018improving}. Each pattern corresponds to a specific dimension in the sentence embedding: when a linguistic pattern is matched, the corresponding dimension is set to 1; otherwise, it is set to 0.

\begin{itemize}
\item \textbf{Comparative adjective}. This feature captures whether a sentence contains a comparative adjective. Sentences with comparative terms (e.g., better, more efficient) often assist users (e.g., developers) in comparing two or more architectural elements suggested as potential design solutions, such as architectural patterns, tactics, or frameworks. For instance, \href{https://stackoverflow.com/a/46166820/12381813}{SO~\#46143618}, “\textit{If you're just starting out, EF Core might be \href{https://stackoverflow.com/a/46166820/12381813}{better than} EF6 mainly because of speed and portability}”. %“If you're looking at a single server, MongoDB is probably a \href{https://stackoverflow.com/a/2894665/12381813}{better} fit. For those more concerned about scaling, Cassandra's no-single-point-of-failure architecture will be \href{https://stackoverflow.com/a/2894665/12381813}{easier} to set up and more reliable”, (see Table \ref{Linguistic_Patterns_for_Sentences}).
%Once you pick the right one, the service to service communication will be the next step and each platform has an approach that works \href{https://stackoverflow.com/questions/51800093/distributed-app-services-in-azure-api-management-service-fabric-ase-applica/51801182#51801182}{better than} others.
%\footnote{\url{https://stackoverflow.com/questions/78015068/how-to-design-aws-architecture-for-my-new-course-selling-website}}, (see Table \ref{Linguistic_Patterns_for_Sentences}).

\item \textbf{Superlative adjective}. This feature captures whether a sentence contains a superlative adjective. Similar to comparative adjectives, sentences with superlative terms (e.g., best, most effective) often indicate a strong inclination for or against an architectural approach, pattern, or solution, e.g., \href{https://stackoverflow.com/q/62688990/12381813}{SO~\#62688990}, “\textit{I would say \href{https://stackoverflow.com/q/62688990/12381813}{the best practice} here would be to not use Flask and go with the API Gateway + Lambda option. This lets you put custom security and checks on your APIs as well as making the application a lot more stable as every request has its own lambda}”.
%I was wondering what is \href{https://stackoverflow.com/questions/32641858/what-is-the-best-practice-for-enterprise-level-application-architecture-using-mv}{the best} practice for enterprise level architecture based on MVC5
%\href{https://stackoverflow.com/questions/75483085/best-practice-of-social-login-with-separate-frontend-and-backend/75483218#75483218}{The best} practice is to use your backend. Here you can manage token more secure and if you use authorization, too, you need the secure token data from google in your backend, too. Here is a link to official Google documentation.
%\href{https://stackoverflow.com/questions/76694434/how-to-store-service-provider-specific-data-without-it-influencing-your-database/76694456#76694456}{The best practice} when it comes to storing external provider's data without it influencing your generic business data model is to use a denormalized data model.

\item \textbf{Imperative sentence}. This feature captures whether a sentence is an imperative sentence. Imperative sentences typically instruct, command, or suggest architectural approaches or solutions to accomplish a task or resolve an issue in architecture. For example, \href{https://stackoverflow.com/a/77606141/12381813}{SO~\#62688990}, “\textit{\href{https://stackoverflow.com/a/77606141/12381813}{You should} use an HTTP request, as they are independent applications/APIs}”.
%Ideally, \href{https://stackoverflow.com/a/77159949/12381813}{you should} have an API Gateway between your Client and all the underlying microservices. 
%I think in your case, \href{https://stackoverflow.com/a/77393299/12381813}{you should} use abstraction (interfaces or abstract classes) that the service layer depends on and the data layer implements.
%Why choose between a traditional database and a NoSQL data store? \href{https://stackoverflow.com/a/10183642/12381813}{Use both!}

\item \textbf{Task-oriented feature}: These features reflect intent or actions related to specific architectural tasks, such as designing, building, or seeking solutions to architectural problems. This feature is based on our observation that certain sentences indicate a task-oriented intent and serve as important sentences. Examples include phrases like \href{https://stackoverflow.com/q/75568037/12381813}{SO \#77068586} “\href{https://stackoverflow.com/q/75568037/12381813}{I am building...}”, “\textit{\href{https://stackoverflow.com/q/75308654/12381813}{I want to design}}”, and \textit{“\href{https://stackoverflow.com/q/76012035/12381813}{How to architecture}}”, among others.
%(see Table \ref{Linguistic_Patterns_for_Sentences}).
\end{itemize}

\begin{table*}
\footnotesize
\caption{Linguistic patterns for sentences that may contain important architectural knowledge}
\label{Linguistic_Patterns_for_Sentences}
\begin{minipage}{0.48\linewidth}
\centering
\begin{tabular}{m{0.3cm} m{2.5cm} m{2.2cm}}
\hline
\textbf{NO.}   &\textbf{Phrase or keyword}  &\textbf{SO Post Example}  \\ \hline

1              & I'm building... 
               &\href{https://stackoverflow.com/questions/886221/has-django-served-an-excess-of-100k-daily-visits/886645}{SO \#886221}\\\cline{1-3}
2              & How to architecture ...
               &\href{https://stackoverflow.com/q/76012035/12381813}{SO \#76012035}  \\\cline{1-3}
3              & I want to design ... 
               &\href{https://stackoverflow.com/q/75308654/12381813}{SO \#75308654}  \\\cline{1-3}

4              & I want to build ... 
               &\href{https://stackoverflow.com/q/1317726/12381813}{SO \#1317726}  \\\cline{1-3}

5              & I am evaluating...
               &\href{https://stackoverflow.com/questions/2892729/mongodb-vs-cassandra}{SO \#2892729} \\\cline{1-3}
6              & how to structure... 
               & \href{https://stackoverflow.com/questions/209641/architectual-design-patterns}{SO \#209641}  \\\cline{1-3}

7              & user should...
               & \href{https://stackoverflow.com/q/78268212/12381813}{SO \#78268212}\\\cline{1-3}

8              & I am developing... &\href{https://stackoverflow.com/q/73290105/12381813}{SO \#73290105}\\\cline{1-3}

9              & I am building... &\href{https://stackoverflow.com/q/75568037/12381813}{SO \#75568037}\\\cline{1-3}

10             & Advise on...  & \href{https://stackoverflow.com/q/78010308/12381813}{SO \#78010308}\\\cline{1-3}

11              & I recommend... & \href{https://stackoverflow.com/a/78268124/12381813}{SO \#77940468}\\\cline{1-3}

12              &I cannot recommend... & \href{https://stackoverflow.com/a/78424239/12381813}{SO \# 78422247}\\\cline{1-3}

\end{tabular}
\end{minipage}
\hfill
\begin{minipage}{0.48\linewidth}
\centering
\begin{tabular}{m{0.3cm} m{2.5cm} m{2.2cm}}
\hline
\textbf{NO.}   &\textbf{Phrase or keyword}  &\textbf{SO Post Example}  \\ \hline

13              &I would recommend... & \href{https://stackoverflow.com/a/78424239/12381813}{SO \# 78422247}\\\cline{1-3}

14              & It is recommended... & \href{https://stackoverflow.com/a/75851917/12381813}{SO \#75851917}\\\cline{1-3}

15             & you don't have to ...& \href{https://stackoverflow.com/a/76178356/12381813}{SO \#76175226}\\\cline{1-3}

16              & In order to  ... & \href{https://stackoverflow.com/a/78296772/12381813}{SO \#78291914}  \\\cline{1-3}

17              & critical ...& \href{https://stackoverflow.com/a/1368346/12381813}{SO \#1368014}\\\cline{1-3}

18              & You should... & \href{https://stackoverflow.com/a/77606141/12381813}{SO \#77604491}\\\cline{1-3}

19              & A good approach...& \href{https://stackoverflow.com/a/3844947/12381813}{SO \#3841569}\\\cline{1-3}

20              & better than...& \href{https://stackoverflow.com/a/46166820/12381813}{SO \#46143618}\\\cline{1-3}

21              & the best...& \href{https://stackoverflow.com/q/62688990/12381813}{SO \#62688990}\\\cline{1-3}

22              & I would suggest...& \href{https://stackoverflow.com/a/511464/12381813}{SO \#509855}\\\cline{1-3}

23              & I suggest...& \href{https://stackoverflow.com/a/3497661/12381813}{SO \#3497623}\\\cline{1-3}

\end{tabular}
\end{minipage}
\end{table*}

\textbf{4) Heuristic Feature Extractor}. Similar to prior work \cite{robillard2015recommending, li2018improving}, we designed a heuristic feature extractor that incorporates non-semantic cues, such as sentence formatting and structure, to enhance the embedding results. This extractor aims to improve the relevance of sentence embeddings by capturing patterns that are not strictly semantic but still carry significant information. The following heuristic features were extracted from our dataset:

\begin{itemize}
    \item \textit{The 5W1H!} refers to the set of question words commonly used when posing questions: “What”, “Why”, “When”, “Who”, “Which”, and “How” along with punctuation, namely the question mark (“?”). The intuition behind this feature is that sentences containing any of these 5W1H question words, especially when elaborating architectural issues, often carry significant informational cues for identifying the focus of the question. 

    \item \textit{Sentence length}. Architectural design discussions often require contextual and elaborate descriptions. Thus, instead of focusing solely on keywords and phrases, we considered entire sentences that describe architectural solutions. Our observation suggests that longer sentences in SO ARPs tend to provide more context and detailed explanations, making them more likely to convey architectural issues or solutions. Similar to prior work \cite{xu2017answerbot}, this feature measures sentence length based on the number of word tokens. However, to mitigate bias toward lengthy but less relevant sentences, it is used in conjunction with other features, such as contextual features extracted by the BERTOverflow model.
\end{itemize}   

\paragraph{III. Similarity and Relevancy Assessment Layer}\label{Similarity_Relevancy_Assessment_Layer}

This layer quantifies the semantic similarity and overall relevancy of each sentence among all candidate sentences within the \textit{Question\_body} and \textit{Answer\_body} of an ARP. First, we calculate context-wise sentence similarity using cosine similarity. Then, we calculate a relevance score for each sentence by integrating multiple features extracted in the preceding Feature Extraction Layer, including BERT embeddings, local TextCNN features, and linguistic pattern cues. These scores are used to rank candidate sentences based on their task-specific significance. The process consists of two main steps:

\begin{enumerate}
    \item \textbf{Similarity Assessment}: This step involves two measures:

    \begin{itemize}
         \item \textit{Question Sentence Similarity}: To assess how each question sentence represents the overall question content, we calculate the cosine similarity between its BERT embedding and the mean BERT embedding of all question sentences. Given the embedding \( \mathbf{qemb}_{Si} \) of the \( i \)-th question sentence and the mean question embedding \( \mathbf{Qemb}_{\text{mean}} \), the similarity is computed as:

            \begin{equation}     
            \text{Similarity\_score}_i = \cos(\mathbf{qemb}_{Si}, \mathbf{Qemb}_{\text{mean}}) = \frac{\mathbf{qemb}_{Si} \cdot \mathbf{Qemb}_{\text{mean}}}{\|\mathbf{qemb}_{Si}\| \cdot \|\mathbf{Qemb}_{\text{mean}}\|}   
            \end{equation}

        \item \textit{Answer-to-Question Similarity}: For Q\&A settings like SO, identifying answer sentences most relevant to the question is essential \cite{iyyer2014neural, unger2012template}. We calculate the cosine similarity between the BERT embedding of each answer sentence and the mean BERT question embedding. Let \( \mathbf{aemb}_{j} \) denote the embedding of the \( j \)-th answer sentence and \( \mathbf{Qemb}_{\text{mean}} \) the mean question embedding. The similarity is computed as:

            \begin{equation} 
            \text{Similarity\_score}_{j} = \cos(\mathbf{aemb}_{j}, \mathbf{Qemb}_{\text{mean}}) = \frac{\mathbf{aemb}_{j} \cdot \mathbf{Qemb}_{\text{mean}}}{\|\mathbf{aemb}_{j}\| \cdot \|\mathbf{Qemb}_{\text{mean}}\|} 
            \end{equation}        
    \end{itemize}

    \item \textbf{Relevancy Assessment}: Inspired by prior work on term weighting in text classification tasks \cite{guo2019improving}, we designed a multi-feature weighting strategy to improve the relevancy and importance of sentence scoring, thereby increasing the likelihood of selecting meaningful sentences in \textit{ArchISPE}. Each sentence in the \textit{Question\_body} and \textit{Answer\_body} is scored independently using features extracted in the previous layer. The final score for a sentence is then computed as:

    \begin{equation}
        \begin{aligned}
        \text{final\_score} &= (W_{\text{bert}} \cdot \text{similarity\_score}) + (W_{\text{textcnn}} \cdot \text{textcnn\_score}) \\
        &\quad + (W_{\text{linguistic}} \cdot \text{linguistic\_score}) + (W_{\text{heuristic}} \cdot \text{heuristic\_score})
        \end{aligned}
    \end{equation}

    \noindent where:
    \begin{itemize}
        \item \( W_{\text{bert}}, W_{\text{textcnn}}, W_{\text{linguistic}}, W_{\text{heuristic}} \) are the weight factors assigned to each feature, constrained such that their sum equals one, i.e., \(\sum W_i = 1\).

        \item \texttt{bert\_score} captures contextual semantic alignment using BERT-derived sentence embeddings, applied in either Question Sentence Similarity or Answer-to-Question Similarity.
        \item \texttt{textcnn\_score} reflects local sentence-level patterns extracted by the TextCNN model.
        \item \texttt{linguistic\_score} captures handcrafted phrases or keywords that may signal important information within the ARP.
        \item \texttt{heuristic\_score} incorporates non-semantic, handcrafted cues that may indicate relevant or critical sentence in the ARP.
    \end{itemize}

The resulting final score integrates all feature signals to estimate sentence significance in the context of the ARP.
\end{enumerate}

\paragraph{IV. Output Layer}\label{Output_Layer}

The output layer consolidates information from the previous layers to rank sentences based on their scores and generate the final set of relevant and important sentences for both the question and the answer, forming issue-solution pairs. Specifically, this layer processes the ranked sentences to extract concise, self-contained issues and solutions. The process involves the following key steps:

\begin{itemize}
    \item \textit{Ranking sentences by score}: For both question and answer sentences, the layer begins by ranking all sentences separately based on their relevance and significance scores. The highest-scoring sentences (i.e., those most relevant and essential to the issue or solution) are prioritized. This ranking is performed by sorting the sentences in descending order, ensuring that the most relevant and significant ones appear first.
    
    \item \textit{Extraction of top-ranked sentences}: Once ranked, the layer extracts the top $n$ sentences. In prior NLP-based text extraction studies, it is common to select a fixed number of sentences~\cite{yang2023techsumbot, uddin2020mining, zhong2020extractive}. In this study, we selected six sentences based on our observation that architectural design discussions often require contextual and detailed descriptions, and SO users tend to elaborate their posts extensively to provide sufficient background and rationale. These sentences, selected based on their relevance scores, form the core content of the issue and solution, respectively. %Once ranked, the layer extracts the top $n$ sentences (we select 6 sentences in this study). In prior work on text extraction from NLP field, it is common to select a fixed number of sentences \cite{zhong2020extractive}. In this study, In we selected six sentences based on our experience that architectural design discussions often require contextual and elaborate descriptions and SO users tend to elaborate their posts in long description in order to ... These sentences, selected based on their relevance scores, form the core content of the issue and solution separately.
    
    \item \textit{Pairing issue-solution}: After extracting key sentences from questions and answers to form separate issue and solution summaries, the layer pairs them back together along with the corresponding \textit{Question\_title} based on their contextual relationship. This pairing is essential because each question on SO is inherently linked to its title and corresponding answer(s). This relationship ensures both coherence and relevance in the final issue-solution pairs. 
\end{itemize}
%####################################################################################################

\section{Practitioner Feedback on ArchISMiner}
To answer RQ3, we conducted a user study to evaluate the quality of our proposed framework, \textit{ArchISMiner}, in identifying ARPs (i.e., \textit{ArchPI}) and extracting architectural issue-solution pairs (i.e., \textit{ArchISPE}). This step corresponds to \textbf{Phase IV} in Figure~\ref{fig_OverviewOftheReseachMethod}. Below, we present the participant recruitment and evaluation procedures.

\fakesection{(1) Participant Recruitment}\label{Participant_Recruitment} 

We recruited seven participants from software companies in Germany, China, and France, all of whom actively engage in architectural discussions or seek solutions on SO. On average, participants had four years of professional experience and a solid understanding of architectural design concepts. 

\fakesection{(2) Quality of Identified ARPs by ArchPI}\label{PractitionersFeedback_On_ArchPI}

%To evaluate the effectiveness of \textit{ArchPI}, we assessed the quality of the identified ARPs. 
To evaluate the quality of \textit{ArchPI}, we assessed the relevancy, comprehensiveness, and usefulness of the identified ARPs. Due to participants’ time and effort constraints, we randomly sampled 40 ARPs identified by \textit{ArchPI} as experimental data and conducted five evaluation sessions. In each session, participants evaluated eight of the 40 ARPs. %Each identified ARP was presented alongside its corresponding ground-truth (original) ARP for reference. %To ensure participants clearly understood the context, we first asked them to review the original ARPs from our dataset before comparing them with the identified ones. %Each predicted ARP was presented alongside its corresponding ground-truth ARP for reference. To ensure adequate context, participants first reviewed the original ARPs from our dataset before comparing the predicted ARPs with the ground truth.

Following the protocol in \cite{yang2023techsumbot,xu2017answerbot}, participants answered four questions using a 5-point Likert scale, ranging from 1 (e.g., ``not relevant at all'', ``not comprehensive at all'', ``not useful at all'', or ``strongly disagree'') to 5 (e.g., ``highly relevant'', ``highly comprehensive'', ``extremely useful'', or ``strongly agree''). Before the formal study, a pilot study involving two participants was conducted to refine the evaluation questions and examples. The final evaluation questions were as follows:
%Following the protocol by Xu \textit{et al.} \cite{xu2017answerbot}, participants responded to three statements using a 5-point Likert scale, ranging from 1 (e.g., “strongly disagree”) to 5 (e.g., “strongly agree”), and the statements were inspired by prior studies \cite{nasab2021automated, khalajzadeh2022supporting}. The Likert response categories included Relevance, Comprehensiveness, and Usefulness. Before the formal study, a pilot study with two participants was conducted to refine the evaluation statements and examples. The final evaluation statements were as follows:
\begin{tcolorbox}[colback=white,colframe=black!75,boxrule=0.5pt, left=0.5pt,right=0.5pt,top=0.5pt,bottom=0.5pt]
\begin{enumerate}[1.]
     \item Are the ARPs identified by the approach relevant to discussions on architectural issues and solutions in real-world software architecture design?
     \item Is the architectural knowledge in the identified ARPs comprehensive?
    \item Is the architectural knowledge in the identified ARPs useful for real-world software architecture design? 
    \item Would you find it useful to have an ARP identifier (e.g., a dedicated label, icon, or visual marker) that explicitly distinguishes architecture-related posts from general programming posts, beyond the existing SO tags?
\end{enumerate}
\end{tcolorbox}

%The Likert response categories included:
\begin{comment}
\begin{itemize}
    \item \textbf{Relevance:} Highly relevant, Mostly relevant, Moderately relevant, Slightly relevant, Not relevant at all
    \item \textbf{Comprehensiveness:} Highly comprehensive, Mostly comprehensive, Moderately comprehensive, Slightly comprehensive, Not comprehensive at all
    \item \textbf{Usefulness:} Extremely useful, Mostly useful, Moderately useful, Slightly useful, Not useful at all
\end{itemize}
\end{comment}

\fakesection{(3) Quality of Extracted Architectural Issue-Solution Pairs by ArchISPE}\label{PractitionersFeedback_On_ArchISPE}

To evaluate the effectiveness of \textit{ArchISPE}, we compared its results against two baseline approaches: \textit{DECA\_PD \& DECA\_SP} and \textit{TextRank}. These baselines were selected based on their best performance in automated evaluations of approaches for extracting architectural issue-solution pairs (see Table~\ref{Table_Performance_Of_Issue_Solution_Extraction_Techniques_SE_Autom} and Table~\ref{Table_Performance_Of_Issue_Solution_Extraction_Techniques_NLP_Autom}).

Due to participants' time and effort constraints, we randomly sampled 10 ARPs as experimental data. For each ARP, we collected architectural issue-solution pairs extracted by \textit{ArchISPE}, \textit{DECA\_PD \& DECA\_SP}, and \textit{TextRank}, resulting in a total of 30 pairs. We conducted five evaluation sessions, and in each session, participants were asked to assess six out of the 30 extracted pairs. Each participant was provided with the original ARP (question and answer), and the issue-solution pairs generated by the three approaches. After reviewing the original posts to understand the context, participants evaluated the extracted issue-solution pairs based on four criteria. Participants responded to four questions using a 5-point Likert scale as in the evaluation for \textit{ArchPI}. To mitigate bias, we anonymized the source of each issue-solution pair and randomly shuffled their presentation order. Additionally, we included an open-ended question at the end of the study to capture participants' further comments and suggestions. The evaluation questions presented to participants were as follows:

\begin{tcolorbox}[colback=white,colframe=black!75,boxrule=0.5pt, left=0.5pt,right=0.5pt,top=0.5pt,bottom=0.5pt]
\begin{enumerate}[1.]
    \item Is the architectural knowledge in the extracted issue-solution pair relevant to real-world software architecture design discussions?
    \item Is the architectural knowledge in the extracted issue-solution pair comprehensive?
    \item Is the architectural knowledge in the extracted issue-solution pair useful for real-world software architecture design?
    \item Would you find it useful to have an architectural issue-solution pair extractor when searching for and reading posts on SO? For example, this could be a key sentence highlighter, a structured summary, or an automated annotation that identifies and extracts key architectural issues and their corresponding solutions within a post.
\end{enumerate}
\end{tcolorbox}

\section{Results}\label{Experiment_Results}
% This section presents our experiment results with corresponding analysis to answer three research questions.
 
\subsection{RQ1: Evaluation of Models for Identifying Architecture-Related Posts in Developer Communities}\label{Evaluation_of_Models_for_IdentifyingARPs}
To address RQ1, we trained 27 models, comprising nine distinct models for identifying ARPs on SO. These included five traditional ML models, one DL model utilizing four different embedding methods (i.e., TF-IDF, word2vec, USE, and GloVe), two PLMs, and one LLM. Specifically, we combined the five ML models with each of the four embedding methods, the DL model with each of the four embedding methods, and included the two standalone PLMs and the standalone LLM. 

We balanced the dataset by selecting an equal number of ARP-labeled and programming-related posts to enhance the generalization capability of the models (see Section~\ref{Data_Collection}) and mitigate class imbalance~\cite{he2009imbalance}, which can significantly impact performance. Posts labeled as ARPs were assigned a target value of 1, while programming-related posts were assigned a value of 0. The dataset was split into 80\% (11,946 posts) for training and 20\% (2,986 posts) for testing. To facilitate a more granular analysis, we categorized the models into four groups: traditional ML-based, DL-based, PLM-based, and LLM-based models (see Table~\ref{Table_Performance_Of_ML_DL_LLM_Models}).

\fakesection{(1) Performance of Traditional ML-based Models}

\begin{figure}
   \centering
    \subfloat[F1-score comparison across five ML models based on four feature extraction methods]{\includegraphics[width=.5\linewidth]{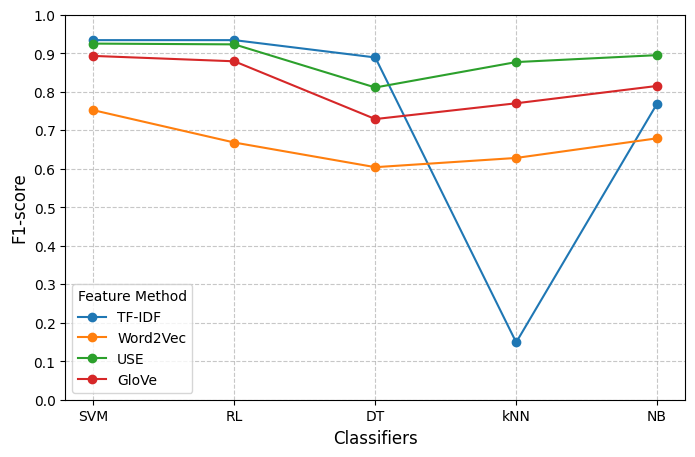}}\hfill
    \subfloat[F1-score comparison across the DL models based on four feature extraction methods]{\includegraphics[width=.5\linewidth]{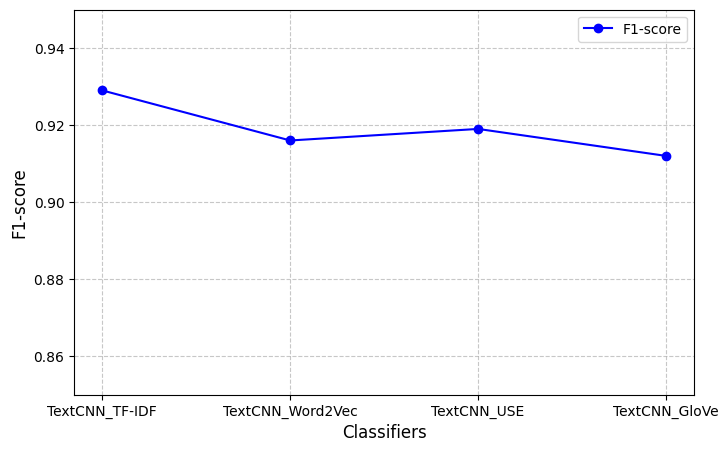}}\hfill
    \caption{F1-score comparison across ML, DL models, and feature extraction methods}
    \label{Fig_F1_Comparison_ML_DL_Classifiers_Feature_Extraction_Methods}
\end{figure}

The upper part of Table~\ref{Table_Performance_Of_ML_DL_LLM_Models} presents the performance of five traditional ML models trained with four embedding methods: TF-IDF, Word2Vec, USE, and GloVe. In total, 20 models were trained (each ML model applied to each embedding method). The highest metric values across all models for a given embedding method are underlined and highlighted in bold. %The best results for each evaluation metric are underlined, while the highest metric values across all models for a given embedding method are highlighted in bold. 
We analyzed the performance of the ML models across four key metrics for each embedding method from two perspectives:

\begin{itemize}
    \item \textbf{Impact of Feature Extraction Methods}. \textbf{TF-IDF} \textit{is optimal for SVM and RL but performs poorly with kNN}: When paired with TF-IDF, both SVM and RL achieve the highest F1-score (0.934), whereas kNN shows a significant performance drop. DT performs moderately well with TF-IDF (0.889 F1-score), but kNN completely fails (0.149 F1-score), as shown in Table \ref{Table_Performance_Of_ML_DL_LLM_Models} and Figure \ref{Fig_F1_Comparison_ML_DL_Classifiers_Feature_Extraction_Methods}(a). TF-IDF struggles with kNN due to its sparse, high-dimensional representation, which negatively affects distance-based computations. kNN with USE (0.877 F1-score) is the only case where kNN performs well, demonstrating that USE embeddings benefit distance-based models. NB performs particularly poorly when paired with Word2Vec, achieving the lowest F1-score (0.679), indicating its sensitivity to feature representation. \textbf{GloVe} \textit{is stable but weaker than USE and TF-IDF}: GloVe maintains a moderate F1-score (mean = 0.862) across SVM, RL, and NB. However, it underperforms in DT (0.729 F1-score) and kNN (0.770 F1-score) compared to TF-IDF and USE. While GloVe is a better alternative to Word2Vec, it remains inferior to TF-IDF and USE in most cases. \textbf{Word2Vec} \textit{is the weakest feature extraction method}: Word2Vec consistently yields the lowest F1-scores across models. SVM with Word2Vec (0.752 F1-score) and RL with Word2Vec (0.668 F1-score) perform poorly compared to USE and TF-IDF in most cases. Word2Vec with DT (0.604 F1-score) ranks among the worst-performing feature extraction methods, indicating that tree-based models/classifiers do not work well with dense Word2Vec embeddings. While kNN with Word2Vec (0.628 F1-score) outperforms kNN with TF-IDF, it still lags behind USE and GloVe. \textbf{USE} \textit{is the most balanced feature extraction method}: USE embeddings provide consistently strong performance across models. USE achieves high F1-scores in SVM (0.925), RL (0.923), and kNN (0.877), demonstrating its versatility. kNN\_USE (0.877 F1-score) surpasses TF-IDF and GloVe, making it the most effective k-Nearest Neighbors model. TF-IDF performs well for SVM and RL but fails for kNN, while Word2Vec generally underperforms. GloVe provides moderate results but is not as strong as USE or TF-IDF. USE embeddings ensure stable performance, consistently achieving F1-scores above 0.85 across models. Overall, \textit{USE is the best feature extraction method for generalization across models}. Figure \ref{Fig_F1_Comparison_ML_DL_Classifiers_Feature_Extraction_Methods}(a) visualizes F1-score comparisons across feature extraction methods, highlighting USE's consistent superiority and Word2Vec's struggles.
   
    \item \textbf{Best Performing Models Across All Metrics}: \noindent Considering the precision metric, RL with TF-IDF as the feature extraction method achieves the highest precision (0.959), indicating that 96\% of the relevant instances (either ARPs or programming posts) are correctly identified. SVM\_TF-IDF, SVM\_USE, RL\_USE, and NB\_USE closely follow with precisions of 0.952, 0.934, 0.932, and 0.926, respectively, demonstrating comparable performance. SVM and RL consistently outperform other models across all feature extraction methods. \noindent Overall, among the traditional ML models, RL\_TF-IDF (Precision: 0.959, Recall: 0.911, F1-score: 0.934, Accuracy: 0.936) and SVM\_TF-IDF (Precision: 0.952, Recall: 0.917, F1-score: 0.934, Accuracy: 0.935) emerged as the top-performing ML models in this study.
\end{itemize}

\begin{table}[htb]
\centering
\small
\caption{Performance comparison of ML, DL, PLM, and LLM models/classifiers}
\label{Table_Performance_Of_ML_DL_LLM_Models}
    \begin{tabular}{p{9.6em}p{8em}p{4em}p{4em}p{4em}p{4em}}\hline
    \textbf{Category} & \textbf{Model/Classifier} & \textbf{Precision} & \textbf{Recall} & \textbf{F1-score} & \textbf{Accuracy}\\\hline
    % \multicolumn{6}{c}{{\bfseries }} \\\hline
    {ML models} & SVM\_TF-IDF & 0.952 & \underline{\textbf{0.917}} & \underline{\textbf{0.934}} & 0.935 \\
    & SVM\_Word2vec & 0.693 & 0.698 & 0.696 & 0.693 \\
    & SVM\_USE & 0.934 & 0.915 & 0.925 & 0.925 \\
    & SVM\_GloVe & 0.893 & 0.893 & 0.893 & 0.893 \\
    \cline{2-6} %\\\\\\\\\\\\\\\\\\\\\\\\\\\\\\\\\\\\\\\\\\\\\\\\\\\\\\\\\\\\\\\\\\\\\\\\\\\\\\\\\\
    & RL\_TF-IDF & \underline{\textbf{0.959}} & 0.911 & \underline{\textbf{0.934}} & \underline{\textbf{0.936}} \\
    & RL\_Word2vec & 0.667 & 0.669 & 0.668 & 0.666 \\
    & RL\_USE & 0.932 & 0.913 & 0.923 & 0.923 \\
    & RL\_GloVe & 0.889 & 0.889 & 0.889 & 0.889 \\
    \cline{2-6}%\\\\\\\\\\\\\\\\\\\\\\\\\\\\\\\\\\\\\\\\\\\\\\\\\\\\\\\\\\\\\\\\\\\\\\\\\\\\\\\\\\\\\\
    & DT\_TF-IDF & 0.891 & 0.856 & 0.873 & 0.875 \\
    & DT\_Word2vec & 0.608 & 0.601 & 0.604 & 0.605 \\
    & DT\_USE & 0.817 & 0.804 & 0.811 & 0.812 \\
    & DT\_GloVe & 0.729 & 0.729 & 0.729 & 0.729 \\
    \cline{2-6} %\\\\\\\\\\\\\\\\\\\\\\\\\\\\\\\\\\\\\\\\\\\\\\\\\\\\\\\\\\\\\\\\\\\\\\\\\\\\\\\\\\\\\\
    & kNN\_TF-IDF & 0.865 & 0.081 & 0.149 & 0.532 \\
    & kNN\_Word2vec & 0.579 & 0.685 & 0.628 & 0.592 \\
    & kNN\_USE & 0.873 & 0.881 & 0.877 & 0.877 \\
    & kNN\_GloVe & 0.782 & 0.772 & 0.770 & 0.772 \\
    \cline{2-6} %\\\\\\\\\\\\\\\\\\\\\\\\\\\\\\\\\\\\\\\\\\\\\\\\\\\\\\\\\\\\\\\\\\\\\\\\\\\\\\\\\\\\\\
    & NB\_TF-IDF & 0.836 & 0.713 & 0.769 & 0.785 \\
    & NB\_Word2vec & 0.604 & 0.775 & 0.679 & 0.632 \\
    & NB\_USE & 0.926 & 0.866 & 0.895 & 0.899 \\
    & NB\_GloVe & 0.815 & 0.815 & 0.815 & 0.815 \\\hline
    
    {DL models} & TextCNN\_TF-IDF & 0.937 & 0.921 & \underline{\textbf{0.929}} & \underline{\textbf{0.930}} \\
    & TextCNN\_Word2vec & 0.883 & \underline{\textbf{0.951}} & 0.916 & 0.913 \\
    & TextCNN\_USE & 0.924 & 0.914 & 0.919 & 0.920 \\
    & TextCNN\_GloVe & \underline{\textbf{0.938}} & 0.887 & 0.912 & 0.914 \\\hline
    
    {PLMs} & RoBERTa & \underline{\textbf{0.963}} & \underline{\textbf{0.956}} & \underline{\textbf{0.960}} & \underline{\textbf{0.960}} \\
    & BERT & 0.949 & 0.920 & 0.934 & 0.935 \\\hline
    
    {LLM} & GPT-4o & 0.913 & 0.840 & 0.875 & 0.880 \\
    \hline
    \end{tabular}
\end{table}

\fakesection{(2) Performance of DL-based Models}

Likewise, we evaluated the DL-based models and compared their performance with different embeddings (see Table~\ref{Table_Performance_Of_ML_DL_LLM_Models}). The best result for each evaluation metric is underlined, while the highest metric value across all models for a given embedding method is shown in bold. Similar to the ML models, we analyzed the performance of the DL models across four key metrics for each embedding method from two perspectives: 

\begin{itemize}
     \item \textbf{Impact of Feature Extraction Methods}. \textbf{TF-IDF} proves to be \textit{surprisingly effective for deep learning}. Despite being a traditional sparse vector method, TF-IDF remains highly relevant for CNN-based models. As shown in Table \ref{Table_Performance_Of_ML_DL_LLM_Models}, the DL model with TF-IDF outperforms others across most metrics, demonstrating that convolutional layers can effectively capture patterns from sparse representations. Its balanced precision and recall contribute to a strong F1-score, highlighting its robust generalization. TF-IDF consistently achieves the highest F1-scores, making it the top choice for TextCNN models. As shown in Figure \ref{Fig_F1_Comparison_ML_DL_Classifiers_Feature_Extraction_Methods}(b), the overall ranking is TextCNN\_TF-IDF > TextCNN\_USE > TextCNN\_Word2Vec > TextCNN\_GloVe in most cases. This ranking underscores the varying impact of feature extraction methods on DL-based classification models, with TF-IDF offering the best balance, USE delivering consistent performance, Word2Vec excelling in recall, and GloVe prioritizing precision. %\textbf{GloVe} \textit{is the weakest feature extraction method}. GloVe embeddings perform the worst in terms of recall, F1-score, and accuracy compared to other DL-based models. While GloVe works in traditional ML classifiers like RL, it underperforms in CNN models because CNNs rely on well-structured feature maps, which GloVe lacks. \textbf{TF-IDF} \textit{is stable but not the best}. Word2Vec performs better than GloVe but not better than USE and Word2Vec. \textit{Overall, If USE is unavailable, Word2Vec is a good alternative. USE boosts TextCNN to achieve the highest F1-score among all DL models}. 
     
    \item \textbf{Best Performing Models Across All Metrics}: As shown in Table \ref{Table_Performance_Of_ML_DL_LLM_Models}, TextCNN\_GloVe achieves the highest precision, indicating that it excels at avoiding false positives. However, this comes at the expense of recall, as it misses more true positives. In contrast, TextCNN\_Word2vec achieves the highest recall, meaning it is better at identifying all true positives, but this results in a tradeoff with precision. TextCNN\_TF-IDF offers a balanced approach, excelling in F1-score and accuracy, demonstrating strong performance overall without the tradeoff between precision and recall. While all models show high accuracy, TextCNN\_TF-IDF is the top performer, closely followed by TextCNN\_USE. \textit{Conclusion}: TextCNN\_TF-IDF stands out as the most well-rounded model, offering optimal performance across precision, recall, F1-score, and accuracy, making it the most reliable overall.
\end{itemize}

\begin{comment}
\small
\begin{longtable}{p{9.6em}p{8em}p{4em}p{4em}p{4em}p{4em}}
\caption{Performance comparison of DL-based models/classifiers} \label{Table_Performance_Of_ML_DL_LLM_Models} \\\hline
\textbf{Category} & \textbf{Model/Classifier} & \textbf{Precision} & \textbf{Recall} & \textbf{F1-score} & \textbf{Accuracy} \\\hline  
\endfirsthead
\multicolumn{6}{c}
{{\bfseries }}\\
\endhead 
\multicolumn{6}{r}{{}} \\ 
\endfoot
\hline
\endlastfoot
{DL models} 
& TextCNN\_TF-IDF & \underline{0.937} & \underline{0.921}
& \underline{\textbf{0.929}} & \underline{\textbf{0.930}} \\\cline{2-6}
& TextCNN\_Word2vec & 0.883 & \underline{\textbf{0.951}} & 0.916 & 0.913 \\ \cline{2-6}
& TextCNN\_USE & 0.924 & 0.914 & 0.919 & \underline{0.920}\\ \cline{2-6}
& TextCNN\_GloVe & \underline{\textbf{0.938}} & 0.887 & 0.912 & 0.914\\ \cline{1-6}                              
\end{longtable}
\normalsize
\end{comment}

\fakesection{(3) Performance of PLMs}

Among the PLMs, the RoBERTa model outperformed BERT, achieving a Precision of 0.981, a Recall of 0.982, an F1-score of 0.981, and an Accuracy of 0.981 (see Table \ref{Table_Performance_Of_ML_DL_LLM_Models}). 

\begin{comment}
\small
\begin{longtable}{p{9.6em}p{8em}p{4em}p{4em}p{4em}p{4em}}
\caption{Performance comparison of PLMs} \label{Table_Performance_Of_ML_DL_LLM_Models} \\\hline
\textbf{Category} & \textbf{Model/Classifier} & \textbf{Precision} & \textbf{Recall} & \textbf{F1-score} & \textbf{Accuracy} \\\hline  
\endfirsthead
\multicolumn{6}{c}
{{\bfseries }}\\
\endhead 
\multicolumn{6}{r}{{}} \\ 
\endfoot
\hline
\endlastfoot
{PLMs} 
& RoBERTa & \underline{0\textbf{.963}} & \underline{\textbf{0.956}} & \underline{\textbf{0.960}} & \underline{\textbf{0.960}} \\ \cline{2-6}
& BERT & 0.949 & 0.920 & 0.934 & 0.935 \\ \cline{1-6}
\end{longtable}
\normalsize
\end{comment}

\fakesection{(4) Performance of LLM}

The bottom part of Table~\ref{Table_Performance_Of_ML_DL_LLM_Models} presents the performance of the LLM, specifically GPT-4o, in identifying ARPs. GPT-4o achieved good results across all evaluation metrics, with a Precision of 0.913, Recall of 0.840, F1-score of 0.875, and Accuracy of 0.880. 

\begin{comment}
\small
\begin{longtable}{p{9.6em}p{8em}p{4em}p{4em}p{4em}p{4em}}
\caption{Performance of LLM} \label{Table_Performance_Of_LLM_Models} \\\hline
\textbf{Category} & \textbf{Model/Classifier} & \textbf{Precision} & \textbf{Recall} & \textbf{F1-score} & \textbf{Accuracy} \\\hline  
\endfirsthead
\multicolumn{6}{c}
{{\bfseries }}\\
\endhead 
\multicolumn{6}{r}{{}} \\ 
\endfoot
\hline
\endlastfoot
{LLM} 
& GPT-4o & 0.913 & 0.840 & 0.875 & 0.880 \\ \cline{2-6} 
\end{longtable}
\normalsize
\end{comment}

\fakesection{(5) Comparison of the Best Models}

Among the ML, DL, PLM, and LLM models, the best-performing models for identifying ARPs from SO were \textit{RL\_TF-IDF}, \textit{SVM\_TF-IDF}, \textit{TextCNN\_TF-IDF}, and \textit{RoBERTa}. These four models consistently outperformed others (see Table~\ref{Table_Performance_Of_ML_DL_LLM_Models}). To determine the overall best-performing model, we compared the top models in Table~\ref{Table_TopThreePerformingClassifiers} by visualizing their performance using confusion matrices.

\begin{figure}
   \centering
    \subfloat[Confusion matrix of SVM\_TF-IDF ]{\includegraphics[width=.45\linewidth]{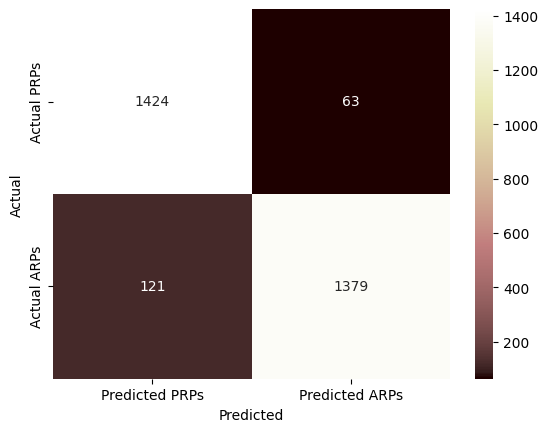}}\hfill
    \subfloat[Confusion matrix of LG\_TF-IDF]{\includegraphics[width=.45\linewidth]{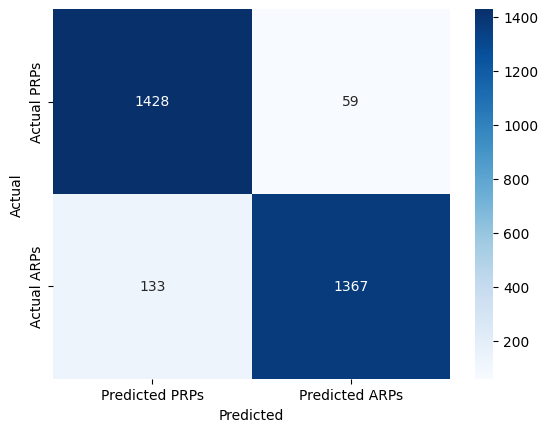}}\hfill
    \caption{Confusion matrices for the best-performing models using traditional ML models.}
    \label{fig_ConfusionMatricesFor_ML_Models}
\end{figure}

\begin{figure}
  \centering
   \subfloat[Confusion matrix of TextCNN\_TF-IDF]{\includegraphics[width=.5\linewidth]{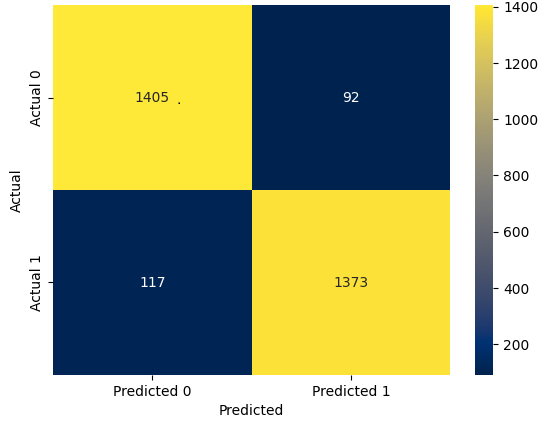}}\hfill
  \subfloat[Confusion matrix for RoBERTa] {\includegraphics[width=.45\linewidth]{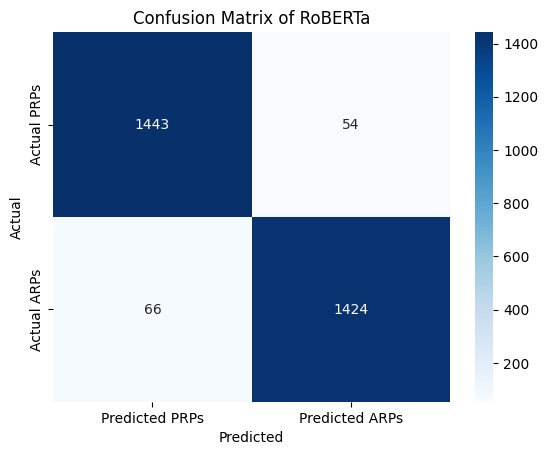}} \hfill 
   \caption{Confusion matrices of the best-performing models using traditional DL models and PLMs.}
   \label{fig_ConfusionMatricesFor_DL_and_PLM}
 \end{figure}

Confusion matrices are widely used in binary classification tasks to report true positives (TP), true negatives (TN), false positives (FP), and false negatives (FN)~\cite{kotsiantis2006machine, caruana2006empirical}. Our dataset maintains a balanced class distribution, with class ``1'' denoting ARPs and class ``0'' representing programming-related posts. Figure~\ref{fig_ConfusionMatricesFor_ML_Models} and Figure~\ref{fig_ConfusionMatricesFor_DL_and_PLM} display the confusion matrices of the top models of ML, DL, and PLMs, respectively.

\begin{itemize}
    \item \textit{True Positives (TP)}: Posts correctly classified as ARPs (label ``1'')
    \item \textit{True Negatives (TN)}: Posts correctly classified as programming-related (label ``0'')
    \item \textit{False Positives (FP)}: Programming-related posts misclassified as ARPs
    \item \textit{False Negatives (FN)}: ARPs misclassified as programming-related posts
\end{itemize}

As the primary goal of the first component \textit{ArchPI} of our framework is to accurately identify ARPs in developer discussions, we selected \textbf{RoBERTa} as the best-performing model (see Table~\ref{Table_TopThreePerformingClassifiers}). This choice is justified by two key findings:

\begin{itemize}
    \item \textit{Consistently Superior Performance}: RoBERTa achieved the highest average scores in Precision, Recall, F1-score, and Accuracy across all evaluated models.
    \item \textit{Strong Classification Results}: As shown in Figure~\ref{fig_ConfusionMatricesFor_DL_and_PLM}(b), RoBERTa correctly identified 1,424 ARPs (TP) and 1,443 programming-related posts (TN) out of 2,987 test posts. It also had the fewest misclassifications, with only 66 FPs and 54 FNs, highlighting its robustness and accuracy.
\end{itemize}

\begin{comment}
For example, RoBERTa successfully identified the following ARP post: 

\textit{``I'm developing a microservices-based system where one of the microservices is a scheduler responsible for handling a large number of dynamic jobs. [...] Is using Kubernetes CronJobs a recommended approach for efficiently managing a large number of dynamic jobs within a microservices architecture?''}  

In contrast, the following was correctly classified as a programming-related post:  

\textit{``I need to convert strings to some form of hash. Is this possible in JavaScript? I'm not utilizing a server-side language so I can't do it that way.''}
\end{comment}

As discussed in Section~\ref{ARPs_Indetification_Component}, we aim to select the most effective model for the \textit{ArchPI} approach, as the first component of our framework. Based on its superior performance, we adopted RoBERTa as the core model for ARP identification in \textit{ArchPI}. 

\begin{table}[htb]
\small
\centering
\caption{Comparison of the performance of the top four best-performing models}
\label{Table_TopThreePerformingClassifiers}
\begin{tabular}{p{10.6em}p{8em}p{4em}p{4em}p{4em}p{4em}}
\hline
\textbf{Category} & \textbf{Model/Classifier} & \textbf{Precision} & \textbf{Recall} & \textbf{F1-score} & \textbf{Accuracy} \\
\hline
\multirow{2}{=}{Best Performing Model in ML models} 
& SVM\_TF-IDF & 0.952 & 0.917 & 0.934 & 0.935 \\
& LG\_TF-IDF & 0.958 & 0.911 & 0.934 & 0.936 \\
\hline
Best Performing Model in DL models 
& TextCNN\_TF-IDF & 0.937 & 0.921 & 0.929 & 0.930 \\
\hline
Best Performing Model in PLMs 
& RoBERTa & \underline{\textbf{0.963}} & \underline{\textbf{0.956}} & \underline{\textbf{0.960}} & \underline{\textbf{0.960}} \\
\hline
\end{tabular}
\normalsize
\end{table}

\begin{comment}
\small
\begin{longtable}{p{10.6em}p{8em}p{4em}p{4em}p{4em}p{4em}}
\caption{Comparison of the performance of the top four best-performing models} \label{Table_TopThreePerformingClassifiers} \\\hline
\textbf{Category} & \textbf{Model/Classifier} & \textbf{Precision} & \textbf{Recall} & \textbf{F1-score} & \textbf{Accuracy} \\\hline  
\endfirsthead
\multicolumn{6}{c}
{{\bfseries }}\\
\endhead 
\multicolumn{6}{r}{{}} \\ 
\endfoot
\hline
\endlastfoot
{Best Performing Model in ML models}                                   
& SVM\_TF-IDF & \underline{0.952} & 0.917 & \underline{0.934} & \underline{0.935} \\ 
& LG\_TF-IDF & \underline{0.958} & 0.911 & \underline{0.934} & \underline{0.936} \\ \cline{1-6}
{Best Performing Model in DL models}                                     & TextCNN\_TF-IDF & 0.937 & \underline{0.921} & 0.929 & 0.930 \\ \cline{1-6}
{Best Performing Model in PLMs}       
& RoBERTa & \underline{0\textbf{.963}} & \underline{\textbf{0.956}} & \underline{\textbf{0.960}} & \underline{\textbf{0.960}}\\ \cline{1-6} 
\end{longtable}
\normalsize
\end{comment}

\noindent
\begin{center}
   \begin{tcolorbox}[colback=black!5, colframe=black!20, width=1.0\linewidth, arc=1mm, auto outer arc, boxrule=1.5pt]
      {\textbf{Key Finding of RQ1}: The RoBERTa model achieved the best performance in identifying ARPs from SO, with a Precision of 0.963, Recall of 0.956, F1-score of 0.960, and Accuracy of 0.960. Therefore, we adopted RoBERTa as the core model for ARP identification in the first component of our proposed approach, \textit{ArchPI}.}
   \end{tcolorbox}
\end{center}
%################################################################################################################

\subsection{RQ2: Evaluating the Performance of Approaches for Extracting Architectural Issue-Solution Pairs}\label{Evaluation_Of_ArchISPE_and_Basilines}
In this section, we evaluate the performance of \textit{ArchISPE} by comparing it with baseline approaches and assessing the extracted issue-solution pairs against a benchmark using established evaluation metrics.

\subsubsection{Benchmark Construction}\label{Benchmark_Construction}
To enable the automatic evaluation of \textit{ArchISPE}, a dedicated benchmark of architectural issue-solution pairs is required. However, to the best of our knowledge, no such dataset currently exists for evaluating techniques that extract architectural issue-solution pairs from online developer communities. To fill this gap, we constructed \textit{ArchISPBench}, an architectural benchmark dataset designed to evaluate and compare different extraction approaches. Specifically, \textit{ArchISPBench} consists of pairs, each containing ground-truth issue sentences (\textit{Issue\_S}) and their corresponding solution sentences (\textit{Solution\_S}), formally represented as \( \text{Bench} = \{ \langle \textit{Issue\_S}, \textit{Solution\_S} \rangle \} \). This benchmark not only enables fair comparison between \textit{ArchISPE} and baseline approaches but also serves as a reusable resource for future research, supporting consistent and replicable evaluation of emerging techniques and tools. We have provided the \textit{ArchISPBench} dataset in our replication package \cite{datasetTOSEM}.  
%To enable the automatic evaluation of ArchISPE, a benchmark of architectural issue–solution pairs is needed. However, to the best of our knowledge, no such dataset currently exists for evaluating techniques that extract architectural issue–solution pairs from online developer communities. To this end, we constructed a benchmark dataset from sentences containing ARPs on SO. It enables fair comparison between ArchISPE and baselines. Furthermore, it can be a reusable resource for future research, supporting consistent and replicable evaluation of emerging techniques and tools.

%\textbf{Benchmark Construction}. We introduce \textit{ArchISPBench}, an architectural benchmark designed for evaluating approaches that extract architectural issue-solution pairs from developer communities. For tasks like issue and solution extraction, an ideal benchmark consists of pairs, each containing ground truth issue sentences (\textit{Issue\_S}) and solution sentences (\textit{Solution\_S}), i.e., \( \text{Bench} = \{ \langle \textit{Issue\_S}, \textit{Solution\_S} \rangle \} \).

\underline{\textit{Data Collection}}. Our framework integrates two complementary components: the output of the first component (\textit{ArchPI}) serves as the input to the second component (\textit{ArchISPE}). \textit{ArchPI}, a fine-tuned RoBERTa model, was selected as the best-performing technique, achieving the highest scores across all evaluation metrics (see Section \ref{Evaluation_of_Models_for_IdentifyingARPs}). As shown in Figure~\ref{fig_ConfusionMatricesFor_DL_and_PLM}(b), \textit{ArchPI} correctly identified 1,424 ARPs (true positives) from a test set of 2,987. \textit{ArchISPE} took these 1,424 ARPs as input to extract architectural issue-solution pairs. To build a benchmark dataset, the first and ninth authors randomly sampled a subset of 367 ARPs (comprising 367 questions and 1,967 answers). Random sampling was employed to ensure topic diversity and avoid bias toward specific architectural concerns~\cite{uddin2020mining}. The selected sample size exceeds the threshold required for a 95\% confidence level with a 5\% margin of error~\cite{israel1992determining}. From each ARP, the first and ninth authors manually extracted the architectural issue (from the question post) and the corresponding solution(s) (from the answer post).

\textit{\underline{Data Cleaning}}. We applied the same preprocessing steps in Section~\ref{ARPs_Indetification_Component} to process the 367 ARPs. We removed HTML tags and other noise, and then used NLTK to tokenize the text into sentences. Hyperlinks embedded in text were replaced with the placeholder \texttt{[external-link]}, while low-frequency tokens such as code, tables, and architectural diagrams were replaced with \texttt{[code-snippet]}, \texttt{[table]}, and \texttt{[figure]}, respectively.
%\textit{\underline{Data Cleaning}}. We applied the same text preprocessing approach as in Section \ref{ARPs_Indetification_Component}, to preprocess the 367 ARPs. Specifically, we utilized the Beautiful Soup\footnote{\url{https://www.crummy.com/software/BeautifulSoup/}} library and applied several heuristics to reduce noise, such as removing HTML tags. Next, we employed the NLTK library to tokenize each text into individual sentences. For sentences containing both hyperlinks and text, we retained the text and replaced the hyperlinks with the placeholder '[external-link]'. Additionally, low-frequency token such as source code, tables, and images or architectural diagrams (if present) in the questions or answers were replaced with placeholders, such as '[code-snippet]', '[table]', and '[figure]', respectively. After this preprocessing, each ARP (question and answer) served as the labeled material for further analysis.

\textit{\underline{Key Architectural Elements in ARPs}}. Our recent empirical study \cite{de2024users} revealed that SO users (e.g., developers) value the explicit presence of certain key architectural elements in architecture-related questions. These elements help clarify the architecture under design and include design context, architectural concerns, system requirements, component types, and architectural decisions. Moreover, we found that developers (e.g., answer seekers) often rely on architectural solutions on SO, such as architectural patterns (e.g., Model-View-Controller), tactics (e.g., Heartbeat), frameworks (e.g., Django), protocols (e.g., OAuth 2.0), and APIs (e.g., REST API) during development \cite{de2024oss, de2023characterizing}. Developers also appreciate when these elements are explicitly stated in answer posts.

Consequently, we manually labeled important sentences in ARPs based on the presence of the aforementioned elements and their roles in architectural discussions. To maintain focus and conciseness, detailed definitions of these elements are provided in the replication package \cite{datasetTOSEM}. Specifically, the elements include \textit{Design Context}, \textit{Architecture Concern}, \textit{System Requirements}, \textit{Architecture Decision}, \textit{Architecture Pattern}, \textit{Architecture Tactic}, \textit{Framework}, \textit{Database System}, \textit{Component Type}, and \textit{Architecture Diagram}.

\textit{\underline{Labeling Process}}. The benchmark of architectural issue-solution pairs was created by two annotators and later inspected and verified by an expert in software engineering and architectural design. Specifically, the first and ninth authors of this study participated in the labeling process. Both annotators have extensive research experience in architecture design. The annotators independently labeled important sentences in 367 ARPs, following established NLP data labeling standards \cite{rodrigues2014sequence, xu2020review}. They identified sentences relevant to the task at hand, ensuring alignment with architectural elements and their roles in architectural discussion. The entire labeling process took approximately 112 hours. The steps for data labeling are outlined below:

\begin{enumerate}
\item Firstly, we prepared labeling materials to guide the annotators through the annotation process. Specifically, we organized the content of each ARP into an MS Excel file containing the question and answer descriptions, the post ID, a hyperlink to the corresponding SO post, and a blank section for labeling.

\item The goal of the labeling process was to select key sentences from each question and answer that explicitly state issues or solutions, forming concise and self-contained issue-solution pairs. 
Following previous NLP studies \cite{parveen2015topical, zhong2020extractive}, the annotators selected \( n \) important sentences (6 sentences in our case) from each ARP thread (question or answer) to construct a pair.

\item To conduct a pilot labeling, 50 ARPs were randomly selected from the set of 367 ARPs. 

\item For each ARP, annotators identified architectural elements and their roles in architectural discussions, then synthesized issue-solution pairs by extracting sentences stating or clarifying the issue or solution. Each sentence was labeled as either “important” or “not important” based on its relevance to the question or answer.
%For each ARP, the annotators examined the post to identify the presence of the aforementioned architectural elements and their role in architectural discussions. They then synthesized an issue-solution pair by extracting sentences that state or clarify the issue or solution from the ARP. The labeling task was structured as a binary classification, where each sentence was categorized as either “important” or “not important” based on its relevance to the question or answer.

\item We compared the issue-solution summaries extracted from the pilot labeling, and any discrepancies were resolved through discussion. We got a Cohen's Kappa value of 0.893 on the pilot labeling results, indicating moderate agreement between the annotators \cite{mchugh2012interrater}.
%\item Disagreements between the annotators were addressed through consultation and consensus building. 

\item An expert reviewed the summaries to check completeness, remove redundancy, and resolve inconsistencies. 

\item Necessary revisions were made based on expert feedback to improve the accuracy and quality of the summaries.
\item The above steps were repeated for the remaining ARPs. 
\end{enumerate}

\textbf{Benchmark Statistics}. The benchmark consists of 5,234 manually labeled key sentences (1,970 from questions and 3,264 from answers) extracted from ARPs on SO. Each issue or solution summary consists of no more than six carefully selected sentences from the corresponding question and answer posts, depending on the available sentences in the question or answer descriptions.

\subsubsection{Baselines} 
Our task can be framed as a text extraction problem. Accordingly, we compared \textit{ArchISPE} against two groups of baselines: (1) State-of-the-art Question-Answer (Q-A) extraction approaches in the SE domain, including \textit{DECA\_PD} and \textit{DECA\_SP} \cite{di2015development} and \textit{ChatEO} \cite{chatterjee2021automatic}; and (2) State-of-the-art text extraction/summarization methods from the NLP domain, including \textit{LexRank} \cite{erkan2004lexrank}, \textit{BertSum} \cite{liu2019fine}, \textit{TextRank} \cite{mihalcea2004textrank}, \textit{Latent Semantic Analysis (LSA)} \cite{haiduc2013automatic}, and \textit{Luhn} \cite{luhn1958automatic}. Note that, these NLP methods were fine-tuned or adapted to extract architectural issue-solution pairs. In total, this comparison involves eight baselines (see Table~\ref{Table_Performance_Of_Issue_Solution_Extraction_Techniques_SE_Autom} and Table~\ref{Table_Performance_Of_Issue_Solution_Extraction_Techniques_NLP_Autom}).

\textbf{Q-A Extraction in SE Domain}. \textit{DECA\_PD \& DECA\_SP} \cite{di2015development} is an approach that uses natural language parsing to analyze email content from development-related discussions. It classifies email sentences into categories such as problem discovery, solution proposal, and information giving, among others. We employed \textit{DECA\_PD}, which was implemented to identify “problem discovery”, as our baseline for architectural issue extraction, whereas \textit{DECA\_SP}, designed for “solution proposal”~\cite{di2015development}, was used as our baseline for architectural solution extraction. \textit{CNC\_PD \& CNC\_SP}~\cite{huang2018automating} is a DL-based approach for classifying sentences in comments extracted from online issue reports. It first learns a matrix representation of each sentence using word embeddings~\cite{mikolov2013distributed}. Then, it builds a CNN ~\cite{haykin1994neural} to categorize sentences into seven intention classes, including \textit{Feature Request}, \textit{Solution Proposal}, and \textit{Problem Discovery}. We used \textit{CNC\_PD}, which predicts \textit{Problem Discovery}, as a baseline for detecting architectural issues. \textit{CNC\_SP}, introduced in the same study, serves as the classifier for \textit{Solution Proposal} identification and was used in our study as a baseline for extracting architectural solutions. %S\&M Q-A \cite{shrestha2004detection} is an approach for question–answer detection within discussion threads. The S\&M algorithm employs RIPPER, a rule-based learning approach, to train a model that classifies each sentence as a question or non-question, and as an answer or non-answer, based on part-of-speech features. Operating at the sentence level, S\&M treats a sentence as a question segment if it contains a sentence annotated by a human as a question. Likewise, a sentence is labeled as an answer segment if it contains at least one sentence marked as an answer to a previously identified question. 
\textit{ChatEO} \cite{chatterjee2021automatic} is an approach for extracting opinion Q\&A pairs from software developer chats. It processes a developer chat history and performs extraction in three major steps: (1) individual conversations are separated from the interleaved chat using conversation disentanglement; (2) conversations that begin with an opinion-seeking question are identified using textual heuristics; and (3) one or more answers to the opinion-seeking question within the conversation are identified using a DL-based technique. 

\textbf{Text Extraction in NLP Domain}. \textit{TextRank} is a graph-based, unsupervised text extraction technique introduced by Mihalcea and Tarau \cite{mihalcea2004textrank}, inspired by Google's PageRank algorithm~\cite{brin1998pagerank}, which ranks web pages. \textit{TextRank} constructs a graph where sentences serve as nodes, and edges represent sentence similarity. The algorithm assigns scores to sentences based on their interconnections. It has been effectively applied in SE research for text extraction tasks, such as summarizing API discussions \cite{naghshzan2021leveraging} and reviews \cite{uddin2017automatic}. In our study, we implemented \textit{TextRank} using the PyTextRank\footnote{\url{https://pypi.org/project/pytextrank/}} Python library and fine-tuned it on ARPs to extract issue-solution pairs in our dataset.

\textit{LexRank Algorithm.} \textit{LexRank} \cite{erkan2004lexrank} is an unsupervised text extraction technique that ranks sentences based on graph centrality. Like TextRank, it constructs a graph where sentences are nodes and edges represent sentence similarity. However, \textit{LexRank} applies a similarity threshold, resulting in a sparser graph that improves sentence selection by filtering weak connections. We implemented \textit{LexRank} using the NetworkX\footnote{\url{https://networkx.org/}} library and fine-tuned it on ARPs to extract issue–solution pairs.

\textit{Latent Semantic Analysis (LSA)} is widely used in information retrieval and extraction. It identifies contextual relationships among words through statistical computations \cite{haiduc2013automatic}. We implemented LSA using the Scikit-learn library and fine-tuned it on our dataset to extract issue-solution pairs.

\textit{Luhn Algorithm} \cite{luhn1958automatic} is a heuristic-based text extraction technique that evaluates sentence scores based on the frequency of significant words. Akin to LSA algorithm, we employed Scikit-learn\footnote{\url{https://scikit-learn.org/stable/}} library in the implementation of Luhn. 

\textit{BertSum} \cite{liu2019fine} is an extractive summarization model that encodes sentences using BERT and employs a transformer-based classifier to select salient sentences. It outperformed several earlier approaches \cite{narayan2018ranking, zhou2018neural} on widely used summarization datasets such as NYT \cite{sandhaus2008new} and CNN/DailyMail \cite{documentsummarization2022}. In our experiment, we used the \textit{BertSum} checkpoint that achieved the best performance on CNN/DailyMail and fine-tuned the model on ARPs to extract architectural issues and their corresponding solutions.

\subsubsection{Evaluation Metrics}. Similar to prior work \cite{kou2023automated}, we employed three metrics: precision, recall, and F1-score, to measure the performance of our \textit{ArchISPE} approach and the comparison baselines in extracting architectural issue-solution pairs from SO. %~\cite{kou2023automated}. 
Each metric is calculated at the sentence level. %Given a set of importance sentences in a set of SO posts (questions or answer) \( Y \), let \( X \) be the set of important sentences extracted by a model. The precision of the model is calculated as \( \frac{|Y \cap X|}{|X|} \). The recall of the model is defined as \( \frac{|Y \cap X|}{|Y|} \). Furthermore, we measure the F1 score, which combines the precision and recall of a model into a single metric by taking their harmonic mean.
Given a set of important sentences in a set of SO posts (questions or answers), denoted as \( Y \), and the set of important sentences extracted by a model, denoted as \( X \), the precision of the model is calculated as 

                            \begin{equation}
                            \frac{|Y \cap X|}{|X|}
                            \end{equation}

\noindent where \( |Y \cap X| \) represents the number of relevant sentences shared between the ground-truth and the sentence extracted by the model, and \( |X| \) is the total number of sentences generated by the model. The recall of the model is defined as

                            \begin{equation}
                            \frac{|Y \cap X|}{|Y|}
                            \end{equation}
                            
\noindent where \( |Y| \) is the total number of relevant sentences in the ground-truth. Additionally, the F1-score is computed, which combines the precision and recall into a single metric by taking their harmonic mean, providing a balanced measure of the model's performance in terms of both relevance and coverage.

\subsubsection{Evaluation Results}

Table~\ref{Table_Performance_Of_Issue_Solution_Extraction_Techniques_SE_Autom} and Table~\ref{Table_Performance_Of_Issue_Solution_Extraction_Techniques_NLP_Autom} present the automatic evaluation results of various approaches for extracting architectural issue-solution pairs. The highest metric values within each domain (SE or NLP) are underlined and highlighted in bold.
%For each approach, the highest scores of each metric are underlined for vertical comparison, while the best results of each metric are highlighted as bold text. \textit{ArchISPE} consistently outperforms all baselines across all three metrics in extracting architectural issues and their corresponding solutions from ARPs on SO. %We presented examples of architectural issue-solution pairs extracted by the top three best-performing models in Figures \ref{fig_Issue_Solution_Extraction_ArchISPE}, \ref{fig_Issue_Solution_Extraction_TextRank}, and \ref{fig_Issue_Solution_Extraction_DECA}. 

\begin{table}[htb]
\small
\centering
\caption{Performance comparison of ArchISPE and baselines in the SE domain}
\label{Table_Performance_Of_Issue_Solution_Extraction_Techniques_SE_Autom}
\begin{tabular}{p{6.5em}p{6em}p{9em}p{4em}p{4em}p{4em}}
\hline
\textbf{Domain} & \textbf{Approach} & \textbf{Question and Solution Extraction} & \textbf{Precision} & \textbf{Recall} & \textbf{F1-score} \\
\hline
\multirow{10}{=}{\textit{Software Engineering Domain}} 

& \textbf{ArchISPE} 
& Issue\_Extraction & \underline{\textbf{0.884}} & \underline{\textbf{0.885}} & \underline{\textbf{0.883}} \\
& & Solution\_Extraction & \underline{\textbf{0.898}} & \underline{\textbf{0.892}} & \underline{\textbf{0.894}} \\
\cline{2-6}

& DECA\_PD \& DECA\_SP \cite{di2015development}
& Issue\_Extraction & 0.682 & 0.556 & 0.540 \\
& & Solution\_Extraction & 0.650 & 0.611 & 0.571 \\
\cline{2-6}

& CNC\_PD \& CNC\_SP \cite{huang2018automating}
& Issue\_Extraction & 0.535 & 0.510 & 0.420 \\
& & Solution\_Extraction & 0.472 & 0.452 & 0.448 \\
\cline{2-6}

& ChatEO \cite{chatterjee2021automatic}
& Issue\_Extraction & 0.535 & 0.510 & 0.500 \\
& & Solution\_Extraction & 0.462 & 0.421 & 0.414 \\
\cline{2-6}

% & S\&M Q-A \cite{shrestha2004detection}
% & Issue\_Extraction & 0.445 & 0.420 & 0.400 \\
% & & Solution\_Extraction & 0.352 & 0.331 & 0.304 \\
\hline
\end{tabular}
\normalsize
\end{table}
%######################################################################
\begin{table}[htb]
\small
\centering
\caption{Performance comparison of ArchISPE and baselines in natural language processing domain}
\label{Table_Performance_Of_Issue_Solution_Extraction_Techniques_NLP_Autom}
\begin{tabular}{p{5em}p{5em}p{12em}p{4em}p{4em}p{4em}}
\hline
\textbf{Domain} & \textbf{Approach} & \textbf{Question and Solution Extraction} & \textbf{Precision} & \textbf{Recall} & \textbf{F1-score} \\
\hline
\multirow{2}{=}{\textit{SE Domain}} 
& \textbf{ArchISPE} & Issue\_Extraction & \underline{\textbf{0.884}} & \underline{\textbf{0.883}} & \underline{\textbf{0.883}} \\
&  & Solution\_Extraction & \underline{\textbf{0.893}} & \underline{\textbf{0.888}} & \underline{\textbf{0.890}} \\
\hline

\multirow{10}{=}{\textit{NLP Domain}} 
& TextRank & Issue\_Extraction & 0.719 & 0.640 & 0.675 \\
&  & Solution\_Extraction & 0.734 & 0.670 & 0.700 \\
\cline{2-6}

& Luhn & Issue\_Extraction & 0.602 & 0.593 & 0.595 \\
&  & Solution\_Extraction & 0.592 & 0.573 & 0.580 \\
\cline{2-6}

& LSA & Issue\_Extraction & 0.573 & 0.540 & 0.553 \\
&  & Solution\_Extraction & 0.554 & 0.517 & 0.531 \\
\cline{2-6}

& Lexrank & Issue\_Extraction & 0.562 & 0.549 & 0.554 \\
&  & Solution\_Extraction & 0.540 & 0.516 & 0.526 \\
\cline{2-6}

& BertSum & Issue\_Extraction & 0.040 & 0.026 & 0.031 \\
&  & Solution\_Extraction & 0.037 & 0.027 & 0.031 \\
\hline
\end{tabular}
\normalsize
\end{table}

\subsubsection{Ablation Study}\label{AblationStudy}
We conducted an ablation study to assess the contribution of individual features integrated into \textit{ArchISPE}. Specifically, we developed three ablated variants of \textit{ArchISPE}, each excluding one or more feature extractors: \textbf{ArchISPE-BERTOverflow}, which omits the contextual feature extractor; \textbf{ArchISPE-TextCNN}, which removes the local feature extractor; and \textbf{ArchISPE-LP-Heu}, which excludes both linguistic and heuristic feature extractors. Table~\ref{Table_Ablation_Study} reports the performance of the full model and its variants.

\textbf{Contextual embeddings} (BERT) were particularly influential in the extraction of architectural issue-solution pairs. Excluding them led to a consistent decline in both precision and recall, with F1-score reductions of about 1.25\% for issue extraction and 2.46\% for solution extraction, highlighting their effectiveness in capturing sentence-level semantics.

\textbf{Local features} captured by TextCNN also contributed meaningfully. Removing this component resulted in moderate F1-score decreases of around 0.91\% and 1.68\% for issue and solution extraction, respectively, suggesting that local semantic patterns enhance - but do not dominate - overall performance.

The largest performance degradation occurred when both \textbf{linguistic patterns} and \textbf{heuristic features} were excluded. This variant exhibited F1-score drops of about 16.3\% for issue extraction and 13.5\% for solution extraction, underscoring their critical role in identifying architectural cues and structural characteristics.

In summary, while all feature types enhance performance, linguistic and heuristic cues are particularly vital, with contextual and local semantic features providing essential complementary signals. Overall, the \textit{ArchISPE} model consistently outperforms all ablated versions across both architectural issue and solution extraction tasks, confirming the complementary value of the different feature types.

\begin{table}[h]
    \centering
    \small
    \caption{Contribution of each feature in ArchISPE}
    \label{Table_Ablation_Study}
    \begin{tabular}{p{3.5cm}p{1.3cm}p{1.3cm}p{1.3cm}}
        \toprule
        \textbf{Model} & \textbf{Precision} & \textbf{Recall} & \textbf{F1-score} \\
        \midrule
        \textbf{ArchISPE} & & & \\
        \hspace{1em}Issue Extraction   & \textbf{\underline{0.884}}  & \textbf{\underline{0.885}}   & \textbf{\underline{0.883}} \\
        \hspace{1em}Solution Extraction & \textbf{\underline{0.898}}  & \textbf{\underline{0.892}}   & \textbf{\underline{0.894}}  \\
        \midrule
        \textit{ArchISPE-BERTOverflow} & & & \\
        \hspace{1em}Issue Extraction    & 0.873  & 0.871  & 0.872 \\
        \hspace{1em}Solution Extraction & 0.875  & 0.871  & 0.872 \\
        \midrule
        \textit{ArchISPE-TextCNN} & & & \\
        \hspace{1em}Issue Extraction    & 0.877  & 0.875  & 0.875 \\
        \hspace{1em}Solution Extraction & 0.883  & 0.878  & 0.879 \\
        \midrule
        \textit{ArchISPE-LP-Heu} & & & \\
        \hspace{1em}Issue Extraction    & 0.741  & 0.739  & 0.739 \\
        \hspace{1em}Solution Extraction & 0.777  & 0.772  & 0.773 \\
        \bottomrule
    \end{tabular}
\end{table}

\noindent\begin{center}
   \begin{tcolorbox}[colback=black!5, colframe=black!20, width=1.0\linewidth, arc=1mm, auto outer arc, boxrule=1.5pt]
             {{\textbf{Key Finding of RQ2}: ArchISPE achieves a Precision, of 0.884, a Recall of 0.885, and an F1-score of 0.883 regarding architectural issue extraction, outperforming the best-performing baseline approaches. It also achieves a Precision of 0.897, a Recall of 0.892, and an F1-score of 0.894 regarding architectural solution extraction, surpassing the best-performing baseline approaches.
             }} 
   \end{tcolorbox}
\end{center}
%##############################################################################################

\subsection{RQ3: User Study Evaluation Results}\label{User_Study_Evaluation}

\subsubsection{Practitioners' Feedback on the Quality of ArchPI} 
\begin{figure}[H] 
 \centering
\includegraphics[width=\textwidth]{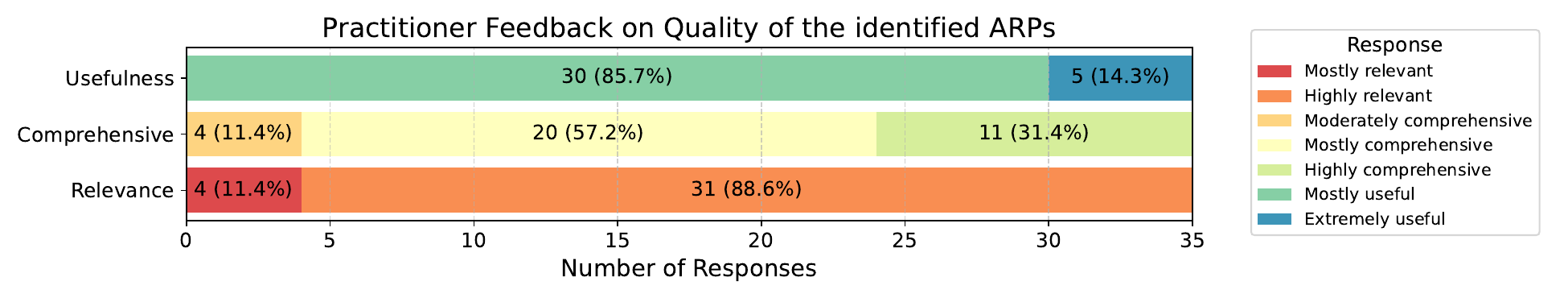}
\caption{Participants' responses on the relevance, comprehensiveness, and usefulness of the identified ARPs}
\label{ArchPI_Participants_Feedback_Relevance_Comprehensiveness_Usefulness}
\end{figure}

Figure~\ref{ArchPI_Participants_Feedback_Relevance_Comprehensiveness_Usefulness} presents the distribution of participants' responses on the perceived relevance, comprehensiveness, and usefulness of the ARPs identified by \textit{ArchPI}. 

As shown in Figure~\ref{ArchPI_Participants_Feedback_Relevance_Comprehensiveness_Usefulness}, a total of 31 responses (88.6\%) from practitioners rated the identified ARPs as \textit{highly relevant}, while 4 responses (11.4\%) classified them as \textit{mostly relevant}. These findings suggest that the \textit{ArchPI} model consistently identifies architectural content considered relevant by practitioners in real-world discussions. Notably, no participants selected \textit{moderately relevant}, \textit{slightly relevant}, or \textit{not relevant at all}, which further demonstrates the perceived relevance of the identified ARPs. Similarly, most of the responses (31 responses, 88.6\%) from participants rated the identified ARPs as either \textit{highly} or \textit{mostly comprehensive}, indicating that the extracted architectural information sufficiently captures the scope and depth of the discussions. Regarding usefulness, 30 responses (85.7\%) from practitioners deemed the ARPs \textit{mostly useful}, while 5 responses (14.3\%) rated them as \textit{moderately useful}. Although none of the practitioners considered the ARPs \textit{extremely useful}, there were also no responses indicating \textit{slightly useful} or \textit{not useful at all}. This distribution suggests that the model effectively captures essential architectural elements that practitioners find applicable and informative.

% \begin{figure}[H]
%  \centering
% \includegraphics[width=\textwidth]{Figures/Agreement_On_ARPsIndentifier_on_SO.pdf}
% \caption{Participants' responses of the usefulness of architectural issue-solution pairs extracted by the three approaches.}
% \label{user_study_Agreement_On_Issue-Solutions_Extractors_on_SO}
% \end{figure}
%########################################################################################
\subsubsection{Practitioners' Feedback on the Quality of ArchISPE}
\begin{figure}[H]
 \centering
\includegraphics[width=\textwidth]{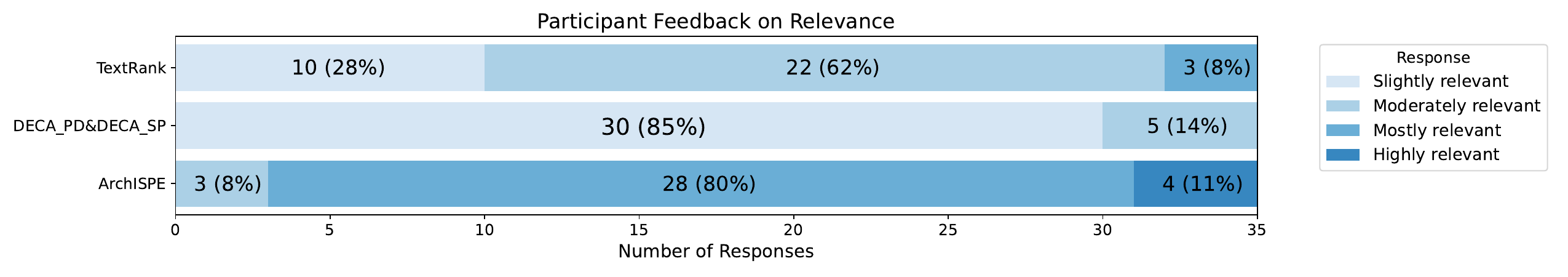}
\caption{Participants' responses on the relevance of architectural issue-solution pairs extracted by the three approaches}
\label{user_study_likert_chart_Relevance}
\end{figure}

\begin{figure}[H]
 \centering
\includegraphics[width=\textwidth]{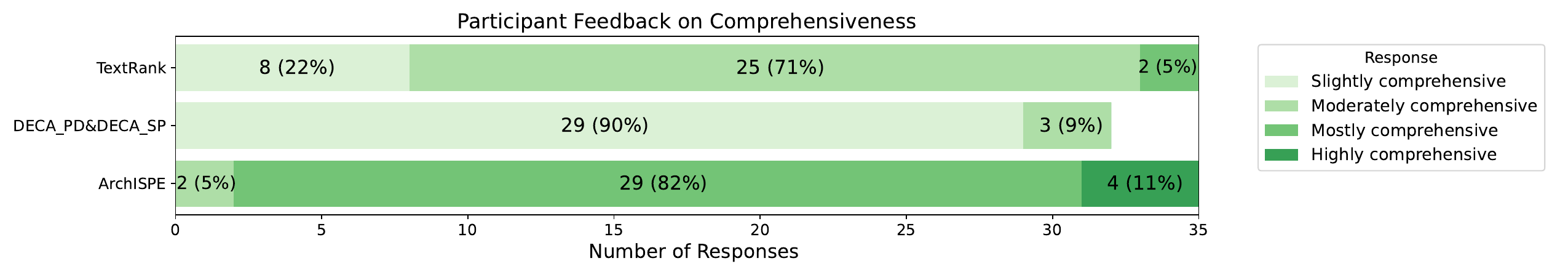}
\caption{Participants' responses on the comprehensiveness of architectural issue-solution pairs extracted by the three approaches.}
\label{user_study_likert_chart_Comprehensiveness}
\end{figure}

\begin{figure}[H]
 \centering
\includegraphics[width=\textwidth]{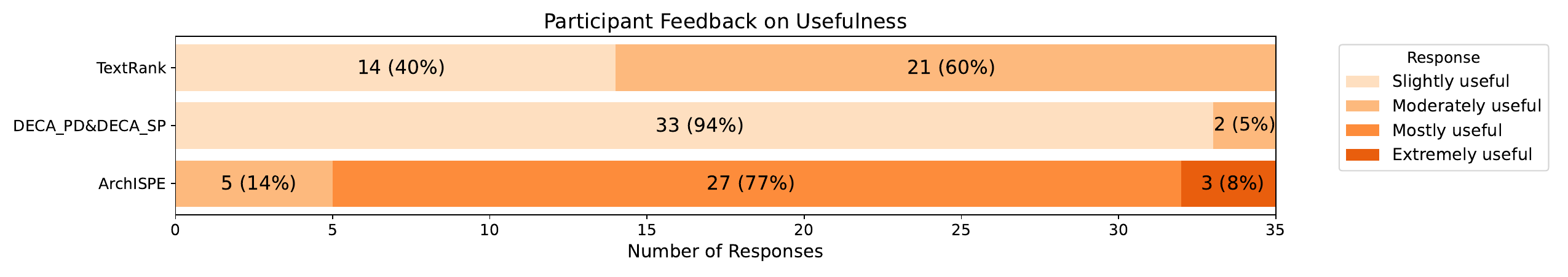}
\caption{Participants' responses to the usefulness of architectural issue-solution pairs extracted by the three approaches.}
\label{user_study_likert_chart_Usefulness}
\end{figure}

Figures~\ref{user_study_likert_chart_Relevance},~\ref{user_study_likert_chart_Comprehensiveness}, and~\ref{user_study_likert_chart_Usefulness} present the distribution of participants' responses on the perceived relevance, comprehensiveness, and usefulness of the architectural issue-solution pairs extracted by the three approaches. 
%The figures reflect participants' feedback on the perceived value of integrating an architectural issue-solution extractor, such as a key sentence highlighter, structured summary, or automated annotation, into SO. 
We summarize and interpret the key findings as follows:

\textbf{(1) Perceived Relevance of Extracted Issue-Solution Pairs}

\textbf{ArchISPE}: From Figure \ref{user_study_likert_chart_Relevance}, the results show that 80\% of the responses considered the extracted issue-solution pairs to be \textit{mostly relevant}, 11\% as \textit{highly relevant}, and 8\% as \textit{moderately relevant}. The results indicate that participants overwhelmingly perceive the pairs extracted by \textit{ArchISPE} as relevant to real-world architecture discussions, with a combined 91\% of responses from \textit{mostly} and \textit{highly relevant}. This reflects good alignment between the results of \textit{ArchISPE} and the expectations of practitioners. The minimal number of \textit{moderately relevant} responses and absence of negative or lower relevance responses further support the consistency and quality of the extracted content. \textbf{TextRank}: 62\% of the responses labeled the results as \textit{moderately relevant}, while only 8\% considered them \textit{mostly relevant}, and 28\% rated them as \textit{slightly relevant}. These findings suggest that \textit{TextRank}'s general-purpose summarization approach lacks domain sensitivity and is less effective in capturing architecture-specific content. \textbf{DECA\_PD \& DECA\_SP}: 80\% of the responses deem the extracted pairs as \textit{slightly relevant}, and 14\% as \textit{moderately relevant}, making this the least effective method among the three evaluated methods. The results are perceived as lacking contextual depth and cohesion, limiting their relevancy in real-world architectural discussions.

\textbf{Summary}. \textit{ArchISPE} is perceived as effective in extracting relevant architectural issue-solution pairs, with 91\% of its results falling into the \textit{mostly} or \textit{highly relevant} categories based on the received responses. In contrast, \textit{TextRank} and \textit{DECA\_PD \& DECA\_SP} received relatively lower feedback ratings, with most results considered only \textit{moderately} or \textit{slightly relevant}. These results demonstrate \textit{ArchISPE}’s ability to extract meaningful architectural content compared to more generic or shallow extraction techniques, such as \textit{TextRank} and \textit{DECA}.

\textbf{(2) Perceived Comprehensiveness of Extracted Issue-Solution Pairs}

Figure~\ref{user_study_likert_chart_Comprehensiveness} shows the distribution of participants' responses regarding the comprehensiveness of the architectural issue-solution pairs extracted by the three approaches.

\textbf{ArchISPE}: 93\% of responses deem the extracted issue-solution pairs either \textit{mostly} (82\%) or \textit{highly comprehensive} (11\%), indicating that \textit{ArchISPE} consistently provides complete and context-rich architectural content. Only 5\% of participants consider the results \textit{moderately comprehensive}, and no negative responses are reported, reflecting a positive perception of the extracted information's completeness. \textbf{TextRank}: 71\% of responses describe the pairs as \textit{moderately comprehensive}, 22\% as \textit{slightly comprehensive}, and only 5\% as \textit{mostly comprehensive}. These results suggest limited depth and incomplete architectural representation, consistent with \textit{TextRank}'s generic summarization approach. 
\textbf{DECA\_PD \& DECA\_SP}: 90\% of responses assess the results as \textit{slightly comprehensive}, and 9\% as \textit{moderately comprehensive}, making this the least comprehensive method. The results highlight a lack of contextual richness and structural completeness in the extracted content.

\textbf{Summary}. ArchISPE is perceived as the most comprehensive, with 93\% of its results falling into the \textit{mostly} or \textit{highly comprehensive} categories. In contrast, \textit{TextRank} and \textit{DECA\_PD \& DECA\_SP} received less favorable responses, indicating worse performance in capturing complete architectural issue-solution pairs. These findings reinforce \textit{ArchISPE}'s strength in delivering results that are both relevant and sufficiently comprehensive for architectural tasks.

\textbf{(3) Perceived Usefulness of Extracted Issue-Solution Pairs}

Figure~\ref{user_study_likert_chart_Usefulness} illustrates participants' responses regarding the usefulness of the architectural issue-solution pairs extracted by the three approaches. \textbf{ArchISPE.} 85\% of the responses deem the pairs \textit{mostly useful} (77\%) or \textit{highly useful} (8\%), with an additional 5\% deeming them \textit{moderately useful}. No participant deems them \textit{slightly useful} or \textit{not useful at all}, reinforcing the positive reception and value of \textit{ArchISPE} for architectural knowledge extraction. \textbf{TextRank.} 60\% of the responses deem the results \textit{moderately useful} and 40\% \textit{slightly useful}, with no \textit{mostly useful} or \textit{highly useful} assessments. This suggests that \textit{TextRank} provides some architectural insight but falls short of delivering actionable, context-specific information. \textbf{DECA\_PD \& DECA\_SP.} 94\% of the responses deem the pairs \textit{slightly useful} and 5\% \textit{moderately useful}, positioning these methods as the least useful among those evaluated. Participants found the extracted results lacking in both structure and contextual relevance, underscoring the need for more targeted architectural extraction techniques.

\textbf{Summary}. In terms of perceived usefulness, \textbf{ArchISPE} outperformed both baseline methods, with 85\% of the responses indicating the results were either \textit{mostly} or \textit{highly useful}. In contrast, \textbf{TextRank} was perceived as only \textit{moderately} or \textit{slightly useful}, while \textbf{DECA\_PD \& DECA\_SP} received overwhelmingly \textit{slightly useful} responses. These results highlight \textit{ArchISPE}’s effectiveness not only in identifying relevant and comprehensive content but also in providing architectural knowledge that is actionable and valuable for supporting architectural design and decision-making activities.

\begin{figure}[H]
 \centering
\includegraphics[width=\textwidth]{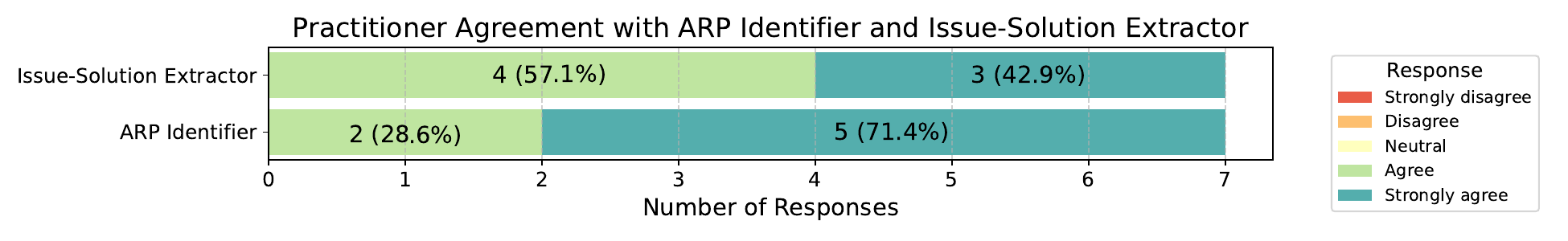}
\caption{Participants' agreement on having ARP identifier and Architectural issue-solution extractor on SO.}
\label{user_study_Agreement_On_Issue_Solutions_Extractors_on_SO}
\end{figure}

\textbf{Practitioners’ Feedback on ARP Identifiers and Issue-Solution Extractors for Stack Overflow}.  
In Figure~\ref{user_study_Agreement_On_Issue_Solutions_Extractors_on_SO}, all participants either \textit{strongly agreed} (5 responses, 71.4\%) or \textit{agreed} (2 responses, 28.6\%) that an automated \textit{ARP identifier}, such as a dedicated label, icon, or visual marker distinguishing ARPs from general programming content, would be useful for browsing SO. Similarly, all participants either \textit{agreed} (4 responses, 57.1\%) or \textit{strongly agreed} (3 responses, 42.9\%) that an automated \textit{architectural issue-solution extractor} (e.g., a key-sentence highlighter, structured summary, or annotation tool) would enhance their experience on the SO platform. These findings highlight the practicality of our proposed framework \textbf{ArchISMiner}, demonstrating its ability to effectively identify ARPs and produce issue-solution pairs that are not only relevant and comprehensive but also potentially impactful in real-world software development.

\noindent\begin{center}
   \begin{tcolorbox}[colback=black!5, colframe=black!20, width=1.0\linewidth, arc=1mm, auto outer arc, boxrule=1.5pt]
   \textbf{Key Finding of RQ3}: The user study results indicate that the ARPs identified by \textit{ArchPI} and the architectural issue-solution pairs extracted by \textit{ArchISPE} are both relevant and valuable for supporting software development tasks. Moreover, all participants expressed agreement and interest in using an ARP identifier and an architectural issue–solution extractor when searching for and reading posts on SO.
   \end{tcolorbox}
\end{center}

\begin{comment}
\subsection{Applicability and Generalizability of the Proposed Framework}
The experiments presented in Sections~\ref{Experiment_Results} and~\ref{User_Study_Evaluation} demonstrated the performance and effectiveness of our proposed framework. In this section, we conduct an additional application study to further illustrate its usefulness and generalizability. Specifically, we applied \textbf{ArchISMiner} to data collected from three additional and highly visited community forums within the Stack Exchange network, \textit{Software Engineering}, \textit{Server Fault}, and \textit{Game Development}. We then employed \textbf{ArchISPE} to extract architectural issue–solution pairs from 7,466 ARPs on Stack Overflow, XXX on Software Engineering, 1,580 on Server Fault, and 1,015 on Game Development. In total, we release an open-source replication package containing a large-scale dataset of over XXXK architectural issue–solution pairs extracted from multiple Q\&A platforms.s
\end{comment}

\section{Implications}\label{Implications}
In this section, we discuss the experimental results and outline several implications (\faLeanpub \hspace{0.5mm}) to guide future research and practice.

\subsection{Implications for Researchers}
\textbf{Choosing an appropriate model for ARPs identification}. In RQ1 (Section~\ref{ARPs_Indetification_Component}), we evaluated a spectrum of models, from traditional ML algorithms to LLMs, to identify the most effective approach for detecting ARPs. Performance varied substantially. For example, \texttt{kNN\_TF-IDF} exhibited poor performance (F1-score of 0.15), whereas \texttt{DT\_Word2vec}, \texttt{kNN\_Word2vec}, and \texttt{RL\_Word2vec} achieved moderate performance (F1-scores between 0.60 and 0.70). A few models, including \texttt{NB\_TF-IDF}, \texttt{kNN\_GloVe}, and \texttt{DT\_GloVe}, demonstrated moderate-to-good performance (F1-scores between 0.70 and 0.80). Good-performing models included \texttt{RL\_GloVe}, \texttt{kNN\_USE}, \texttt{NB\_USE}, and GPT-4o (F1-scores between 0.80 and 0.90). The top-performing models (including \texttt{SVM\_TF-IDF}, \texttt{SVM\_USE}, \texttt{RL\_TF-IDF}, \texttt{RL\_USE}, \texttt{TextCNN\_TF-IDF}, \texttt{TextCNN\_USE}, BERT, and RoBERTa) achieved F1-scores between 0.90 and 0.96. However, identifying the best-performing model was challenging due to the trade-offs between precision and recall observed in our experiments. While some models exhibited conservative results (high Precisions, low Recalls), others favored inclusiveness (high recall, lower precision), as shown in Table \ref{Table_Performance_Of_ML_DL_LLM_Models}. This complexity required a careful balance aligned with the intended use case rather than simply choosing the model with the highest F1-score. Ultimately, we adopted fine-tuned RoBERTa as the Architectural Post Identifier (\textit{ArchPI}) for its \textit{robust classification capabilities} and consistently best performance across multiple metrics. Its adoption substantially reduced noise in downstream data mining, thereby improving the performance (e.g., Precision, F1-score) and reliability of architectural issue-solution pair extraction (Section~\ref{AutomaticExtractionArchitecturalIssue_solution}). Although our study did not systematically evaluate training efficiency, we observed that DL, PLM, and LLMs require substantially more time and computational resources due to their large number of parameters. In many cases, the performance gains (e.g., Accuracy, F1-score) over traditional ML models were modest, indicating that the additional computational cost does not always yield proportionally higher performance. We argue that the widespread adoption of pre-trained LLMs should not overshadow traditional ML models, which may outperform more complex models on certain datasets. \faLeanpub \hspace{0.5mm} Our results suggest that researchers should systematically evaluate multiple models when identifying ARPs from textual artifacts in Q\&A threads, carefully weighing trade-offs between precision, recall, and computational cost (i.e., required CPU/GPU resources, memory, and training/inference time). High computational cost can be a limiting factor when deploying models in real-time systems or on resource-constrained machines. While LLMs are particularly promising due to their capabilities, rigorous comparisons with traditional and intermediate approaches remain essential for software development tasks such as ARP identification and issue-solution pair extraction.

\textbf{Feature integration and model architecture matter for performance}. On Q\&A platforms such as SO, questions and answers are authored by different users, leading to variations in terminology and  conceptual framing of architectural discussions despite their logical connection. Prior research shows that methods based solely on text similarity or TF–IDF fail to capture latent semantic relationships, and Word2Vec models trained on small datasets lack rich semantic representation \cite{lan2023btlink}. To address these limitations, we employed a pre-trained model (\textit{BERTOverflow} \cite{tabassum2020code}) trained on a large corpus of SO posts for mining architectural issue-solution pairs. However, our RQ2 results reveal that leveraging semantic association alone is insufficient. Findings from the ablation study (see Table~\ref{Table_Ablation_Study}) indicate that performance gains depend not only on a strong feature extractor but also on a well-structured model architecture. Removing key components from \textit{ArchISPE} consistently degraded performance, underscoring their role in capturing semantic nuances and structural patterns in architectural discussions.
\faLeanpub \hspace{0.5mm} These findings suggest that pre-trained models alone do not guarantee strong results when extracting architectural issue-solution pairs from SO. Researchers should combine effective feature extractors with thoughtfully designed model architectures and integration strategies to enhance semantic representation and achieve superior performance in development tasks such as architectural issue-solution pair extraction.

\textbf{Toward open-source benchmarks for architectural knowledge extraction}. In RQ2, we employed an iterative and multifaceted process to construct a ground-truth benchmark dataset comprising sentence-level annotations from ARPs. This benchmark offered nuanced insights into the complexity of architectural knowledge shared on Q\&A threads, including SO. Moreover, the resulting benchmark dataset formed the basis for evaluating our \textit{ArchISPE} model and the selected baselines, wherein \textit{ArchISPE} achieved great performance for extracting architectural issues and solutions (see Table \ref{Table_Performance_Of_Issue_Solution_Extraction_Techniques_SE_Autom}). Furthermore, we conducted a human evaluation of the identified ARPs and extracted issue-solution pairs with industry practitioners to gain more insights on ArchISMiner's performance. By assessing our framework via this multifaceted evaluation, we aim to provide empirical evidence of ArchISMiner’s capability on effectively automating ARPs identification and issue-solution pair extraction, benchmarking its performance, and validating its potential to enhance efficiency and scalability in real-world software architecture contexts. Our experience in RQ2 shows that constructing a sentence-level annotated benchmark not only supports robust and reproducible evaluation of issue-solution extraction techniques but also enriches our understanding of how architectural knowledge is communicated in Q\&A forums. While prior studies have examined the extraction and classification of architectural knowledge from Q\&A platforms such as SO \cite{bi2021mining, wijerathna2022mining, soliman2018improving}, they often lack openly available benchmark datasets tailored to architecture-specific tasks. In this study, we proposed an open access architectural benchmark for automatic evaluation of approaches for extracting architectural issue-solution pairs from developer community forums. Our dataset, comprising SO posts (i.e., ARPs and programming-related posts) and sentence-level annotations from ARPs, partially addresses this gap by enabling the automatic evaluation of techniques for identifying ARPs and extracting architectural issue-solution pairs from SO. However, it currently covers only a subset of software architecture design domains. %\faLeanpub \hspace{0.5mm} We advocate for community-wide efforts within the software architecture research field to build and maintain diverse, open datasets that can serve as shared benchmarks to accelerate progress in this area. 
\faLeanpub \hspace{0.5mm} We advocate for community-wide efforts within the software architecture community to build, maintain, and share diverse open datasets that can serve as common benchmarks for architectural knowledge extraction. Such collaborative initiatives would not only accelerate progress in this area but also enhance model generalizability and enable more consistent comparison of results across future studies.

\subsection{Implications for SO Owners}
\textbf{Leveraging architectural issue-solution pairs for enhanced support}. Despite the popularity and success of technical Q\&A sites like SO, the “solution-hungry” problem, where many questions remain unanswered, persists \cite{gao2020technical}. In RQ2, we used our proposed model, \textit{ArchISPE}, to extract architectural issue-solution pairs from ARPs related to task-specific architecture problems (e.g., performance bottlenecks in real-time systems) on SO. These pairs represent a valuable opportunity for SO owners to address the persistent “solution-hungry” problem. \faLeanpub \hspace{0.5mm} Our findings suggest that integrating mined architectural issue-solution pairs into SO can significantly improve user experience by automated solution recommendations. When users post new questions similar to previously discussed architectural problems, the system could automatically suggest relevant solutions, reducing duplicate effort and speeding up problem resolution. Furthermore, such a knowledge base could support intelligent features like automated question linking and answer retrieval, fostering more efficient knowledge sharing and enhancing SO’s role as a reliable source of architectural design guidance during software development.

\textbf{The need for ARP identifiers and issue-solution extractors in online developer communities}. Practitioners often struggle to locate architecture-relevant content amid the vast and noisy discussions on SO \cite{de2024oss, de2022developerssearch, de2024users}. In RQ3, our user study revealed similar frustrations: current search features on SO lack architectural context awareness. Participants found ARPs identified by \textit{ArchPI} and issue-solution pairs extracted by \textit{ArchISPE} both relevant and accurate, particularly valuing \textit{ArchPI} for filtering irrelevant content and simplifying knowledge discovery. They also appreciated \textit{ArchISPE}’s capability to extract meaningful architectural content. Moreover, participants expressed strong interest in platform-level features such as dedicated ARP identifiers or markers (e.g., labels, icons) and architectural issue-solution extractors (e.g., key-sentence highlighters, structured summaries). \faLeanpub \hspace{0.5mm} These results suggest that SO owners could significantly improve user experience by integrating intelligent ARP identifiers and extractors, enabling faster discovery and extraction of architecture-related knowledge and more efficient navigation of architectural discussions.
%Practitioners often struggle to locate architecture-relevant content amid the vast and noisy discussions on SO \cite{de2024oss, de2022developerssearch, de2024users}. In RQ3, our user study participants voiced similar frustrations, noting that current search functionalities lack architectural context awareness. They found the ARPs identified by \textit{ArchPI} to be relevant, useful, and accurately categorized, and particularly appreciated the ArchPI’s performance in filtering out irrelevant content, reducing the effort to locate architectural knowledge on SO. Participants agreed that an automated \textit{ARP identifier}, such as a dedicated label, icon, or visual marker distinguishing ARPs from general programming content, would be useful for browsing SO. They also agreed  that an automated \textit{architectural issue–solution extractor} (e.g., a key-sentence highlighter, structured summary, or annotation tool) would enhance their experience on the SO platform. \faLeanpub \hspace{0.5mm} These findings imply that SO owners could integrate intelligent ARP identifiers into the platform to substantially improve developers’ efficiency in discovering architecture-related knowledge and navigating architectural discussions. 

\subsection{Implications for Tool Designers}

While recent advances in AI-powered retrieval systems (e.g., Google’s AI-based code understanding and generative search tools) can surface relevant architectural information, such as solutions for performance bottlenecks in microservices, they primarily perform \textit{text-level retrieval and summarization}. These systems lack the capability to explicitly identify, model, and link architectural issues with their corresponding solutions. Without this explicit linkage, the architectural rationale remains implicit and fragmented, limiting its usefulness for systematic reuse and decision support. To the best of our knowledge, no prior work has explicitly addressed the \textit{automated extraction and linkage} of architectural issue-solution pairs from developer discussions. Our study partially bridges this gap by introducing a framework that identifies ARPs and mines issue-solution pairs from diverse textual artifacts, such as Q\&A threads. This dual focus enables the capture of both problems and resolutions, producing more context-rich architectural knowledge for practitioners. However, our approach currently covers only a subset of architecture domains and data source. \faLeanpub \hspace{0.5mm} We advocate for future community-wide efforts to design more comprehensive tools that leverage multiple sources, including issue-tracking systems \cite{soliman2021exploratory}, developer chats \cite{shi2020detection}, and user stories \cite{rodeghero2017detecting}, and integrate AI-based retrieval with structured extraction techniques. Such hybrid tools would move beyond generic AI summarization toward \textit{explicit, traceable extraction and reuse of architectural knowledge}, ultimately enhancing developers' ability to reason about architectural trade-offs and decisions. 

\section{Threats to Validity}\label{ThreatsValidity}
In this section, we discuss the potential threats to the validity of our study, along with the measures we adopted to mitigate these threats.

\textbf{Internal Validity}. A key internal threat concerns the labeling of key sentences in SO ARPs, an inherently subjective process where annotators may select different sentences from the same post. To mitigate this threat, we employed a rigorous labeling procedure involving two annotators and an expert in software architecture. All annotators had extensive research experience in architecture design. Disagreements were resolved through discussion, and only sentences agreed upon by both annotators were retained. This process yielded a Cohen’s Kappa score of 0.893, indicating substantial agreement~\cite{fleiss1981measurement}. Another threat relates to the re-implementation of baseline approaches, which might not fully replicate the original methods. To address this, we thoroughly reviewed the original studies, verified implementation details, and ensured correctness through code quality assurance mechanisms such as peer code reviews. These measures significantly reduced the risk of implementation errors.

\textbf{Construct Validity}. One threat to construct validity concerns the selection of learning models, feature extraction methods, and evaluation metrics. As noted by Peter \textit{et al.}~\cite{peters2017text}, exhaustively testing all models is infeasible. To mitigate this threat, we selected a diverse set of techniques - including ML, DL, PLMs, and LLMs - and four widely adopted embedding models, chosen for their proven effectiveness in text classification and extraction tasks. For evaluation, we employed standard metrics such as Recall, Precision, F1-score, and Accuracy, which are widely accepted for assessing automated techniques (see Section~\ref{ARPs_Indetification_Component}). These choices minimize the threat associated with metric selection. Another threat lies in the user study design, as participants were asked to assess the quality of identified ARPs and extracted architectural issue-solution pairs based on the ground-truth benchmark and original SO posts. To reduce bias, we evaluated our framework, \textit{ArchISMiner}, across multiple dimensions, including relevancy, comprehensiveness, and usefulness - criteria commonly used in SE summarization tasks~\cite{yang2022answer, xu2017answerbot, Sorbo2016}. Through the above measures, we substantially mitigated the construct validity threat.

\textbf{External Validity}. External validity relates to the generalizability of our proposed framework, \textit{ArchISMiner}. We evaluated the framework on a dataset of architecture-related discussions extracted from SO. However, the choice of dataset and experimental settings may influence generalizability. Specifically, the selected online community posts may not fully represent all ARPs. To mitigate this threat, future work could incorporate data from diverse sources and obtain direct feedback from practitioners (e.g., Open sources software developers) to better understand how architectural issue-solution pairs are mined in real-world settings. Such efforts would provide deeper insights and strengthen applicability. %Additionally, to enable broader validation and replication, we have made our dataset publicly available~\cite{datasetTOSEM}, allowing researchers and practitioners to replicate, reuse, and extend our study.

\textbf{Reliability}. Reliability concerns the consistency and reproducibility of our results. To enhance it, we ensured that all experimental steps, including data preprocessing, key sentence labeling, model training, hyperparameter configuration, and evaluation, were conducted under controlled and well-documented conditions. The full implementation of \textit{ArchISMiner}, comprising both the \textit{ArchPI} and \textit{ArchISPE} modules, is publicly available together with the dataset and replication package~\cite{datasetTOSEM}. This facilitates replication, verification, and extension of our proposed framework \textit{ArchISMiner} by other researchers. Additionally, we performed multiple runs with fixed random seeds to reduce stochastic variation, following best practices in ML-related research~\cite{madhyastha2019model}, thereby supporting the reproducibility of our results. The high inter-annotator agreement (Cohen’s Kappa values of 0.887 and 0.893) further strengthens the reliability of the manual labeling results of SO posts (see Section \ref{Data_Collection}) and sentences (see Section \ref{Benchmark_Construction}.

\section{Conclusions and Future Work} \label{Conclusion}
In this work, we proposed \textbf{ArchISMiner}, a framework for mining architectural knowledge from online developer communities. \textit{ArchISMiner} integrates two complementary components: \textit{ArchPI} and \textit{ArchISPE}. \textit{ArchPI} is designed to train and evaluate multiple models, including conventional ML/DL techniques, PLMs, and LLMs, selecting the best-performing model to automatically distinguish ARPs from general programming discussions. \textit{ArchISPE}, an indirectly supervised method, leverages a combination of BERT embeddings, local features extracted via TextCNN, linguistic patterns, and heuristic cues to address the heterogeneous nature of SO posts. This enriched representation substantially improves the performance of the \textit{ArchISPE} model in extracting architectural issue–solution pairs. Results from our ablation study further confirm that these complementary feature types jointly enhance the effectiveness of \textit{ArchISPE} within the overall \textbf{ArchISMiner} framework. %This enriched data representation improves performance in extracting architectural issue-solution pairs. Results from the ablation study demonstrate that these complementary feature types jointly enhance the performance of the proposed model on extracting architectural issue-solution pairs from SO posts. 

Given the absence of a benchmark dataset for evaluating the extraction of architectural issue-solution pairs, we found the need to develop such a benchmark. We manually constructed the first architectural issue-solution pairs benchmark dataset from SO, \textbf{ArchISPBench}, which comprises 367 architectural issues and 1,967 solutions, extracted from SO posts, with 5,234 candidate sentences (1,970 from questions and 3,264 from answers). We evaluated \textit{ArchISMiner} using both automatic (by our benchmark) and manual (by a user study) evaluation methods. Results from both evaluations consistently demonstrate that \textit{ArchISMiner} significantly outperforms the baseline approaches from both SE and NLP domains. Moreover, the user study results demonstrate that both identified ARPs and extracted architectural issue-solution pairs are relevant, comprehensive, and useful in practice. We applied \textit{ArchISMiner} to three additional forums, releasing a dataset of over 18K architectural issue-solution pairs. Overall, \textit{ArchISMiner} helps architects and developers identify ARPs and extract succinct, relevant, and useful architectural knowledge from developer communities more accurately and efficiently.

In the future, we plan to integrate our approach into Integrated Development Environments (IDEs) as a plugin or tool to support developers in efficiently identifying ARPs and extracting architectural issue-solution pairs from Q\&A platforms. We also aim to extend the approach to other developer community forums beyond Q\&A sites, such as GitHub, enabling broader applicability across diverse sources of architectural knowledge. Ultimately, our goal is to provide developers with effective tools for the search, retrieval, and synthesis of relevant architectural knowledge, thereby enhancing their efficiency to address architectural design challenges.

\section*{Data Availability}\label{Data_Availability}
The replication package, including the code and the dataset used in this work, has been made available at \cite{datasetTOSEM}.

%% The acknowledgments section is defined using the “acks” environment
\begin{acks}
This work has been partially supported by the National Natural Science Foundation of China (NSFC) with Grant No. 62172311 and 62402348. The numerical calculations in this paper have been done on the supercomputing system in the Supercomputing Center of Wuhan University.
\end{acks}

%\clearpage
% \bibliographystyle{ACM-Reference-Format}
\bibliographystyle{IEEEtran}
\bibliography{basebib}

\begin{thebibliography}{100}
\providecommand{\url}[1]{#1}
\csname url@samestyle\endcsname
\providecommand{\newblock}{\relax}
\providecommand{\bibinfo}[2]{#2}
\providecommand{\BIBentrySTDinterwordspacing}{\spaceskip=0pt\relax}
\providecommand{\BIBentryALTinterwordstretchfactor}{4}
\providecommand{\BIBentryALTinterwordspacing}{\spaceskip=\fontdimen2\font plus
\BIBentryALTinterwordstretchfactor\fontdimen3\font minus \fontdimen4\font\relax}
\providecommand{\BIBforeignlanguage}[2]{{%
\expandafter\ifx\csname l@#1\endcsname\relax
\typeout{** WARNING: IEEEtran.bst: No hyphenation pattern has been}%
\typeout{** loaded for the language `#1'. Using the pattern for}%
\typeout{** the default language instead.}%
\else
\language=\csname l@#1\endcsname
\fi
#2}}
\providecommand{\BIBdecl}{\relax}
\BIBdecl

\bibitem{naghshzan2021leveraging}
A.~Naghshzan, L.~Guerrouj, and O.~Baysal, ``Leveraging unsupervised learning to summarize apis discussed in stack overflow,'' in \emph{Proceedings of the 21st IEEE International Working Conference on Source Code Analysis and Manipulation (SCAM)}, 2021, pp. 142--152.

\bibitem{chen2024empirical}
X.~Chen, F.~Xu, Y.~Huang, X.~Zhou, and Z.~Zheng, ``An empirical study of code reuse between github and stack overflow during software development,'' \emph{Journal of Systems and Software}, p. 111964, 2024.

\bibitem{huang2022towards}
Y.~Huang, F.~Xu, H.~Zhou, X.~Chen, X.~Zhou, and T.~Wang, ``Towards exploring the code reuse from stack overflow during software development,'' in \emph{Proceedings of the 30th IEEE/ACM International Conference on Program Comprehension (ICPC)}, 2022, pp. 548--559.

\bibitem{alnusair2018utilizing}
A.~Alnusair, M.~Rawashdeh, M.~A. Hossain, and M.~F. Alhamid, ``Utilizing semantic techniques for automatic code reuse in software repositories,'' in \emph{Proceedings of the Workshop on Formal Methods Integration (FMI)}, 2018, pp. 42--62.

\bibitem{soliman2021exploring}
M.~Soliman, M.~Wiese, Y.~Li, M.~Riebisch, and P.~Avgeriou, ``Exploring web search engines to find architectural knowledge,'' in \emph{Proceedings of the 18th IEEE International Conference on Software Architecture (ICSA)}, 2021, pp. 162--172.

\bibitem{uddin2020mining}
G.~Uddin, F.~Khomh, and C.~K. Roy, ``Mining api usage scenarios from stack overflow,'' \emph{Information and Software Technology}, vol. 122, p. 106277, 2020.

\bibitem{silva2019recommending}
R.~F. Silva, C.~K. Roy, M.~M. Rahman, K.~A. Schneider, K.~Paixao, and M.~de~Almeida~Maia, ``Recommending comprehensive solutions for programming tasks by mining crowd knowledge,'' in \emph{Proceedings of the 27th IEEE/ACM International Conference on Program Comprehension (ICPC)}, 2019, pp. 358--368.

\bibitem{mahajan2020recommending}
S.~Mahajan, N.~Abolhassani, and M.~R. Prasad, ``Recommending stack overflow posts for fixing runtime exceptions using failure scenario matching,'' in \emph{Proceedings of the 28th ACM Joint Meeting on European Software Engineering Conference and Symposium on the Foundations of Software Engineering (ESEC/FSE)}, 2020, pp. 1052--1064.

\bibitem{barua2014developers}
A.~Barua, S.~W. Thomas, and A.~E. Hassan, ``What are developers talking about? an analysis of topics and trends in {Stack Overflow},'' \emph{Empirical Software Engineering}, vol.~19, no.~3, pp. 19--32, 2014.

\bibitem{treude2011programmers}
C.~Treude, O.~Barzilay, and M.-A. Storey, ``How do programmers ask and answer questions on the web?(nier track),'' in \emph{Proceedings of the 33rd International Conference on Software Engineering (ICSE)}, 2011, pp. 804--807.

\bibitem{de2023characterizing}
M.~J. de~Dieu, P.~Liang, M.~Shahin, and A.~A. Khan, ``Characterizing architecture related posts and their usefulness in stack overflow,'' \emph{Journal of Systems and Software}, vol. 198, p. 111608, 2023.

\bibitem{soliman2017developing}
M.~Soliman, M.~Galster, and M.~Riebisch, ``Developing an ontology for architecture knowledge from developer communities,'' in \emph{Proceedings of the 14th IEEE International Conference on Software Architecture (ICSA)}, 2017, pp. 89--92.

\bibitem{de2024users}
M.~J. de~Dieu, P.~Liang, M.~Shahin, and A.~A. Khan, ``How do users revise architectural related questions on stack overflow: An empirical study,'' \emph{Empirical Software Engineering}, vol.~30, pp. 1--42, 2025.

\bibitem{bi2021mining}
T.~Bi, P.~Liang, A.~Tang, and X.~Xia, ``Mining architecture tactics and quality attributes knowledge in stack overflow,'' \emph{Journal of Systems and Software}, vol. 180, p. 111005, 2021.

\bibitem{jean2024mining}
M.~J. de~Dieu, P.~Liang, M.~Shahin, C.~Yang, and Z.~Li, ``Mining architectural information: A systematic mapping study,'' \emph{Empirical Software Engineering}, vol.~29, no.~4, pp. 1--59, 2024.

\bibitem{feng2024empirical}
Q.~Feng, S.~Liu, H.~Ji, X.~Ma, and P.~Liang, ``An empirical study of untangling patterns of two-class dependency cycles,'' \emph{Empirical Software Engineering}, vol.~29, no.~2, pp. 1--26, 2024.

\bibitem{baldwin2000design}
C.~Y. Baldwin and K.~B. Clark, \emph{Design Rules: The Power of Modularity}.\hskip 1em plus 0.5em minus 0.4em\relax MIT Press, 2000.

\bibitem{de2024oss}
M.~J. de~Dieu, P.~Liang, and M.~Shahin, ``How do oss developers reuse architectural solutions from q\&a sites: An empirical study,'' \emph{IEEE Transactions on Software Engineering}, vol.~51, no.~7, pp. 2015--2043, 2025.

\bibitem{SA2012}
L.~Bass, P.~Clements, and R.~Kazman, \emph{Software Architecture in Practice}, 3rd~ed.\hskip 1em plus 0.5em minus 0.4em\relax Addson-Wesley Professional, 2012.

\bibitem{datasetTOSEM}
M.~J. de~Dieu, R.~Li, P.~Liang, M.~Shahin, M.~Waseem, Z.~Li, B.~Wang, A.~A. Khan, and M.~S. Aktar, ``Replication package for the paper: Archisminer: A framework for automatic mining of architectural issue-solution pairs from online developer communities,'' 2025, \url{https://github.com/JeanMusenga/ArchISMiner}.

\bibitem{xu2017answerbot}
B.~Xu, Z.~Xing, X.~Xia, and D.~Lo, ``Answerbot: Automated generation of answer summary to developers' technical questions,'' in \emph{Proceedings of the 32nd IEEE/ACM International Conference on Automated Software Engineering (ASE)}, 2017, pp. 706--716.

\bibitem{nadi2020essential}
S.~Nadi and C.~Treude, ``Essential sentences for navigating stack overflow answers,'' in \emph{Proceedings of the 27th IEEE International Conference on Software Analysis, Evolution and Reengineering (SANER)}, 2020, pp. 229--239.

\bibitem{yang2023techsumbot}
C.~Yang, B.~Xu, J.~Liu, and D.~Lo, ``Techsumbot: A stack overflow answer summarization tool for technical query,'' in \emph{Proceedings of the 45th IEEE/ACM International Conference on Software Engineering: Companion Proceedings}, 2023, pp. 132--135.

\bibitem{chinnappan2021architectural}
K.~Chinnappan, I.~Malavolta, G.~A. Lewis, M.~Albonico, and P.~Lago, ``Architectural tactics for energy-aware robotics software: A preliminary study,'' in \emph{Proceedings of the 15th European Conference on Software Architecture (ECSA)}, 2021, pp. 164--171.

\bibitem{wijerathna2022mining}
L.~Wijerathna, A.~Aleti, T.~Bi, and A.~Tang, ``Mining and relating design contexts and design patterns from stack overflow,'' \emph{Empirical Software Engineering}, vol.~27, no.~1, pp. 1--53, 2022.

\bibitem{tian2019developers}
F.~Tian, P.~Liang, and M.~A. Babar, ``How developers discuss architecture smells? {An} exploratory study on {Stack Overflow},'' in \emph{Proceedings of the 16th IEEE International Conference on Software Architecture (ICSA)}, 2019, pp. 91--100.

\bibitem{li2021understanding}
R.~Li, P.~Liang, M.~Soliman, and P.~Avgeriou, ``Understanding architecture erosion: The practitioners’ perceptive,'' in \emph{Proceeding of the 29th IEEE/ACM International Conference on Program Comprehension (ICPC)}, 2021, pp. 311--322.

\bibitem{aktar2025architecture}
M.~S. Aktar, P.~Liang, M.~Waseem, A.~Tahir, A.~Ahmad, B.~Zhang, and Z.~Li, ``Architecture decisions in quantum software systems: An empirical study on stack exchange and github,'' \emph{Information and Software Technology}, vol. 177, p. 107587, 2025.

\bibitem{guo2020caspar}
H.~Guo and M.~P. Singh, ``Caspar: Extracting and synthesizing user stories of problems from app reviews,'' in \emph{Proceedings of the 42nd ACM/IEEE International Conference on Software Engineering (ICSE)}, 2020, pp. 628--640.

\bibitem{gao2018online}
C.~Gao, J.~Zeng, M.~R. Lyu, and I.~King, ``Online app review analysis for identifying emerging issues,'' in \emph{Proceedings of the 40th International Conference on Software Engineering (ICSE)}, 2018, pp. 48--58.

\bibitem{gao2019emerging}
C.~Gao, W.~Zheng, Y.~Deng, D.~Lo, J.~Zeng, M.~R. Lyu, and I.~King, ``Emerging app issue identification from user feedback: Experience on wechat,'' in \emph{Proceedings of the 41st IEEE/ACM International Conference on Software Engineering: Software Engineering in Practice (ICSE-SEIP)}, 2019, pp. 279--288.

\bibitem{shrestha2004detection}
L.~Shrestha and K.~McKeown, ``Detection of question-answer pairs in email conversations,'' in \emph{Proceedings of the 20th International Conference on Computational Linguistics (COLING)}, 2004, pp. 889--895.

\bibitem{henss2012semi}
S.~Hen{\ss}, M.~Monperrus, and M.~Mezini, ``Semi-automatically extracting faqs to improve accessibility of software development knowledge,'' in \emph{Proceedings of the 34th International Conference on Software Engineering (ICSE)}, 2012, pp. 793--803.

\bibitem{wang2021automatic}
H.~Wang, X.~Xia, D.~Lo, J.~Grundy, and X.~Wang, ``Automatic solution summarization for crash bugs,'' in \emph{Proceedings of the 43rd IEEE/ACM International Conference on Software Engineering (ICSE)}.\hskip 1em plus 0.5em minus 0.4em\relax IEEE, 2021, pp. 1286--1297.

\bibitem{UnderstQuestQuali2014}
L.~Ponzanelli, A.~Mocci, A.~Bacchelli, and M.~Lanza, ``Understanding and classifying the quality of technical forum questions,'' in \emph{Proceedings of the 14th IEEE International Conference on Quality Software (QoSA)}, 2014, pp. 343--352.

\bibitem{cohen1960}
J.~Cohen, ``A coefficient of agreement for nominal scales,'' \emph{Educational and Psychological Measurement}, vol.~20, no.~1, pp. 37--46, 1960.

\bibitem{campbell2013coding}
J.~L. Campbell, C.~Quincy, J.~Osserman, and O.~K. Pedersen, ``Coding in-depth semistructured interviews: Problems of unitization and intercoder reliability and agreement,'' \emph{Sociological Methods \& Research}, vol.~42, no.~3, pp. 294--320, 2013.

\bibitem{israel1992determining}
G.~D. Israel, ``Determining sample size,'' Florida, USA, Fact Sheet PEOD-6, 1992.

\bibitem{he2009imbalance}
H.~He and E.~A. Garcia, ``Learning from imbalanced data,'' \emph{IEEE Transactions on Knowledge and Data Engineering}, vol.~21, no.~9, pp. 1263--1284, 2009.

\bibitem{ezzini2022automated}
S.~Ezzini, S.~Abualhaija, C.~Arora, and M.~Sabetzadeh, ``Automated handling of anaphoric ambiguity in requirements: a multi-solution study,'' in \emph{Proceedings of the 44th International Conference on Software Engineering (ICSE)}, 2022, pp. 187--199.

\bibitem{abdalkareem2020machine}
R.~Abdalkareem, S.~Mujahid, and E.~Shihab, ``A machine learning approach to improve the detection of ci skip commits,'' \emph{IEEE Transactions on Software Engineering}, vol.~47, no.~12, pp. 2740--2754, 2020.

\bibitem{aggarwal2012survey}
C.~C. Aggarwal and C.~Zhai, ``A survey of text classification algorithms,'' \emph{Mining Text Data}, pp. 163--222, 2012.

\bibitem{caruana2006empirical}
R.~Caruana and A.~Niculescu-Mizil, ``An empirical comparison of supervised learning algorithms,'' in \emph{Proceedings of the 23rd International Conference on Machine Learning (ICML)}, 2006, pp. 161--168.

\bibitem{kotsiantis2006machine}
S.~B. Kotsiantis, I.~D. Zaharakis, and P.~E. Pintelas, ``Machine learning: A review of classification and combining techniques,'' \emph{Artificial Intelligence Review}, vol.~26, no.~3, pp. 159--190, 2006.

\bibitem{bird2009natural}
S.~Bird, E.~Klein, and E.~Loper, \emph{Natural Language Processing with Python: Analyzing Text with the Natural Language Toolkit}.\hskip 1em plus 0.5em minus 0.4em\relax O'Reilly Media, Inc., 2009.

\bibitem{wolf2020transformers}
T.~Wolf, L.~Debut, V.~Sanh, J.~Chaumond, C.~Delangue, A.~Moi, P.~Cistac, T.~Rault, R.~Louf, M.~Funtowicz \emph{et~al.}, ``Transformers: State-of-the-art natural language processing,'' in \emph{Proceedings of the 25th Conference on Empirical Methods in Natural Language Processing (EMNLP)}, 2020, pp. 38--45.

\bibitem{balakrishnan2014stemming}
V.~Balakrishnan and E.~Lloyd-Yemoh, ``Stemming and lemmatization: A comparison of retrieval performances,'' \emph{Lecture Notes on Software Engineering}, vol.~2, no.~3, p. 262, 2014.

\bibitem{ramos2003using}
J.~Ramos \emph{et~al.}, ``Using tf-idf to determine word relevance in document queries,'' in \emph{Proceedings of the 1st Instructional Conference on Machine Learning (ICML)}, 2003, pp. 29--48.

\bibitem{uddin2017mining}
G.~Uddin and F.~Khomh, ``Mining api aspects in api reviews,'' \emph{Technical Report}, 2017.

\bibitem{mikolov2013distributed}
T.~Mikolov, I.~Sutskever, K.~Chen, G.~S. Corrado, and J.~Dean, ``Distributed representations of words and phrases and their compositionality,'' \emph{Advances in Neural Information Processing Systems}, vol.~26, 2013.

\bibitem{pennington2014glove}
J.~Pennington, R.~Socher, and C.~D. Manning, ``Glove: Global vectors for word representation,'' in \emph{Proceedings of the 19th Conference on Empirical Methods in Natural Language Processing (EMNLP)}, 2014, pp. 1532--1543.

\bibitem{cer2018universal}
D.~Cer, Y.~Yang, S.-y. Kong, N.~Hua, N.~Limtiaco, R.~S. John, N.~Constant, M.~Guajardo-Cespedes, S.~Yuan, C.~Tar \emph{et~al.}, ``Universal sentence encoder,'' \emph{arXiv preprint arXiv:1803.11175}, 2018.

\bibitem{yang2022predictive}
Y.~Yang, X.~Xia, D.~Lo, T.~Bi, J.~Grundy, and X.~Yang, ``Predictive models in software engineering: Challenges and opportunities,'' \emph{ACM Transactions on Software Engineering and Methodology}, vol.~31, no.~3, pp. 1--72, 2022.

\bibitem{han2022data}
J.~Han, J.~Pei, and H.~Tong, \emph{Data Mining: Concepts and Techniques}.\hskip 1em plus 0.5em minus 0.4em\relax Morgan kaufmann, 2022.

\bibitem{treude2016augmenting}
C.~Treude and M.~P. Robillard, ``Augmenting {API} documentation with insights from {Stack Overflow},'' in \emph{Proceedings of the 38th International Conference on Software Engineering (ICSE)}, 2016, pp. 392--403.

\bibitem{minaee2021deep}
S.~Minaee, N.~Kalchbrenner, E.~Cambria, N.~Nikzad, M.~Chenaghlu, and J.~Gao, ``Deep learning--based text classification: a comprehensive review,'' \emph{ACM Computing Surveys}, vol.~54, no.~3, pp. 1--40, 2021.

\bibitem{guo2019improving}
B.~Guo, C.~Zhang, J.~Liu, and X.~Ma, ``Improving text classification with weighted word embeddings via a multi-channel textcnn model,'' \emph{Neurocomputing}, vol. 363, pp. 366--374, 2019.

\bibitem{kim2014cnn}
Y.~Kim, ``Convolutional neural networks for sentence classification,'' in \emph{Proceedings of the 19th Conference on Empirical Methods in Natural Language Processing (EMNLP)}, 2014, pp. 1746--1751.

\bibitem{liu2018recurrent}
T.~Liu, S.~Yu, B.~Xu, and H.~Yin, ``Recurrent networks with attention and convolutional networks for sentence representation and classification,'' \emph{Applied Intelligence}, vol.~48, pp. 3797--3806, 2018.

\bibitem{abadi2016tensorflow}
M.~Abadi, P.~Barham, J.~Chen, Z.~Chen, A.~Davis, J.~Dean, M.~Devin, S.~Ghemawat, G.~Irving, M.~Isard \emph{et~al.}, ``Tensorflow: A system for large-scale machine learning,'' in \emph{Proceedings of the 12th USENIX Symposium on Operating Systems Design and Implementation (OSDI)}, 2016, pp. 265--283.

\bibitem{devlin2018bert}
J.~Devlin, M.-W. Chang, K.~Lee, and K.~Toutanova, ``Bert: Pre-training of deep bidirectional transformers for language understanding,'' in \emph{Proceedings of Association for Computational Linguistics: Human Language Technologies (NAACL-HLT)}, 2018, pp. 4171--4186.

\bibitem{liu2019roberta}
Y.~Liu, M.~Ott, N.~Goyal, J.~Du, M.~Joshi, D.~Chen, O.~Levy, M.~Lewis, L.~Zettlemoyer, and V.~Stoyanov, ``Roberta: A robustly optimized bert pretraining approach,'' \emph{arXiv preprint arXiv:1907.11692}, 2019.

\bibitem{hurst2024gpt}
A.~Hurst, A.~Lerer, A.~P. Goucher, A.~Perelman, A.~Ramesh, A.~Clark, A.~Ostrow, A.~Welihinda, A.~Hayes, A.~Radford \emph{et~al.}, ``Gpt-4o system card,'' \emph{arXiv preprint arXiv:2410.21276}, 2024.

\bibitem{tanzil2024chatgpt}
M.~H. Tanzil, J.~Y. Khan, and G.~Uddin, ``Chatgpt incorrectness detection in software reviews,'' in \emph{Proceedings of the 46th IEEE/ACM International Conference on Software Engineering (ICSE)}, 2024, pp. 1--12.

\bibitem{ding2023parameter}
N.~Ding, Y.~Qin, G.~Yang, F.~Wei, Z.~Yang, Y.~Su, S.~Hu, Y.~Chen, C.-M. Chan, W.~Chen \emph{et~al.}, ``Parameter-efficient fine-tuning of large-scale pre-trained language models,'' \emph{Nature Machine Intelligence}, vol.~5, no.~3, pp. 220--235, 2023.

\bibitem{Li2025urc}
R.~Li, P.~Liang, Y.~Wang, Y.~Cai, W.~Sun, and Z.~Li, ``Unveiling the role of chatgpt in software development: Insights from developer-chatgpt interactions on github,'' \emph{arXiv preprint arXiv:2505.03901}, 2025.

\bibitem{uddin2019automatic}
G.~Uddin and F.~Khomh, ``Automatic mining of opinions expressed about apis in stack overflow,'' \emph{IEEE Transactions on Software Engineering}, vol.~47, no.~3, pp. 522--559, 2019.

\bibitem{tabassum2020code}
J.~Tabassum, M.~Maddela, W.~Xu, and A.~Ritter, ``Code and named entity recognition in stackoverflow,'' in \emph{In Proceedings of the 58th Annual Meeting of the Association for Computational Linguistics (ACL)}, 2020, pp. 4913--4926.

\bibitem{zhou2024large}
X.~Zhou, S.~Cao, X.~Sun, and D.~Lo, ``Large language model for vulnerability detection and repair: Literature review and the road ahead,'' \emph{ACM Transactions on Software Engineering and Methodology}, vol.~34, no.~5, pp. 1--31, 2025.

\bibitem{he2022ptm4tag}
J.~He, B.~Xu, Z.~Yang, D.~Han, C.~Yang, and D.~Lo, ``Ptm4tag: sharpening tag recommendation of stack overflow posts with pre-trained models,'' in \emph{Proceedings of the 30th IEEE/ACM International Conference on Program Comprehension (ICPC)}, 2022, pp. 1--11.

\bibitem{von2022validity}
J.~Von~der Mosel, A.~Trautsch, and S.~Herbold, ``On the validity of pre-trained transformers for natural language processing in the software engineering domain,'' \emph{IEEE Transactions on Software Engineering}, vol.~49, no.~4, pp. 1487--1507, 2022.

\bibitem{dos2014deep}
C.~Dos~Santos and M.~Gatti, ``Deep convolutional neural networks for sentiment analysis of short texts,'' in \emph{Proceedings of the 25th International Conference on Computational Linguistics (COLING)}, 2014, pp. 69--78.

\bibitem{young2018recent}
T.~Young, D.~Hazarika, S.~Poria, and E.~Cambria, ``Recent trends in deep learning based natural language processing,'' \emph{IEEE Computational Intelligence Magazine}, vol.~13, no.~3, pp. 55--75, 2018.

\bibitem{robillard2015recommending}
M.~P. Robillard and Y.~B. Chhetri, ``Recommending reference api documentation,'' \emph{Empirical Software Engineering}, vol.~20, no.~6, pp. 1558--1586, 2015.

\bibitem{li2018improving}
H.~Li, S.~Li, J.~Sun, Z.~Xing, X.~Peng, M.~Liu, and X.~Zhao, ``Improving api caveats accessibility by mining api caveats knowledge graph,'' in \emph{Proceedings of the IEEE International Conference on Software Maintenance and Evolution (ICSME)}, 2018, pp. 183--193.

\bibitem{iyyer2014neural}
M.~Iyyer, J.~Boyd-Graber, L.~Claudino, R.~Socher, and H.~Daum{\'e}~III, ``A neural network for factoid question answering over paragraphs,'' in \emph{Proceedings of the 19th Conference on Empirical Methods in Natural Language Processing (EMNLP)}, 2014, pp. 633--644.

\bibitem{unger2012template}
C.~Unger, L.~B{\"u}hmann, J.~Lehmann, A.-C. Ngonga~Ngomo, D.~Gerber, and P.~Cimiano, ``Template-based question answering over rdf data,'' in \emph{Proceedings of the 21st International Conference on World Wide Web (WWW)}, 2012, pp. 639--648.

\bibitem{zhong2020extractive}
M.~Zhong, P.~Liu, Y.~Chen, D.~Wang, X.~Qiu, and X.~Huang, ``Extractive summarization as text matching,'' in \emph{Proceedings of the 58th Annual Meeting of the Association for Computational Linguistics (ACL)}, 2020, pp. 6197--6208.

\bibitem{rodrigues2014sequence}
F.~Rodrigues, F.~Pereira, and B.~Ribeiro, ``Sequence labeling with multiple annotators,'' \emph{Machine Learning}, vol.~95, pp. 165--181, 2014.

\bibitem{xu2020review}
J.~Xu, W.~Zuo, S.~Liang, and X.~Zuo, ``A review of dataset and labeling methods for causality extraction,'' in \emph{Proceedings of the 28th International Conference on Computational Linguistics (COLING)}, 2020, pp. 1519--1531.

\bibitem{parveen2015topical}
D.~Parveen, H.-M. Ramsl, and M.~Strube, ``Topical coherence for graph-based extractive summarization,'' in \emph{Proceedings of the 20th International Conference on Empirical Methods in Natural Language Processing (EMNLP)}, 2015, pp. 1949--1954.

\bibitem{mchugh2012interrater}
M.~L. McHugh, ``Interrater reliability: the kappa statistic,'' \emph{Biochemia Medica}, vol.~22, no.~3, pp. 276--282, 2012.

\bibitem{di2015development}
A.~Di~Sorbo, S.~Panichella, C.~A. Visaggio, M.~Di~Penta, G.~Canfora, and H.~C. Gall, ``Development emails content analyzer: Intention mining in developer discussions (t),'' in \emph{Proceedings of the 30th IEEE/ACM International Conference on Automated Software Engineering (ICSE)}, 2015, pp. 12--23.

\bibitem{chatterjee2021automatic}
P.~Chatterjee, K.~Damevski, and L.~Pollock, ``Automatic extraction of opinion-based q\&a from online developer chats,'' in \emph{Preedings of the 43rd IEEE/ACM International Conference on Software Engineering (ICSE)}, 2021, pp. 1260--1272.

\bibitem{erkan2004lexrank}
G.~Erkan and D.~R. Radev, ``Lexrank: Graph-based lexical centrality as salience in text summarization,'' \emph{Journal of Artificial Intelligence Research}, vol.~22, pp. 457--479, 2004.

\bibitem{liu2019fine}
Y.~Liu, ``Fine-tune bert for extractive summarization,'' \emph{arXiv preprint arXiv:1903.10318}, 2019.

\bibitem{mihalcea2004textrank}
R.~Mihalcea and P.~Tarau, ``Textrank: Bringing order into text,'' in \emph{Proceedings of the 9th Internaltional Conference on Empirical Methods in Natural Language Processing (EMNLP)}, 2004, pp. 404--411.

\bibitem{haiduc2013automatic}
S.~Haiduc, G.~Bavota, A.~Marcus, R.~Oliveto, A.~De~Lucia, and T.~Menzies, ``Automatic query reformulations for text retrieval in software engineering,'' in \emph{Proceedings of the 35th International Conference on Software Engineering (ICSE)}, 2013, pp. 842--851.

\bibitem{luhn1958automatic}
H.~P. Luhn, ``The automatic creation of literature abstracts,'' \emph{Journal of Research and Development}, vol.~2, no.~2, pp. 159--165, 1958.

\bibitem{huang2018automating}
Q.~Huang, X.~Xia, D.~Lo, and G.~C. Murphy, ``Automating intention mining,'' \emph{IEEE Transactions on Software Engineering}, vol.~46, no.~10, pp. 1098--1119, 2018.

\bibitem{haykin1994neural}
S.~Haykin, \emph{Neural Networks: A Comprehensive Foundation}.\hskip 1em plus 0.5em minus 0.4em\relax Prentice Hall PTR, 1994.

\bibitem{brin1998pagerank}
L.~Page and S.~Brin, ``The pagerank citation ranking: bringing order to the web,'' in \emph{Proceedings of the 7th International World Wide Web Conference (WWW)}, 1998, pp. 161--172.

\bibitem{uddin2017automatic}
G.~Uddin and F.~Khomh, ``Automatic summarization of api reviews,'' in \emph{Proceedings of the 32nd IEEE/ACM International Conference on Automated Software Engineering (ICSE)}, 2017, pp. 159--170.

\bibitem{narayan2018ranking}
S.~Narayan, S.~B. Cohen, and M.~Lapata, ``Ranking sentences for extractive summarization with reinforcement learning,'' \emph{arXiv preprint arXiv:1802.08636}, 2018.

\bibitem{zhou2018neural}
Q.~Zhou, N.~Yang, F.~Wei, S.~Huang, M.~Zhou, and T.~Zhao, ``Neural document summarization by jointly learning to score and select sentences,'' \emph{arXiv preprint arXiv:1807.02305}, 2018.

\bibitem{sandhaus2008new}
E.~Sandhaus, ``The new york times annotated corpus,'' \emph{Linguistic Data Consortium, Philadelphia}, vol.~6, no.~12, p. e26752, 2008.

\bibitem{documentsummarization2022}
``Document summarization on cnn/daily mail,'' 2022, retrieved March 29, 2022, from \url{https://paperswithcode.com/sota/document-summarization-on-cnn-daily-mail}.

\bibitem{kou2023automated}
B.~Kou, M.~Chen, and T.~Zhang, ``Automated summarization of stack overflow posts,'' in \emph{Peoceedings of the 45th IEEE/ACM International Conference on Software Engineering (ICSE)}, 2023, pp. 1853--1865.

\bibitem{lan2023btlink}
J.~Lan, L.~Gong, J.~Zhang, and H.~Zhang, ``Btlink: automatic link recovery between issues and commits based on pre-trained bert model,'' \emph{Empirical Software Engineering}, vol.~28, no.~4, p. Article No. 103, 2023.

\bibitem{soliman2018improving}
M.~Soliman, A.~R. Salama, M.~Galster, O.~Zimmermann, and M.~Riebisch, ``Improving the search for architecture knowledge in online developer communities,'' in \emph{Proceedings of the 15th IEEE International Conference on Software Architecture (ICSA)}, 2018, pp. 186--18\,609.

\bibitem{gao2020technical}
Z.~Gao, X.~Xia, D.~Lo, and J.~Grundy, ``Technical q8a site answer recommendation via question boosting,'' \emph{ACM Transactions on Software Engineering and Methodology}, vol.~30, no.~1, pp. 1--34, 2020.

\bibitem{de2022developerssearch}
M.~J. de~Dieu, P.~Liang, and M.~Shahin, ``How do developers search for architectural information? an industrial survey,'' in \emph{Proceedings of the 19th IEEE International Conference on Software Architecture (ICSA)}, 2022, pp. 58--68.

\bibitem{soliman2021exploratory}
M.~Soliman, M.~Galster, and P.~Avgeriou, ``An exploratory study on architectural knowledge in issue tracking systems,'' in \emph{Processings of the European Conference on Software Architecture}, 2021, pp. 117--133.

\bibitem{shi2020detection}
L.~Shi, M.~Xing, M.~Li, Y.~Wang, S.~Li, and Q.~Wang, ``Detection of hidden feature requests from massive chat messages via deep siamese network,'' in \emph{Proceedings of the 42nd ACM/IEEE International Conference on Software Engineering}, 2020, pp. 641--653.

\bibitem{rodeghero2017detecting}
P.~Rodeghero, S.~Jiang, A.~Armaly, and C.~McMillan, ``Detecting user story information in developer-client conversations to generate extractive summaries,'' in \emph{Proceedings of the 39th IEEE/ACM International Conference on Software Engineering}, 2017, pp. 49--59.

\bibitem{fleiss1981measurement}
J.~L. Fleiss, B.~Levin, M.~C. Paik \emph{et~al.}, ``The measurement of interrater agreement,'' \emph{Statistical Methods for Rates and Proportions}, vol.~2, no. 212-236, pp. 22--23, 1981.

\bibitem{peters2017text}
F.~Peters, T.~T. Tun, Y.~Yu, and B.~Nuseibeh, ``Text filtering and ranking for security bug report prediction,'' \emph{IEEE Transactions on Software Engineering}, vol.~45, no.~6, pp. 615--631, 2017.

\bibitem{yang2022answer}
C.~Yang, B.~Xu, F.~Thung, Y.~Shi, T.~Zhang, Z.~Yang, X.~Zhou, J.~Shi, J.~He, D.~Han \emph{et~al.}, ``Answer summarization for technical queries: Benchmark and new approach,'' in \emph{Proceedings of the 37th IEEE/ACM International Conference on Automated Software Engineering (ASE)}, 2022, pp. 1--13.

\bibitem{Sorbo2016}
A.~Di~Sorbo, S.~Panichella, C.~V. Alexandru, J.~Shimagaki, C.~A. Visaggio, G.~Canfora, and H.~C. Gall, ``What would users change in my app? summarizing app reviews for recommending software changes,'' in \emph{Proceedings of the 24th ACM SIGSOFT International Symposium on Foundations of Software Engineering}, 2016, p. 499–510.

\bibitem{madhyastha2019model}
P.~Madhyastha and A.~Jain, ``On model stability as a function of random seed,'' in \emph{Proceedings of the 23rd Conference on Computational Natural Language Learning (CoNLL)}.\hskip 1em plus 0.5em minus 0.4em\relax Association for Computational Linguistics, 2019, pp. 929--939.

\end{thebibliography}

%%
% %% If your work has an appendix, this is the place to put it.
% \appendix









\end{document}